\documentclass[a4paper,11pt,twoside]{report}
\usepackage{latexsym,fancyheadings,amsmath,amsfonts,wrapfig,epsfig,fancybox,amssymb}
\usepackage[small,it]{caption}
\usepackage[latin1]{inputenc}

\newcommand{\D}{\ensuremath{{\not\negthickspace D}}}
\newcommand{\pa}{\ensuremath{\!{\not\!\partial}}}
\newcommand{\p}{\ensuremath{\!{\not\! p}}}
\newcommand{\q}{\ensuremath{\!{\not\! q}}}
\newcommand{\K}{\ensuremath{\!{\not\! k}}}
\newcommand{\A}{\ensuremath{\!{\not\negthickspace A}}}

\title{\textbf{\huge{(4+1)-Dimensional Quantum Hall Effect\\
       \& \\
       Applications to Cosmology}}\\[2cm]
       {Diplomarbeit}\\[2cm]
       {Philipp Werner}\\[6cm]
       {\normalsize{Ausgeführt an der ETH Zürich bei Prof. Dr. J. Fröhlich}\\[2cm]
       {Februar 2000}}}
\date{}
\begin{document}

\thispagestyle{empty}
\maketitle

\vspace{3cm}


\newpage
\thispagestyle{empty}
\textit{\\[2cm]}
\newpage
\thispagestyle{empty}

\vspace{5cm}
\textit{\\[2.2cm]Ich möchte an dieser Stelle Herrn Prof. J. Fröhlich herzlich danken, dass er mir die Möglichkeit geboten hat, an einem faszinierenden Forschungsprojekt mitzuwirken. Ich bin froh, dass Herr Fröhlich trotz grosser zeitlicher Belastung bereit war, diese Diplomarbeit zu betreuen und bedanke mich für die zahlreichen Besprechungen von hohem Informations- und nicht geringem Unterhaltungswert.\\ 
Ferner gilt mein Dank Herrn B. Pedrini, dessen Einführungen in die Quantenfeldtheorie und Differenzialgeometrie mir den Start zu dieser Diplomarbeit erleichterten, sowie Herrn Dr. N. Macris für dessen Bereitschaft, die Arbeit als Verantwortlicher seitens der ETH Lausanne zu begleiten.}

\newpage

\tableofcontents

\newpage

\chapter{Introduction and summary of the results}

Today, magnetic fields are present throughout the universe and play an important role in a multitude of astrophysical situations. Our Galaxy and many other spiral Galaxies are endowed with magnetic fields which are dynamically important \cite{Turner}. How do these cosmic magnetic fields arise? Many astrophysicists believe that galactic magnetic fields are generated and maintained by some non-linear dynamo-mechanism, whereby the energy associated with the rotation of spiral galaxies is converted into magnetic field energy \cite{Parker}. The dynamo mechanism only serves as a means of amplification, so that the latter scenario requires the presence of seed magnetic fields, whose origin today is still uncertain.

\begin{figure}[htbp]
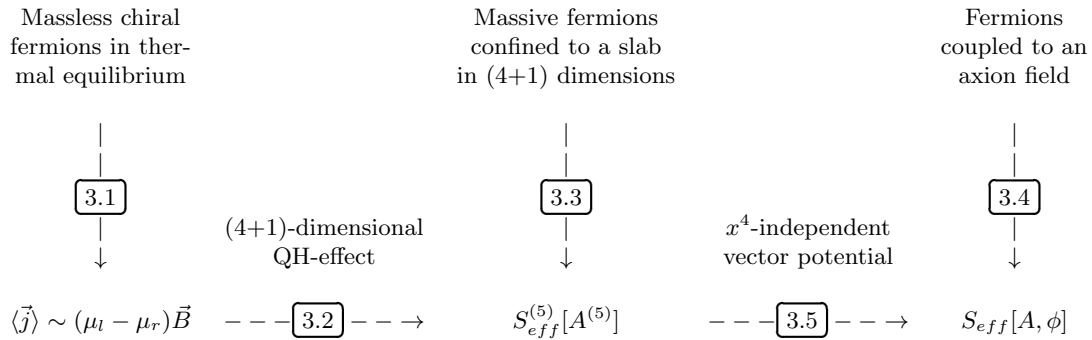

\centering
\footnotesize
\begin{tabular}{ccccc}
Massless chiral&&Massive fermions&&Fermions\\
fermions in ther-&&confined to a slab&&coupled to an\\
mal equilibrium&&in (4+1) dimensions&&axion field\\
&&&&\\
$\arrowvert$&&$\arrowvert$&&$\arrowvert$\\
$\arrowvert$&&$\arrowvert$&&$\arrowvert$\\
$\Ovalbox{3.1}$&&$\Ovalbox{3.3}$&&$\Ovalbox{3.4}$\\
$\arrowvert$&(4+1)-dimensional&$\arrowvert$&$x^4$-independent&$\arrowvert$\\
$\downarrow$&QH-effect&$\downarrow$&vector potential&$\downarrow$\\
&&&&\\
$\langle\vec{j}\rangle\sim(\mu_l-\mu_r)\vec{B}$&$---\Ovalbox{3.2}--\rightarrow$&$S_{eff}^{(5)}[A^{(5)}]$&$---\Ovalbox{3.5}--\rightarrow$&$S_{eff}[A,\phi]$\\
\end{tabular}
\vspace{5mm}
\caption{Illustration of the structure of chapter 3, showing the different models discussed and how they are interrelated.}
\label{Schema}
\end{figure}

This diploma thesis is devoted to the study of models and mechanisms which could explain instabilities towards the generation of such seed magnetic fields. The structure of this report is the following: In chapter 2 we discuss the chiral (abelian) anomaly and related questions. The purpose of these developments is to lay the theoretical ground work for the calculations in chapter 3. The latter one constitutes the main body of this report and is divided into five sections. In sections 3.1, 3.3 and 3.4 we present different mechanisms for producing seed fields and derive equations of motion, while sections 3.2 and 3.5 will shed some light upon the relations between the individual approaches (see figure \ref{Schema}).

The calculation in section 3.1 assumes that the early universe is filled with a hot plasma of charged fermions, whose mass may be neglected, so that chirality is approximately conserved. This plasma is supposed to be in a state of thermal equilibrium and we shall admit that there existed a slight asymmetry in the chemical potentials $\mu_l$ and $\mu_r$ corresponding to fermions of left and right chirality. One may then use a formula - derived by A. Y. Alekseev, V. V. Cheianov and J. Fröhlich in 1998, \cite{Alekseev} - which relates the expectation value for the electric current $\vec{j}=\vec{j}_l+\vec{j}_r$ to the magnetic field $\vec{B}$ and the difference in chemical potentials $\mu_l-\mu_r$:

\begin{equation}
\langle \vec{j}\rangle_{\mu_l,\mu_r}=-\frac{1}{4\pi^2}(\mu_l-\mu_r)\vec{B}.
\label{1.1}
\end{equation}

By substituting expression (\ref{1.1}) for the current into Maxwells equations and making some simple assumption concerning  the charge density, such as $\langle j^0\rangle_{\mu_l,\mu_r}=0$, we obtain a system of equations which can be solved by means of Fourier transformation. One finds that the modes $\vec{b}(t,\vec{k})=\int d^3xe^{-i\vec{k}\cdot\vec{x}}\vec{B}(t,\vec{x})$ for

\begin{equation}
|\vec{k}|<\frac{\alpha}{4\pi^2}|\mu_l-\mu_r|
\label{1.2}
\end{equation}

grow exponentially with time ($\alpha=e^2$ is the feinstructure constant, we shall set $\hbar=c=1$ througout this report). Hence an equilibrium state with $\mu_l\ne\mu_r$ is unstable with respect to the generation of (electro-)magnetic fields. The reason, why we always talk about the generation of \textit{magnetic} fields is that large electric fields rapidly die out if dissipative processes are allowed for \cite{J-F}. 

Some of the hypotheses underlying the calculations of section 3.1 seem quite unnatural. Chirality is not really conserved, since fermions are massive. The very early universe is not really an equilibrium state and the chemical potentials of left- and righthanded fermions neither have an unambiguous meaning, nor would they be space- and time-independent.

In section 3.2 we will show a way out of these difficulties. Based on an analogy with the quantum Hall effect we develop a (4+1)-dimentional theory which leads to an effective action functional

\begin{equation}
S_{eff}^{(5)}[A^{(5)}]=S_{EM}^{(5)}[A^{(5)}]-S_{CS}^{(5)}[A^{(5)}]+\Gamma_{\partial\Lambda}[A^{(5)}|_{\partial_\Lambda}].
\label{1.3}
\end{equation}

In the above formula, $S_{EM}^{(5)}[A^{(5)}]$ denotes the (4+1)-dimensional analogue of the Maxwell term, $S_{CS}^{(5)}[A^{(5)}]$ is proportional to the five dimensional Chern-Simons action and the boundary term $\Gamma_{\partial\Lambda}[A^{(5)}|_{\partial_\Lambda}]$ must be introduced in order to assure the gauge invariance of the effective action. The equations of motion derived from (\ref{1.3}) generalise those obtained in section 3.1. In particular they yield an equation for the time evolution of the (now space-time dependent) difference in chemical potentials.

While the introduction of a fourth space dimension in section 3.2 could be regarded as a mathematical ``trick'', we take the (4+1)-dimensional point of view more seriously in the following section and imagine that there exists a slab $\Lambda$ of with $L$ in (4+1)-dimensional Minkowski space-time. This slab is filled with massive, (4+1)-dimensional fermions and the latter ones are coupled to an external vector potential $A^{(5)}$. Choosing a step function for the $x^4$-dependent mass one should then be able to derive the effective action (\ref{1.3}) up to terms $O(\frac{1}{m})$, where $m$ is the fermion mass. The massive bulk modes produce the Chern-Simons term, whereas $\Gamma_{\partial\Lambda}[A^{(5)}|_{\partial_\Lambda}]$ corresponds to the effective action for the massless, chiral boundary modes identified with the (3+1)-dimensional left- and right-handed fermions filling the early universe.

The calculation of $\text{det}(\D_A^{(5)}+im(x^4))$ for an $x^4$-dependent mass is rather involved. We shall therefore ignore the existence of domain walls in sections 3.3.1 to 3.3.5 and calculate the fermion determinant for a constant mass $m$. To this end one has to compute one-loop Feynman diagrams with $n=1,\ldots,5$ vertices. The vacuum polarisation graph ($n=2$) yields a contribution which may be absorbed into the Maxwell term $S_{EM}$ by an appropriate redefinition of the bare coupling constant (charge renormalization), whereas the triangle graph ($n=3$) is shown to produce a Chern-Simons term. The two remaining potentially divergent diagrams ($n=4$ and $n=5$) will not be considered. We end section 3.3 by showing how chiral boundary modes occur in the presence of domain walls.

Equations of motion similar to the ones derived from the (4+1)-dimensional theory are obtained by coupling (3+1)-dimensional fermions to an \textit{axion field}. The axion field \mbox{approach}, which appears to be the the most satisfactory, is presented in section 3.4 and yields the effective action

\begin{equation}
S_{eff}[A,\phi]=S_{EM}[A]-\frac{l}{32\pi^2}\int\phi(F\wedge F)+\int d^4x\frac{1}{2}(\partial^\mu\phi)(\partial_\mu\phi)+W[A],
\label{1.4}
\end{equation}

where $\phi$ denotes the axion field, $l$ some length parameter and $W[A]=-i\ln\text{det}(\D_A)$ is the fermionic effective action.

The action functional (\ref{1.4}) is equivalent to (\ref{1.3}) in the case of an $x^4$-independent vector potential $A^{(5)}=(A_0,\ldots,A_4)$. This can be shown by setting 

\begin{equation}
A_4=\phi,
\label{1.5}
\end{equation}

as will be done in section 3.5. The above identity and the relation $E_4=\partial_0A_4\sim(\mu_l-\mu_r)$ suggested by the analogy with the quantum Hall effect yield an interpretation for the axion field: \textit{Its time derivative plays the role of a space-time dependent ``difference in chemical potentials''} between fermions of left and right chirality.     

Another advantage of the axion field approach is that it does no longer rely on any implausible assumptions, such as ``masslessness'' of the fermions or a universe in thermal equilibrium. One can replace $W[A]$ in (\ref{1.4}) by the effective action for massive fermions.

Since the equations of motion derived in sections 3.1 and 3.2 may also be obtained from the axion field theory, we will restrict our attention to the system of equations involving the axion. Of course, the study of these highly \textit{non-linear} equations is a difficult task and the results presented in chapter 4 are only a beginning. We shall try to find special solutions and gain some insights into the dynamics by linearising the system of equations around these special solutions. 

The form of instability which seems the most appealing to us is the growth of electromagnetic fields by \textit{parametric resonance}. The latter mechanism requires an oscillating axion field and is made possible by the presence of a periodic \textit{axionic potential} $U[\phi]$. Such an additional term in the effective action (\ref{1.4}) is obtained by evaluating the path integral over $A$ using a semi-classical expansion based on the stationary phase method. The potential $U[\phi]$ might also be useful for finding nontrivial special solutions of finite energy.

All the calculations in chapters 3 and 4 are based on the assumption that space-time is flat. Physically more relevant results would probably be obtained by considering an expanding Friedman-Robertson-Walker universe. While the transcription of our equations of motion to the latter model does not appear to present much difficulties, time has not permitted us to analyse this new situation and we shall renounce entering into a discussion of this subject.

\chapter{The chiral anomaly}

\section{Massless fermions and conservation of chirality at the classical level}

Before turning to physics, let us present a certain number of results related to the chiral (abelian) anomaly - the phenomenon which is at the origin of the mechanisms for producing large scale magnetic fields proposed in chapter 3.

The purpose of this first section is to review some basic notions and to fix the notation. We consider fermions in (3+1) dimensions which are described by four-component spinor fields $\psi(x)$. For free particles of mass $m$ the spinor satisfies the Dirac equation

\begin{equation}
(i\gamma^\mu\partial_\mu-m)\psi=0,
\label{2.1}
\end{equation}

where the $\gamma^\mu$, $\mu=0,\ldots,3$ are a set of $4\times 4$ matrices satisfying $\{\gamma^\mu,\gamma^\nu\}=2g^{\mu\nu}$. For the metric in (3+1)-dimensional Minkowski space we choose $g=\text{diag}(1,-1,-1,-1)$. In the \textit{chiral representation}, the $\gamma$-matrices are

\begin{equation}
\gamma^0=\left(\begin{array}{cc}0&I\\I&0\end{array}\right),\hspace{.5cm}\gamma^i=\left(\begin{array}{cc}0&\sigma^i\\-\sigma^i&0\end{array}\right),\hspace{.5cm}\gamma^5\equiv i\gamma^0\gamma^1\gamma^2\gamma^3=\left(\begin{array}{cc}-I&0\\0&I\end{array}\right).
\label{2.2}
\end{equation}

$I\equiv\sigma^0$ is the $2\times 2$ unit matrix and the $\sigma^i$'s denote the Pauli matrices. The four-component field $\psi$ may then be written as a bispinor in terms of two-component spinors $u$ and $v$

\begin{equation}
\psi=\left(\begin{array}{c}u\\v\end{array}\right),
\label{2.3}
\end{equation}

and one can introduce the projectors $\gamma_l$ and $\gamma_r$ such that $\gamma_l\psi=u$ and $\gamma_r\psi=v$:

\begin{equation}
\gamma_l=\frac{1}{2}(1-\gamma^5),\hspace{.5cm}\gamma_r=\frac{1}{2}(1+\gamma^5).
\label{2.4}
\end{equation}

The reason for choosing the subscripts $l$ (=left) and $r$ (=right) will emerge shortly. In momentum space the Dirac equation (\ref{2.1}) reads $(\gamma^\mu p_\mu-m)\psi=0$, or more explicitly

\begin{equation}
\left(\begin{array}{cc}0&p\cdot\sigma\\\hat p\cdot\sigma&0\end{array}\right)\psi=m\psi,
\label{2.5}
\end{equation}

where $p\cdot\sigma=p_\mu\sigma^\mu$, $\hat p\cdot\sigma=\hat p_\mu\sigma^\mu$ and $p=(p^0,\vec{p})$, $\hat p=(p^0,-\vec{p})$. If the fermions are massless, then (\ref{2.5}) decouples into two independent equations for $u$ and $v$:

\begin{equation}
\left\{\begin{array}{ccc}
(\hat p\cdot \sigma)u&=&0\\
(p\cdot \sigma)v&=&0
\end{array}\right.
\hspace{5mm}\Longleftrightarrow\hspace{5mm}
\left\{\begin{array}{ccc}
(\vec p\cdot \vec\sigma)u&=&-p_0u\\
(\vec p\cdot \vec\sigma)v&=&p_0v
\end{array}\right.
\label{2.6}
\end{equation}

Since $p_0=|\vec p|$ for massless (positive energy) particles and $(\vec p\cdot \vec\sigma)$ is the helicity operator, we conclude that $u$ and $v$ describe fermions of left and right chirality respectively. Furthermore chirality is conserved, since the dynamics of $u$ and $v$ is decoupled.

We now couple the massless Dirac fermions to an external electromagnetic field. The system obtained thereby is described by the Lagrangian density 

\begin{equation}
\mathcal{L}=-\frac{1}{4\alpha}F^{\mu\nu}F_{\mu\nu}+\bar\psi i\gamma^\mu(\partial_\mu-iA_\mu)\psi,
\label{2.7}
\end{equation}

where $-\frac{1}{4\alpha}F^{\mu\nu}F_{\mu\nu}$ is the kinetic energy term for the gauge field and $\bar\psi=\psi^\dag\gamma^0$ denotes the conjugate spinor field. In fact the Euler-Lagrange equations of motion derived from (\ref{2.7}) are $\partial_\mu F^{\mu\nu}=\alpha\bar\psi\gamma^\nu\psi$ and $i\gamma^\mu(\partial_\mu-iA_\mu)\psi=0$, in which we recognise the Maxwell-Dirac electrodynamics. 

According to Noether's theorem, the invariance of a Lagrangian $\mathcal{L}$ under global infinitesimal transformations implies the existence of a conserved current. Given the field transformation

\begin{equation}
\psi(x)\rightarrow\psi(x)+\delta\psi(x),\hspace{.5cm}\delta\psi(x)=f(x)\delta\alpha
\label{2.8}
\end{equation}

depending on the infinitesimal parameter $\delta\alpha$, this current is 

\begin{equation}
j^\mu(x)=\frac{\partial L}{\partial(\partial_\mu\psi)}\frac{\delta\psi}{\delta a},
\label{2.9}
\end{equation}

and the charge $Q=\int d^3x j^0$ is then the generator of the transformation (\ref{2.8}) through the Poisson bracket operation

\begin{equation}
{\{Q,\psi\}}_{PB}=\frac{\delta\psi}{\delta a}.
\label{2.10}
\end{equation}

The Lagrangian density $\mathcal{L}$ written in (\ref{2.7}) is invariant under the local gauge transformation

\begin{equation}
\psi(x)\rightarrow e^{i\alpha(x)}\psi(x),\hspace{0.5cm}\bar\psi(x)\rightarrow e^{-i\alpha(x)}\bar\psi(x),\hspace{0.5cm}A\rightarrow A+d\alpha.
\label{2.11}
\end{equation}

This invariance - for constant $\alpha$ - leads to the conserved current

\begin{equation}
j^\mu=\frac{\partial L}{\partial(\partial_\mu\psi)}\Big(-i\frac{\delta\psi}{\delta \alpha}\Big)=\bar\psi\gamma^\mu\psi.
\label{2.12}
\end{equation}

In addition, the \textit{massless} theory is invariant under local chiral rotations

\begin{equation}
\psi(x)\rightarrow e^{i\theta(x)\gamma^5}\psi(x),\hspace{0.5cm}\bar\psi(x)\rightarrow \bar\psi(x) e^{i\theta(x)\gamma^5},\hspace{0.5cm}A\rightarrow A+\gamma^5d\theta.
\label{2.13}
\end{equation}

In particular, if $\theta$ is constant the transformations (\ref{2.13}) are a symmetry of the classical system and the corresponding conserved current - called \textit{axial current} - is

\begin{equation}
j_5^\mu=\frac{\partial L}{\partial(\partial_\mu\psi)}\Big(-i\frac{\delta\psi}{\delta \theta}\Big)=\bar\psi\gamma^\mu\gamma^5\psi.
\label{2.14}
\end{equation}

\newpage

\section{Anomalies from the path integral in Euclidean space}

Anomalies occur when the quantum mechanical vacuum functional of a field theory fails to have all the symmetries of the classical Lagrangian from which it is derived. The example of interest to us concerns local chiral rotations (\ref{2.13}), which do not leave quantum mechanical transition amplitudes invariant. As a consequence the axial current $\langle j_5^\mu\rangle=\langle\bar{\psi}\gamma^\mu\gamma^5\psi\rangle$ is not conserved for arbitrary external electromagnetic fields. This phenomenon is called the \textit{chiral anomaly}.

In this section we discuss a technique of anomaly calculation using path integrals in Euclidean space, the so called Fujikawa method \cite{Fujikawa} (a useful review on the subject can also be found in \cite{A-G}). As we shall see, the chiral anomaly can be understood as a non-invariance of the path integral measure under local chiral transformations on the fermion fields.

It was mentioned in the previous section that the Lagrangian density $\mathcal{L}=\bar\psi i\gamma^\mu(\partial_\mu-iA_\mu)\psi$ leads to the Dirac equation for massless fermions coupled to an external gauge field $A$. The fermion effective action functional $S_{eff}[A]$ is obtained by performing the integral over the fermion fields

\begin{equation}
e^{iS_{eff}[A]}=\int \mathcal D\psi \mathcal D\bar\psi e^{i\int d^4x\bar\psi i\gamma^\mu(\partial_\mu-iA_\mu)\psi},\hspace{5mm}\mu=0,\ldots,3.
\label{2.15}
\end{equation}

$S_{eff}[A]$ is the generating functional for the connected Greens functions of the vector currents

\begin{equation}
\langle j^{\mu_1}(x_1)\ldots j^{\mu_n}(x_n)\rangle_A^c=(-i)\frac{\delta}{\delta A_{\mu_1}(x_1)}\ldots(-i)\frac{\delta}{\delta A_{\mu_n}(x_n)}S_{eff}[A].
\label{2.16}
\end{equation}

After the Wick rotation

\begin{equation}
x^0=ix^4,\hspace{0.5cm}A_0=-iA_4,\hspace{0.5cm}\gamma^0=i\gamma^4
\label{2.17}
\end{equation}

the metric becomes $g=\text{diag}(-1,-1,-1,-1)$. This is somewhat unusual and we prefer to redefine the $\gamma$-matrices according to 

\begin{equation}
\gamma^\mu\rightarrow -i\gamma^\mu
\label{2.18}
\end{equation}

in order to obtain $g=\text{diag}(1,1,1,1)$. The integral for the effective action in Euclidean space then reads

\begin{equation}
e^{-S_{eff}^E[A]}=\int \mathcal D\psi \mathcal D\bar\psi e^{-\int d^4x\bar\psi\gamma^\mu(\partial_\mu-iA_\mu)\psi},
\label{2.19}
\end{equation}

and the Euclidean $\gamma$-matrices $\gamma^\mu$, $\mu=1,\ldots,4$ are hermitian, so that 

\begin{align}
i\D_A^E &\equiv i\gamma^\mu(\partial_\mu-iA_\mu)
\label{2.20}
\end{align}

is a hermitian operator with real eigenvalues. Under the local chiral rotations

\begin{equation}
\psi'(x)=e^{i\alpha(x)\gamma^5}\psi(x),\hspace{5mm}\bar\psi'(x)=\bar\psi(x) e^{i\alpha(x)\gamma^5}
\label{2.21}
\end{equation}

the exponent in (\ref{2.19}) is transformed for infinitesimal $\alpha(x)$ as

\begin{align}
\int d^4x\bar\psi\D_A^E \psi=\int d^4x\bar\psi'\D_A^E \psi'+i\int d^4x\alpha(x)\partial_\mu {j'}_5^\mu+O(\alpha^2).
\label{2.22}
\end{align}

Since $\gamma^5\equiv i\gamma^0\gamma^1\gamma^2\gamma^3=\gamma^1\gamma^2\gamma^3\gamma^4$ anticommutes with $\gamma^\mu$, $\mu=1,\ldots,4$ the axial current ${j}_5^\mu$ is invariant under chiral rotations (\ref{2.21}):
${j'}_5^\mu=\bar{\psi'}\gamma^\mu\gamma^5\psi'=\bar{\psi}\gamma^\mu\gamma^5\psi={j}_5^\mu$. The rule for the transformation of the measure $\mathcal D\psi \mathcal D\bar\psi$ in terms of the standard Jacobian $J$ reads \cite{Zuber}

\begin{equation}
\mathcal D\psi \mathcal D\bar\psi=J^{-1}\mathcal D\psi'\mathcal D\bar\psi'. 
\label{2.23}
\end{equation}

Note that for Grassmann fields, the Jacobian appears inverted. The latter one can be calculated by means of the formula

\begin{align}
\text{det}X=e^{\text{trln}X},
\label{2.24}
\end{align}

where $X$ denotes some operator. In our case the operators involved are 
those
which perform the chiral transformation (\ref{2.21}) on the fermion fields $\psi$ and $\bar\psi$. The Jacobian $J$ is therefore 

\begin{align}
J=e^{-2i\text{tr}\alpha\gamma^5}.
\label{2.25}
\end{align}

If we assume that $i\D_A^E$ possesses a discrete spectrum and introduce the eigenfunctions $\psi_n$ corresponding to the eigenvalues $\lambda_n$,

\begin{align}
i\D_A^E \psi_n&=\lambda_n\psi_n,
\label{2.26}
\end{align}

then the trace in (\ref{2.25}) can be written as

\begin{align}
\text{tr}\alpha\gamma^5&=\sum_n\langle\psi_n|\alpha\gamma^5|\psi_n\rangle=\int d^4x\alpha(x)\mathcal{A}(x).
\label{2.27}
\end{align}

The function $\mathcal{A}(x)$ appearing in the above formula explicitly reads

\begin{align}
\mathcal{A}(x)&=\sum_n\psi_n^\dag(x)\gamma^5\psi_n(x)
\label{2.28}
\end{align}

and is called the \textit{anomaly}. From (\ref{2.22}), (\ref{2.25}) and (\ref{2.27}) one obtains

\begin{align}
\int \mathcal D\psi \mathcal D\bar\psi e^{-\int d^4x \bar\psi \not D_A^E \psi}&=\int \mathcal D\psi' \mathcal D\bar\psi' e^{-\int d^4x\{ \bar\psi' \not D_A^E \psi'+i\alpha(x)[\partial_\mu{j'}_5^\mu-2\mathcal{A}]\}}.
\label{2.29}
\end{align}

The right hand side of (\ref{2.29}) can be expanded to first order in $\alpha$ and the entire equation divided by $\int \mathcal D\psi \mathcal D\bar\psi e^{-\int d^4x \bar\psi \not D_A^E \psi}$:

\begin{align}
1&=1-i\frac{\int \mathcal D\psi' \mathcal D\bar\psi' \int d^4x\alpha(x)(\partial_\mu {j'}_5^\mu-2\mathcal{A})e^{-\int d^4x \bar\psi' \not D_A^E \psi'}}{\int \mathcal D\psi \mathcal D\bar\psi e^{-\int d^4x \bar\psi \not D_A^E \psi}}+O(\alpha^2).
\label{2.30}
\end{align}

 Using the invariance of the path integral under the change of variable as well as ${j'_5}^\mu=j_5^\mu$, this finally yields

\begin{align}
\partial_\mu \langle j_5^\mu\rangle_A&=2\mathcal{A}.
\label{2.31}
\end{align}

$\mathcal{A}(x)$ as it stands in (\ref{2.28}) is an ill-defined quantity. We may evaluate it by regularizing the large eigenvalues $\lambda_n$ (which are real, since $i\D_A^E $ is hermitian) with a Gaussian cut-off and changing the basis vectors to plane waves as

\begin{align}
\mathcal{A}(x)&=\lim_{M\rightarrow\infty}\sum_n\psi_n^\dag(x)\gamma^5\psi_n(x)e^{-\frac{\lambda_n^2}{M^2}}=\lim_{M\rightarrow\infty}\sum_n\psi_n^\dag(x)\gamma^5e^{-\frac{(i\not D_A^E )^2}{M^2}}\psi_n(x)\nonumber\\
&=\lim_{M\rightarrow\infty}\int\frac{d^4k}{(2\pi)^4}\text{Tr}\Big[\gamma^5e^{ikx}e^{-\frac{(i\not D_A^E )^2}{M^2}}e^{-ikx}\Big]\nonumber\\
&=-\frac{1}{32\pi^2}\epsilon^{\mu\nu\rho\lambda}F_{\mu\nu}F_{\rho\lambda}=-\frac{1}{32\pi^2}\ast(F\wedge F)(x).
\label{2.32}
\end{align} 

The symbol ``$\wedge$'' denotes the exterior product and ``$\ast$'' the Hodge dual. Upon substitution of (\ref{2.32}) into (\ref{2.31}) we find that in Euclidean space

\begin{align}
\partial_\mu \langle j_5^\mu\rangle_A&=-\frac{1}{16\pi^2}\ast(F\wedge F).
\label{2.33}
\end{align}

The Minkowski-space version of (\ref{2.33}) is obtained by undoing the Wick rotation:

\begin{equation}
\boxed{\partial_\mu \langle j_5^\mu\rangle_A=\frac{1}{16\pi^2}\ast(F\wedge F).}
\label{2.34}
\end{equation}

\newpage

\section{Chiral currents}

\subsection{Conserved version of the chiral currents}

The chiral currents $j_l^\mu$ and $j_r^\mu$ corresponding to fermions of left and right chirality are defined as

\begin{equation}
j_l^\mu=\bar\psi\gamma^\mu\frac{1}{2}(1-\gamma^5)\psi,\hspace{5mm}
j_r^\mu=\bar\psi\gamma^\mu\frac{1}{2}(1+\gamma^5)\psi.
\label{2.35}
\end{equation}

They are related to the electric current $j^\mu$ and to the axial current $j_5^\mu$ by

\begin{equation}
j^\mu\equiv \bar\psi\gamma^\mu\psi=j_r^\mu+j_l^\mu,\hspace{5mm}
j_5^\mu\equiv \bar\psi\gamma^\mu\gamma^5\psi=j_r^\mu-j_l^\mu.
\label{2.36}
\end{equation}

The chiral currents $j_l^\mu$ and $j_r^\mu$ are gauge invariant, but not conserved because of the chiral anomaly:

\begin{equation}
\partial_\mu j_{l,r}^\mu=\mp\frac{1}{8\pi^2}\epsilon^{\mu\nu\rho\sigma}\partial_\mu A_\nu\partial_\rho A_\sigma.
\label{2.37}
\end{equation}

In certain situations it will be useful to introduce the currents

\begin{equation}
\tilde j_{l,r}^\mu=j_{l,r}^\mu\pm\frac{1}{8\pi^2}\epsilon^{\mu\nu\rho\sigma} A_\nu\partial_\rho A_\sigma,
\label{2.38}
\end{equation}

which are \textit{conserved}, but fail to be gauge invariant. However, the corresponding charges

\begin{equation}
\tilde Q_{l,r}=\int d^3x\tilde j_{l,r}^0
\label{2.39}
\end{equation}

are not only conserved, but also gauge invariant. More precisely the gauge variation of $\tilde Q_{l,r}$ amounts to a surface term, which may be dropped.

\subsection{Gauge variation of the chiral determinant}

In analogy with (\ref{2.15}) and (\ref{2.16}) the functionals $\Gamma_{l,r}[A]$ defined through

\begin{equation}
e^{i\Gamma_{r,l}[A]}=\int \mathcal D\psi \mathcal D\bar\psi e^{i\int d^4x\bar\psi i\gamma^\mu(\partial_\mu-iA_\mu)\frac{1}{2}(1\pm\gamma^5)\psi},
\label{2.40}
\end{equation}

could be regarded as the generating functionals of the Greens functions for the righthanded and lefthanded currents respectively. The interpretation of the right hand side of (\ref{2.40}) as $\text{det}(\D_A\gamma_{r,l})$, as suggested by the formula for Gaussian Berezin integrals \cite{Zuber}

\begin{equation}
\int\prod_kd\eta_kd\bar\eta_k \exp[-\sum_{k,l}\bar\eta_kA_{k,l}\eta_l]=\text{det}A,
\label{2.41}
\end{equation}

confronts us with a problem though. In fact the determinants of the operators $\D_A\gamma_{r,l}$ are formally zero. These difficulties may be overcome by redefining $e^{i\Gamma_{r,l}[A]}$ as $\text{det}(\D_A^{r,l})$, where the operators $\D_A^{r,l}$ act on 4-component spinors, but couple the gauge field to positive/negative chirality components only:

\begin{eqnarray}
\D_A^r&=&\gamma^\mu(\partial_\mu-iA_\mu\frac{1}{2}(1+\gamma^5))=\left(\begin{array}{cc} 0&\pa_A\\\hat{\pa}&0\end{array}\right),\label{2.42}\\
\D_A^l&=&\gamma^\mu(\partial_\mu-iA_\mu\frac{1}{2}(1-\gamma^5))=\left(\begin{array}{cc} 0&\pa\\\hat{\pa}_A&0\end{array}\right).\label{2.43}
\end{eqnarray} 

In equations (\ref{2.42}) and (\ref{2.43}) we used the notation $\hat{\pa}=\hat\sigma\cdot\partial=-\sigma^0\partial_0-\vec{\sigma}\cdot\vec{\nabla}$, $\pa=\sigma\cdot\partial=\sigma^0\partial_0-\vec{\sigma}\cdot\vec{\nabla}$ and a subscript $A$ denotes a covariant derivative as usual.

Under an infinitesimal gauge transformation $A\rightarrow A+d\theta$, the generating functionals $\Gamma_{r,l}[A]$ change as

\begin{align}
\Gamma_{r,l}[A+d\theta]&=\Gamma_{r,l}[A]-\int d^4x\theta(x)\partial_\mu\frac{\delta \Gamma_{r,l}[A]}{\delta A_{\mu}(x)}.
\label{2.44}
\end{align}

Hence gauge invariance would require $\partial_\mu\frac{\delta \Gamma_{r,l}[A]}{\delta A_{\mu}}=0$, which is equivalent to the current conservation conditions $\partial_\mu\langle j_{l,r}^\mu\rangle_A=0$, since

\begin{align}
\frac{\delta\Gamma_{r,l}[A]}{\delta A_\mu}&=\langle \bar\psi\gamma^\mu\frac{1}{2}(1\pm\gamma^5)\psi\rangle_A=\langle j_{r,l}^\mu \rangle_A.
\label{2.45}
\end{align}

But in the preceeding section we have shown that chiral currents are not conserved. A theory of massless chiral fermions coupled to an external electromagnetic field is anomalous in the sense that it fails to be gauge invariant. 

From (\ref{2.36}), $j_{r,l}=\frac{1}{2}(j\pm j_5)$ and the conservation of the electric current we find

\begin{align}
\partial_\mu\frac{\delta\Gamma_{r,l}[A]}{\delta A_\mu}&=\partial_\mu\langle j_{r,l}^\mu\rangle_A=\pm\frac{1}{2}\partial_\mu\langle j_5^\mu\rangle_A=\pm\frac{1}{32\pi^2}\ast(F\wedge F),
\label{2.46}
\end{align}

where in the last step we have replaced the anomalous divergence of the axial current by its explicit expression obtained in (\ref{2.34}). The variation of the generating functionals $\Gamma_{r,l}[A]$ under gauge transformations is obtained by substituting (\ref{2.46}) into (\ref{2.44}):

\begin{equation}
\boxed{\Gamma_{r,l}[A+d\theta]=\Gamma_{r,l}[A]\mp\frac{1}{32\pi^2}\int \theta(F\wedge F).}
\label{2.47}
\end{equation}

\newpage

\section{Anomalous commutators}

In this section we would like to determine the equal-time commutators of the current components $j_{l,r}^0$. The following argument is due to J. Fröhlich \cite{J-F}. It does not claim the status of a proof, but hopefully provides a reasonably clear idea about the origin of the anomalous commutator and its relation with the anomalous divergence of the chiral currents. For a more mathematical approach based on methods of group theory cohomology see for example \cite{Faddeev} and \cite{Zumino}.

Let $\mathcal{V}$ denote the the space of configurations of external electromagnetic vector potentials $A$ corresponding to static electromagnetic fields. We consider the Hilbert bundle $\mathcal{H}$ over $\mathcal{V}$ whose fibre $\mathcal{F}_A$ at a point $A\in\mathcal{V}$ is the Fock space of state vectors of chiral (e.g. left-handed) fermions coupled to the vector potential $A$.

\begin{figure}[htbp]
\centering
\includegraphics[width=0.5\textwidth]{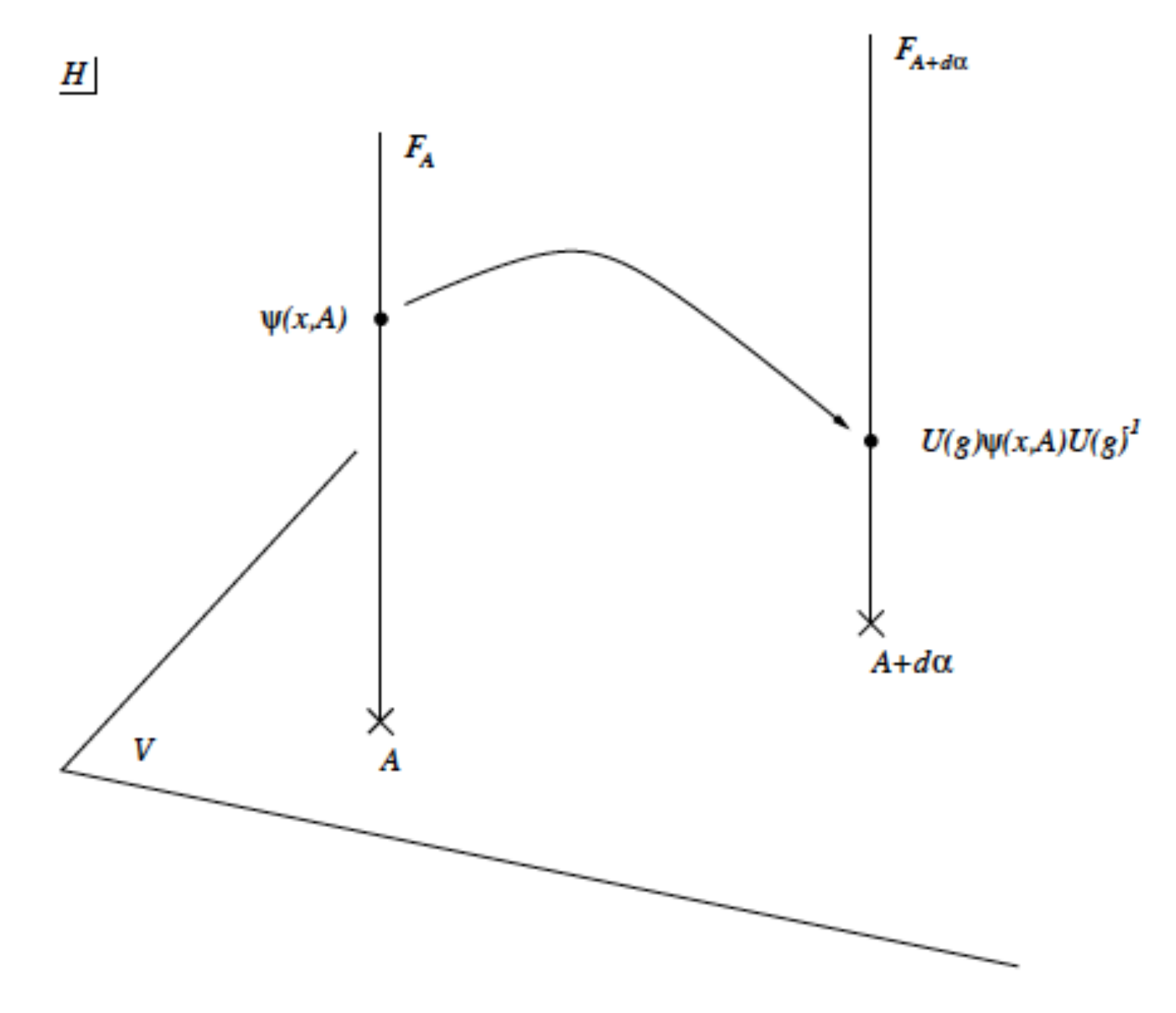}
\caption{Illustration of the Hilbert bundle $\mathcal{H}$ and the gauge transformation $U(g)=U(g^\alpha(x))$.}
\label{}
\end{figure}

$\mathcal{H}$ carries a projective representation $U$ of the group $\mathcal{G}$ of time-independent electromagnetic gauge transformations $g=(g^\alpha(x))$ - $g^\alpha(x)=e^{i\alpha(x)}$, $\alpha(x)=\alpha(\vec{x})$ independent of $x^0=t$ - with the following properties

\begin{enumerate}
\item $U(g): \mathcal{F}_{A}\longrightarrow\mathcal{F}_{A+d\alpha}$,
\item $U(g)\psi(x,A)U(g)^{-1}=e^{i\alpha(x)}\psi(x,A+d\alpha)$,
\end{enumerate}

where $\psi(x,A)$ is the Dirac spinor field acting on $\mathcal{F}_A$. The gauge transformation $U(g^\alpha(.))$ may be written in terms of the generator $G$ of $\mathcal{G}$ as

\begin{equation}
U(g^\alpha(.))=e^{-i\alpha\cdot G}.
\label{2.48} 
\end{equation}

We used the notation $\alpha\cdot G=\int d^3x\alpha(x)G(x)$, where the integration is  over space. The explicit expression for $G$ reads

\begin{equation}
G(x)=-i\vec{\nabla}\cdot \frac{\delta}{\delta \vec A(x)}+j_l^0(x,A).
\label{2.49}
\end{equation}

The first term in (\ref{2.49}) generates the gauge transformation $A\rightarrow A+d\alpha$ on $A$, while the zeroth component of the current $j_l^0$ generates a rotation of the fermion fields. The left-handed current appears because we have chosen to consider left-handed fermions.

Locally the phase factor of the projective representation $U$ of $\mathcal{G}$ can be made trivial by redefining the generators $G$ as
$G(x)\rightarrow \hat G(x)=-i\vec{\nabla}\cdot \frac{\delta}{\delta \vec A(x)}+\hat j_l^0(x,A)$.

The operators $\hat G(x)$ generate a representation of the group $\mathcal{G}$ of gauge transformations on $\mathcal{H}$ if and only if

\begin{equation}
[\hat G(t,\vec{x}),\hat G(t,\vec{y})]=0
\label{2.50}
\end{equation}

for all times $t$. We pretend that the right choice for the redefined generator compatible with (\ref{2.50}) is

\begin{align}
\tilde G(x)&=-i\vec{\nabla}\cdot \frac{\delta}{\delta \vec A(x)}+\tilde j_l^0(x,A),\hspace{5mm}
\tilde j_l^0=j_l^0+\frac{1}{8\pi^2}\epsilon^{ijk}A_i\partial_j A_k.
\label{2.51}
\end{align}

This follows, heuristically, from the fact that $\tilde j_l^\mu$ is a conserved current (see section 2.3.1). Furthermore, since the current $j_l^\mu(x,A)$ is gauge invariant we have

\begin{equation}
\left[\vec{\nabla}\cdot \frac{\delta}{\delta\vec{A}},j_l^0(x,A)\right]=0.
\label{2.52}
\end{equation}

This enables us to compute the anomalous commutator of the left-handed currents as follows (an integration over $x$ and $y$ of the form $\int dx \alpha(x)\int dy\beta(y)\ldots$ is implicitely understood):

\begin{align}
0&=[\tilde G(x),\tilde G(y)]=\left[-i\vec{\nabla}\cdot \frac{\delta}{\delta \vec A(x)}+\tilde j_l^0(x),-i\vec{\nabla}\cdot \frac{\delta}{\delta \vec A(y)}+\tilde j_l^0(y)\right]\nonumber\\
&=[j_l^0(x),j_l^0(y)]-\left[i\vec{\nabla}\cdot \frac{\delta}{\delta \vec A(x)},\frac{1}{8\pi^2}\epsilon^{ijk}A_i\partial_j A_k(y)\right]-
\left[\frac{1}{8\pi^2}\epsilon^{ijk}A_i\partial_j A_k(x),i\vec{\nabla}\cdot \frac{\delta}{\delta \vec A(y)}\right]\nonumber\\
&=[j_l^0(x),j_l^0(y)]-\frac{i}{8\pi^2}\epsilon^{ijk}\Big(\frac{\partial}{\partial x_i}\delta(x-y)\Big)\frac{\partial}{\partial y_j}A_k(y)+\frac{i}{8\pi^2}\epsilon^{ijk}\Big(\frac{\partial}{\partial y_i}\delta(x-y)\Big)\frac{\partial}{\partial x_j}A_k(x)\nonumber\\
&=[j_l^0(x),j_l^0(y)]+\frac{i}{4\pi^2}\frac{\partial}{\partial x_k}[B_k(x)\delta(x-y)].
\label{2.53}
\end{align}

$B_i=-\epsilon^{ijk}\partial_j A_k$ denotes the magnetic field strenght. The calculation for the right-handed current is similar. All that changes is the plus sign on the right hand side of (\ref{2.51}) and in the subsequent calculations. We therefore obtain the anomalous commutators 

\begin{equation}
\boxed{[j_{l,r}^0(x),j_{l,r}^0(y)]=\mp\frac{i}{4\pi^2}\frac{\partial}{\partial x_k}[B_k(x)\delta(x-y)].}
\label{2.54}
\end{equation}

\chapter{The generation of seed magnetic fields in the early universe}

\section{First attempt: Equilibrium statistical mechanics}

This chapter is devoted to the study of several models and mechanisms which could explain the generation of magnetic fields in the early universe. In a first attempt we will assume that the early universe is a hot plasma of charged fermions and that chirality flips constitute a dynamical process slower than the expansion rate of the universe. Under these assumptions the chiral charges $\tilde Q_{l,r}$ defined in (\ref{2.39}) are approximately conserved and we may introduce the chemical potentials $\mu_l$ and $\mu_r$ canonically conjugate to these conserved charges.

In this first section we shall attempt to show that if there exists an asymmetry in the chemical potentials of left- and right-handed fermions, this could lead to the generation of large cosmic magnetic fields.

\subsection{Current expectation value in thermal equilibrium}

The starting point of our investigations is a formula relating the current expectation value in thermal equilibrium to the magnetic field strength, which was obtained by \mbox{A. Y.} Alekseev, V. V. Cheinaov and J. Fröhlich in 1998, \cite{Alekseev}. We will now go through its derivation, that is compute the expectation value of the electric current $\vec{j}$ in the background electromagnetic field $A_\mu$.

The continuity equation for the electric current reads $\partial_\mu j^\mu=0$. It can be solved in terms of a 3-vector field $\vec{a}$:

\begin{equation}
j^0=\vec{\nabla}\cdot \vec{a},\hspace{1cm} \vec{j}=-\partial_0\vec{a}.
\label{3.1}
\end{equation}

In fact we can define $\vec{a}(x)$ as 

\begin{equation}
\vec{a}(x^0,\vec{x})=-\int_0^{x^0}d\tau\vec{j}(\tau,\vec{x})-\vec{\nabla}_{\vec{x}}\int d^3y\frac{1}{4\pi|\vec{x}-\vec{y}|}j^0(0,\vec{y}),
\label{3.2}
\end{equation}
 
as one may easily verify. A derivation of this result - although not really necessary - is given in appendix A.

The thermal state of the system characterized by the chemical potentials $\vec{\mu}=(\mu_l,\mu_r)$ and the inverse temperature $\beta$ is given by the density matrix

\begin{equation}
\Sigma_{\beta,\vec{\mu}}=\frac{e^{-\beta \mathcal{H}_\mu}}{\mathcal{Z}_\mu},\hspace{1cm}\mathcal{H}_\mu=\mathcal{H}-\mu_l\tilde Q_l-\mu_r\tilde Q_r,
\label{3.3}
\end{equation}

where $\mathcal{H}$ is the Hamiltonian and $\mathcal{Z}_\mu=\text{Tr}e^{-\beta \mathcal{H}_\mu}$. In this equilibrium state the expectation value of the current is

\begin{align}
\langle \vec{j}(x)\rangle_{\beta,\vec{\mu}}&=-\langle \partial_0\vec{a}(x)\rangle_{\beta,\vec{\mu}}=-i\langle [\mathcal{H},\vec{a}(x)]\rangle_{\beta,\vec{\mu}}\nonumber\\
&=-\frac{i}{\mathcal{Z}_{\beta,\vec{\mu}}}\text{Tr}\{e^{-\beta \mathcal{H}_\mu}[\mathcal{H}_\mu,\vec{a}(x)]\}-\frac{i}{\mathcal{Z}_{\beta,\vec{\mu}}}\text{Tr}\{e^{-\beta \mathcal{H}_\mu}[\mu_l\tilde Q_l+\mu_r\tilde Q_r,\vec{a}(x)]\}
\label{3.4}
\end{align}

The first trace on the right hand side of (\ref{3.4}) vanishes by cyclicity of the trace. If we furthermore use $[\vec{\nabla}\cdot\vec{a},\vec{a}]=0$, that is $[j^0,\vec{a}]=0$, we finally obtain

\begin{align}
\langle \vec{j}(x)\rangle_{\beta,\vec{\mu}}&=-i\langle[\mu_l\tilde Q_l+\mu_r\tilde Q_r,\vec{a}(x)]\rangle_{\beta,\vec{\mu}}\nonumber\\
&=\frac{i}{2}(\mu_l-\mu_r)\int d^3y\langle[\tilde j_5^0(y),\vec{a}(x)]\rangle_{\beta,\vec{\mu}}.
\label{3.5}
\end{align}

The commutators of the densities of the left- and right-handed fermions have been calculated in (\ref{2.54}):

\begin{equation}
[j_{l,r}^0(x),j_{l,r}^0(y)]=[\tilde j_{l,r}^0(x),\tilde j_{l,r}^0(y)]=\mp\frac{i}{4\pi^2}\frac{\partial}{\partial x_k}[B_k(x)\delta(x-y)],
\label{3.6}
\end{equation}

whereas the commutator of the left-handed and right-handed current is zero. Hence

\begin{equation}
[\tilde j_5^0(y),\partial_k a_k(x)]=[\tilde j_r^0(y)-\tilde j_l^0(y),\tilde j_r^0(x)+\tilde j_l^0(x)]=\frac{i}{2\pi^2}\frac{\partial}{\partial x_k}[B_k(x)\delta(x-y)].
\label{3.7}
\end{equation}

We remove the divergence in (\ref{3.7}):

\begin{equation}
[\tilde j_5^0(y),\vec{a}(x)]=\frac{i}{2\pi^2}\vec{B}(x)\delta(x-y)+\vec{\nabla}_x\times\vec{\Pi}(x-y),
\label{3.8}
\end{equation}

and substitute this result into (\ref{3.5}). The second term on the right hand side of (\ref{3.8}) drops out after integration over $y$. This finally yields

\begin{equation}
\boxed{\langle \vec{j}(x)\rangle_{\beta,\vec{\mu}}=-\frac{1}{4\pi^2}(\mu_l-\mu_r)\vec{B}(x).}
\label{3.9}
\end{equation}

\subsection{Equations of motion}

By substituting expression (\ref{3.9}) for the current expectation value into Maxwells equations, we obtain the following system of equations 

\begin{eqnarray}
\vec{\nabla}\cdot \vec{E}&=&\alpha\langle j^0\rangle_{\beta, \vec{\mu}}\label{3.10}\\
\vec{\nabla}\times \vec{B} - \partial_0\vec{E}&=&\alpha\langle \vec{j}\rangle_{\beta, \vec{\mu}}=-\frac{\alpha}{4\pi^2}(\mu_l-\mu_r)\vec{B}\label{3.11}\\
\vec{\nabla}\cdot \vec{B}&=&0\label{3.12}\\
\vec{\nabla}\times \vec{E} + \partial_0\vec{B}&=&0,\label{3.13}
\end{eqnarray}

which is supposed to govern the evolution of the electromagnetic field in the early universe. The feinstructure constant $\alpha=e^2$ appears on the right hand side of (\ref{3.10}) and (\ref{3.11}), because we chose to absorb a factor of $e$ into the definition of the vector potential:

\begin{equation}
eA \rightarrow A.
\label{3.14}
\end{equation}

As long as the chemical potentials $\mu_l$ and $\mu_r$ remain constant, the above equations are linear in the fields $\vec{E}$ and $\vec{B}$ with constant coefficients (at least if we make a simple assumption for $j^0$ such as $\langle j^0\rangle_{\beta, \vec{\mu}}=0$). The time evolution can thus be calculated by means of Fourier transformation, as will be done in chapter 4.

We should, however, be aware that some of the hypotheses underlying the derivation of formula (\ref{3.9}) appear quite unnatural in the present context. The chiral charges $\tilde Q_{l,r}$ are not really conserved since fermions are massive. The very early universe is not really an equilibrium state and the chemical potentials $\mu_l$ and $\mu_r$ of left- and righthanded fermions neither have an unambiguous meaning, nor would they be space- and time-independent. 

In the following sections, we shall try to generalise the system of equations (\ref{3.10})-(\ref{3.13}) in order to obtain an equation of motion describing the evolution of a (space-)time dependent ``difference in chemical potentials''. This generalisation will ultimately lead to a new formulation of the theory in terms of an axion field, which no longer relies on the implausible assumptions mentioned above, but could still provide an explanation for the generation of cosmic magnetic fields in the early universe.

\newpage

\section{Second attempt: (4+1)-dimensional quantum Hall effect}

One idea is to imagine that the difference in the chemical potentials $\mu_l-\mu_r$ is generated by an electric field $E_4$ pointing in a direction perpendicular to our (3+1)-dimensional world. This idea is based on an analogy with the (1+1)-dimensional quantum Hall effect, which we shall now discuss.

\subsection{The (2+1)-dimensional quantum Hall effect}

It might be useful to summarize some key features of the (2+1)-dimensional quantum Hall effect first. This introduction again closely follows the review by J. Fröhlich and B. Pedrini \cite{J-F}.

A quantum Hall fluid (QHF) is an interacting electron gas confinded to some domain $D$ in a two-dimensional plane, subject to a constant magnetic field $\vec{B}^{(0)}$ transversal to the confinement plane. For $D$ we choose a strip of width $L$ in the 1,2-plane, which is infinitely extended along the 1-direction.

Among the experimental control parameters is the filling factor, $\nu$, defined by

\begin{equation}
\nu=2\pi\frac{n^{(0)}}{B^{(0)}},
\label{3.15}
\end{equation} 

where $n^{(0)}$ is the (constant) electron density, $B^{0}$ the component of the magnetic field $\vec{B}^{(0)}$ perpendicular to the plane of the fluid and $2\pi$ the quantum of magnetic flux (in units where $c=\hbar=1$).

Transport properties of a QHF in an external electric field are described by the equation

\begin{equation}
\vec{j}(x^0,\vec{x})=\left(\begin{array}{cc}\sigma_L&-\sigma_H\\\sigma_H&\sigma_L\end{array}\right)\vec{E}(x^0,\vec{x}).
\label{3.16}
\end{equation}

In the above formula $\vec{x}$ is a point in $D$, $\vec{j}$ the bulk electric current and $\vec{E}$ the component of the external electric field parallel to the sample plane. Furthermore, $\sigma_L$ denotes the longitudinal conductivity and $\sigma_H$ the transverse or Hall conductivity.

Experimenally, one observes that the longitudinal conductivity, $\sigma_L$, vanishes  when the filling factor $\nu$ belongs to certain small intervals. At the same time, the Hall conductivity $\sigma_H$ is a rational multiple of $\frac{1}{2\pi}$. Such a QHF is called ``incompressible''.

We will now describe the basic equations describing the electromagnetics of an incompressible QHF. To this end it is useful to combine the two-dimensional space of the fluid and time to a three-dimensional space-time. The field strength tensor of the system is given by

\begin{equation}
F_{\mu\nu}=\left(\begin{array}{ccc}0&E_1&E_2\\-E_1&0&-B\\-E_2&B&0\end{array}\right),
\label{3.17}
\end{equation}

where $E_1$ and $E_2$ are the 1- and 2-components of an external electric field and $B$ is the component of an external magnetic field, $\vec{B}$, perturbing the constant field $\vec{B}^{(0)}$ perpendicular to the sample plane ($\vec{B}_{tot}=\vec{B}^{(0)}+\vec{B}$). We define $j^0$ to denote the sum of the electron charge density at the space-time point $x=(x^0,\vec{x})$ and the uniform background charge density $n^{(0)}$. From the continuity equation for the electric current density $j^\mu=(j^0,\vec{j})$, $\partial_\mu j^\mu=0$, the three-dimensional homogeneous Maxwell equations, $dF=0$, and from the transport equations (\ref{3.16}) with $\sigma_L=0$ it follows that 

\begin{equation}
j^0=-\sigma_HB.
\label{3.18}
\end{equation}

Equations (\ref{3.16}), for $\sigma_L=0$, and (\ref{3.18}) can be combined to the covariant expression

\begin{equation}
j^\mu=\sigma_H\epsilon^{\mu\nu\rho}\partial_\nu A_\rho=\frac{\sigma_H}{2}(\ast F)^\mu,
\label{3.19}
\end{equation}

which describes the response of an incompressible QHF to an external electromagnetic field perturbing the constant magnetic field $\vec{B}^{(0)}$. In (\ref{3.19}), $A$ denotes the vector potential of this external eletromagnetic field ($F_{\mu\nu}=\partial_\mu A_\nu-\partial_\nu A_\mu$). 

The finite extension of the sample, confined to a space-time region $\Lambda=D\times\mathbb{R}$, is taken into account by setting $\sigma_H(.)$ to zero outside $\Lambda$,

\begin{equation}
\sigma_H(x)=\sigma_H\Omega_\Lambda(x).
\label{3.20}
\end{equation}

In the above equation, $\sigma_H$ is the (constant) value of the Hall conductivity inside the sample and $\Omega_\Lambda$ the characteristic function of the space-time domain $\Lambda$. Taking the divergence of (\ref{3.19}) we get

\begin{equation}
\partial_\mu j^\mu=\sigma_H\epsilon^{\mu\nu\rho}\partial_\mu\Omega_\Lambda\partial_\nu A_\rho=\frac{\sigma_H}{2}\epsilon^{\mu\nu\rho}\partial_\mu\Omega_\Lambda F_{\nu\rho}.
\label{3.21}
\end{equation} 

Thus $\partial_\mu j^\mu$ fails to vanish on the boundary $\partial \Lambda$ of the sample, unless $F_{\mu\nu}|_{\partial\Lambda}=0$. For arbitrary external electromagnetic fields, there must exist an electric current density $j_{\partial\Lambda}$ localized on the boundary $\partial\Lambda=\{(x|x^2=L)\}\cup\{(x|x^2=0)\}$ of the sample space-time such that the total electric current density

\begin{equation}
j_{tot}^\mu=j^\mu+j_{\partial\Lambda}^\mu
\label{3.22}
\end{equation}

satisfies the continuity equation. These edge currents are chiral, a property which is correctly predicted by the naive classical picture of electrons bouncing off the domain walls. The current density $I_l$ - localized on the upper boundary $x^2=L$ - is produced by left-moving modes (for an appropriate choice of the direction of $\vec{B}_{tot}$) and $I_r$ - localized on the lower boundary $x^2=0$ - by right movers. If $\mathcal{I}_{l,r}^\alpha$ denote the corresponding quantum mechanical current operators, then the edge currents are given by the quantum mechanical expectation value $\langle \mathcal{I}_{l,r}^\alpha\rangle_A$.

The effective action $S_{eff}[A]$ can be found from equation (\ref{3.19}) relating the current expectation value to the external electromagnetic field 

\begin{equation}
-\frac{\delta S_{eff}[A]}{\delta A_\mu(x)}=j^\mu(x)=\sigma_H\epsilon^{\mu\nu\rho}\partial_\nu A_\rho(x).
\label{3.23}
\end{equation}

The solution to the above equation is 

\begin{equation}
-S_{eff}[A]=S_{CS}[A]=-\frac{\sigma_H}{2}\int d^3x\epsilon^{\mu\nu\rho}A_\mu\partial_\nu A_\rho=-\frac{\sigma_H}{4}\int A\wedge F.
\label{3.24}
\end{equation}

We denote this action functional by $S_{CS}$ - for Chern-Simons action - since it is proportional to the integral over the Chern-Simons 3-form. Since the latter one is not invariant under gauge transformations of $A$ that do not  vanish on the boundary $\partial\Lambda$ of the sample, the effective action must be corrected by a boundary term

\begin{equation}
S_{eff}[A]=-S_{CS}[A]+\Gamma_{\partial\Lambda}[A|_{\partial\Lambda}].
\label{3.25}
\end{equation}

The boundary term $\Gamma_{\partial\Lambda}[A|_{\partial\Lambda}]$ is the effective action of the charged chiral modes propagating along $\partial\Lambda$. The explicit form of (\ref{3.25}) and expressions for the boundary currents can be found in appendix B.

\subsection{Generalisation to (4+1) dimensions}

Let us consider the (2+1)-dimensional Hall sample of width $L$ extended along the \mbox{$1$-direction}. The total Hall current $I^1$ is related to the potential difference $V=(\mu_l-\mu_r)$ between the upper and lower boundary by the formula \mbox{$I^1=\sigma_H(\mu_l-\mu_r)$}. If the electric field in the bulk of the Hall sample vanishes, then $I^1$ is the sum of two contributions $I_l^1$ and $I_r^1$ supported by so called \textit{edge states} localized respectively near the upper and lower boundary. These edge currents are \textit{chiral}: $I_l^1$ is produced by leftmoving modes and $I_r^1$ by rightmovers, $\mu_l$ and $\mu_r$ being the chemical potentials of their respective reservoirs.

The (1+1)-dimensional system obtained by considering the boundaries of the (2+1)-dimensional Hall sample corresponds to a quantum wire in which the left- and rightmoving electrons are coupled to reservoirs with chemical potentials $\mu_l$ and $\mu_r$ respectively. The total current through the wire is then given by the formula for the Hall current, which in turn may easily be derived by considering the bulk of the (2+1)-dimensional Hall sample. In fact, if we admit that the entire potential difference between the two edges is generated within the bulk of the sample, then (\ref{3.16}) yields the formula for the total Hall current:

\begin{equation}
I^1=\int_0^Ldx^2j^1(x^2)=-\sigma_H\int_0^Ldx^2E_2(x^2)=\sigma_H(\mu_l-\mu_r).
\label{3.26}
\end{equation} 

It seems plausible that formula (\ref{3.9}) for the current $\vec{j}$, which is also the sum of two contributions $\vec{j}_l$ and $\vec{j}_r$ corresponding to left- and righthanded fermions, could be ``derived'' in a similar way. To this end we consider a slab of thickness $L$ in (4+1) dimensions, which is extended in the $1,2,3$-directions. The upper and lower surfaces represent two copies of our (3+1)-dimensional world. Inspired by the analogy with the (2+1)-dimensional Hall sample, we place the lefthanded fermions on the top and the righthanded ones on the bottom. The potential difference between the two surfaces, which is generated by the 4-component of an electric field, will be denoted by $\mu_l-\mu_r$.

\begin{figure}[h!]
\noindent
\begin{minipage}[b]{0.48\linewidth}
\centering\epsfig{figure=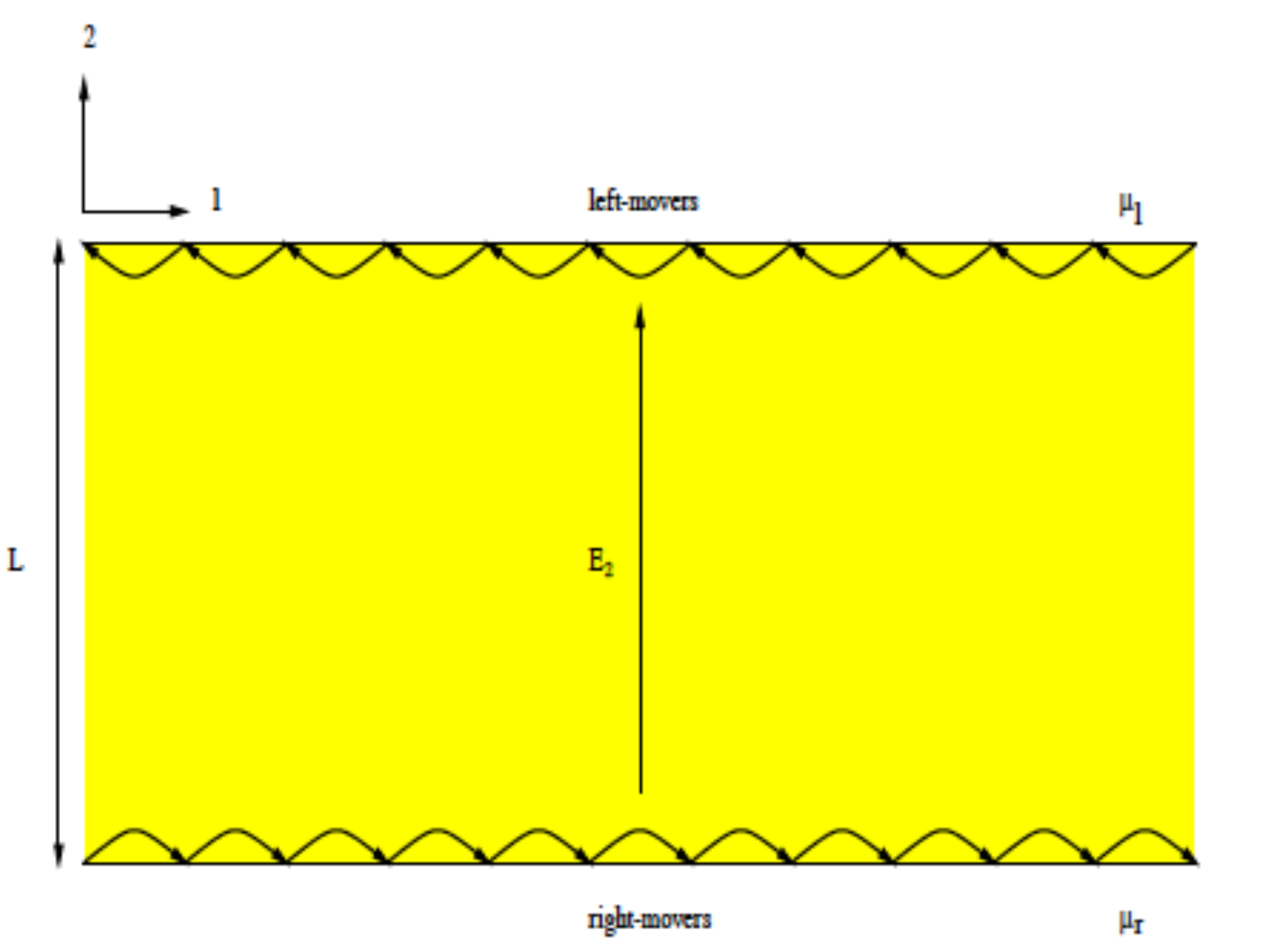,width=\linewidth,height=5.5cm}
\small{(2+1)-dimensional QH sample}
\end{minipage} \hfill
\begin{minipage}[b]{0.48\linewidth}
\centering\epsfig{figure=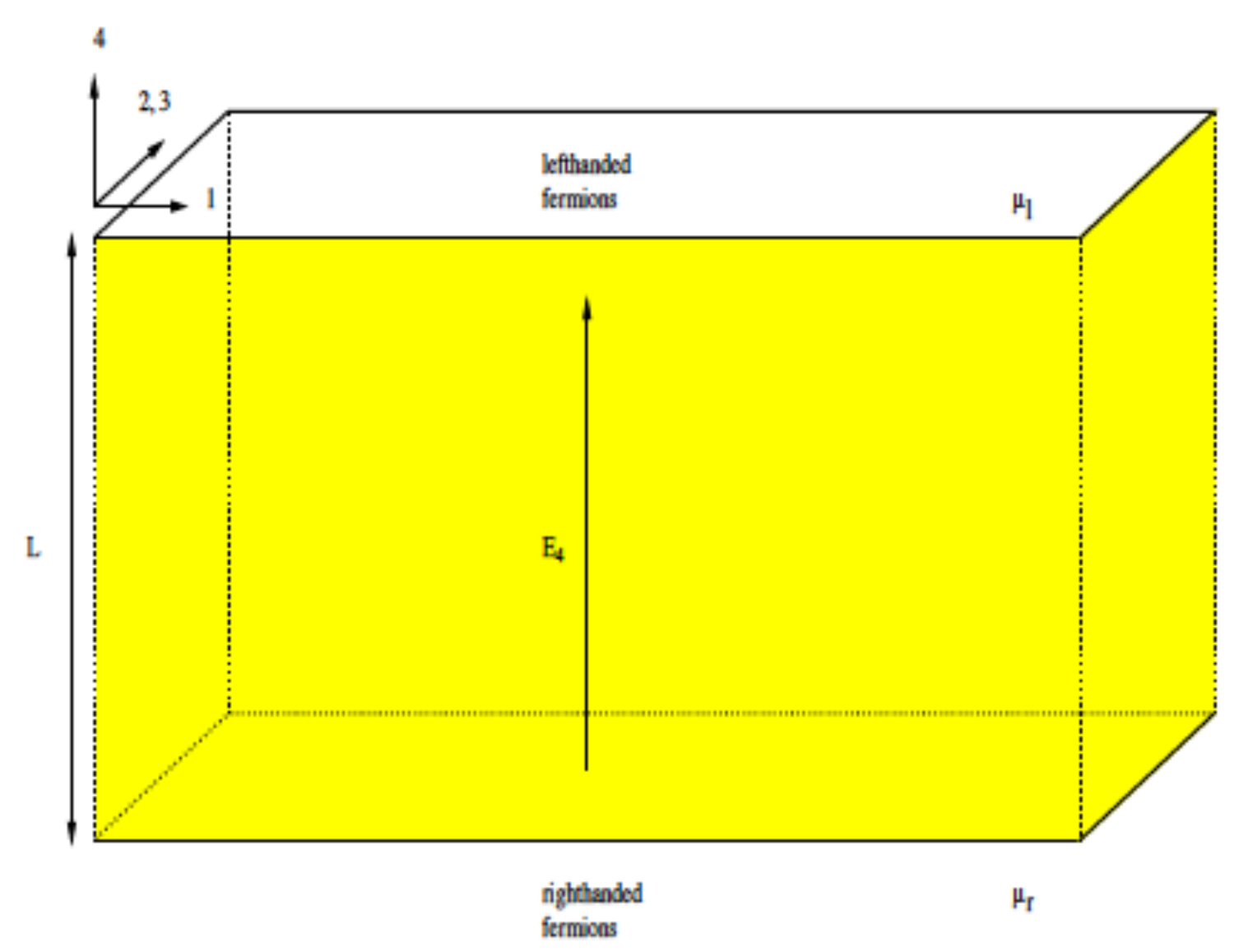,width=\linewidth,height=5.5cm}
\small{Slab in (4+1) dimensions}
\end{minipage}
\caption{In order to obtain an equation for the time evolution of $\mu_l-\mu_r$ one imagines that the difference in the chemical potentials is generated by an electric field pointing in a direction perpendicular to our (3+1)-dimensional world and exploits an analogy with the situation encountered in a (2+1)-dimensional Quantum Hall sample.}
\label{slab}
\end{figure}

This analogy is illustrated in figure \ref{slab}. It can thus far be summarized by the following equations:

(1+1)-dimensional quantum wire:\hspace{1cm}Massless fermions coupled to an electromagnetic
\mbox{\hspace{6.6cm}} field in (3+1) dimensions:
\begin{xalignat}{2}
I^1&=I_l^1+I_r^1&\qquad \vec{j}&=\vec{j}_l+\vec{j}_r\label{3.27}\\
I^1&=\sigma_H(\mu_l-\mu_r)&\qquad \vec{j}&=-\frac{1}{4\pi^2}(\mu_l-\mu_r)\vec{B}\label{3.28}\\
\intertext{The above analogies and equation (\ref{3.26}) suggest to introduce a current density $J^\mu(x^0,\vec{x},x^4)$ in (4+1) dimensions and to proceed as follows:}
\intertext{(2+1)-dimensional Hall sample:\hspace{1.38cm}(4+1)-dimensional formulation:}
I&=\int_0^Ldx^2j^1(x^0,x^1,x^2)&\qquad \vec{j}&=\int_0^Ldx^4\vec{J}(x^0,\vec{x},x^4)\label{3.29}\\
&=\sigma_H(\mu_l-\mu_r)&\qquad &\equiv-\frac{1}{4\pi^2}(\mu_l-\mu_r)\vec{B}\label{3.30}\\
&=-\sigma_H\int_0^Ldx^2E_2(x^0,x^1,x^2)&\qquad &=\frac{1}{4\pi^2}\int_0^Ldx^4E_4(x^0,\vec{x},x^4)\vec{B}(x^0,\vec{x})\label{3.31}
\end{xalignat}

Looking at (\ref{3.29}) and (\ref{3.31}) we realise that the (4+1)-dimensional current density $J$ satisfies $J^k(x^0,\vec{x},x^4)=\frac{1}{4\pi^2} E_4(x^0,\vec{x},x^4)B_k(x^0,\vec{x})$. The covariant version of this equation reads

\begin{equation}
J^\mu=\frac{1}{32\pi^2}\epsilon^{\mu\nu\rho\lambda\sigma} F_{\nu\rho}F_{\lambda\sigma}=\frac{1}{32\pi^2}\left(\ast(F\wedge F)\right)^\mu,
\label{3.32}
\end{equation}

that is

\begin{equation}
\boxed{J=\frac{1}{32\pi^2}\ast(F\wedge F).}
\label{3.33}
\end{equation}

The analogue of the field strength tensor in (4+1) dimensions becomes

\begin{equation}
F_{\mu\nu}=\left(\begin{array}{ccccc}0&E_1&E_2&E_3&E_4\\
                            -E_1&0&-B_3&B_2&V_1\\
                            -E_2&B_3&0&-B_1&V_2\\
                            -E_3&-B_2&B_1&0&V_3\\
                            -E_4&-V_1&-V_2&-V_3&0
                            \end{array}\right).
\label{3.34}
\end{equation}

where $V_j=F_{j4}$ denotes some vector field, for which there exists no experimental evidence. We shall nevertheless keep the terms in $V$ throughout the following calculations, because they will turn out to be related to the axion field which we will introduce in section 3.4. 

In (\ref{3.31}) it has been assumed that the magnetic field does not depend on $x^4$. This condition is satisfied if $\vec{\nabla}\times\vec{V}=0$, as will be shown below. Our ``derivation by analogy'' therefore only works in that case.

\subsection{Effective action}

Our aim is to find an effective action functional $S_{eff}[A]$ from which the equations of motion for the electromagnetic field in (4+1) dimensions can be calculated by means of the formula 

\begin{equation}
\frac{\delta S_{eff}[A]}{\delta A_\mu}=0,\text{ }\mu=0,\ldots,4.
\label{3.35}
\end{equation}

Of course we know what these equations should be, namely the inhomogeneous Maxwell equations for the current density $J$ written in (\ref{3.33}). It is again possible to find an action functional whose functional derivative with respect to the vector potential $A$ yields the latter current. The result is the Chern-Simons action

\begin{equation}
S_{CS}[A]=\frac{1}{24\pi^2}\int d^5x\epsilon^{\mu\nu\rho\lambda\sigma}A_\mu\partial_\nu A_\rho\partial_\lambda A_\sigma=\frac{1}{96\pi^2}\int A\wedge F\wedge F.
\label{3.36}
\end{equation} 

The apparent solution to our problem is then readily found by adding a Maxwell term

\begin{equation}
S_{EM}=-\frac{1}{4L\alpha}\int d^5xF^{\mu\nu}F_{\mu\nu}
\label{3.37}
\end{equation}

to the Chern-Simons action $S_{CS}$. The factor of $\frac{1}{L}$ has been introduced in (\ref{3.37}) for reasons of dimensionality and $\alpha=e^2$ denotes the four-dimensional feinstructure constant. In fact, setting

\begin{equation}
S_{eff}=S_{EM}-S_{CS},
\label{3.38}
\end{equation}

and introducing (\ref{3.38}) into (\ref{3.35}) yields 

\begin{equation}
\partial_\nu F^{\nu\mu}=L\alpha J^\mu,\text{ }\mu=0,\ldots,4.
\label{3.39}
\end{equation}

An additional set of equations - corresponding to the homogeneous Maxwell equations - follows from 

\begin{equation}
dF=0.
\label{3.40}
\end{equation} 

The latter condition assures that $F$ can be written as the exterior derivative of some vector potential $A$: $F=dA$.

However, the action functional $S_{eff}$ written in (\ref{3.38}) suffers from a serious deficiency: lack of gauge invariance. The electromagnetic part $S_{EM}$ obviously \textit{is} gauge invariant. But the Chern-Simons action $S_{CS}$ transforms under a gauge transformation $A\rightarrow A+d\theta$ like

\begin{eqnarray}
S_{CS}[A+d\theta]&=&\frac{1}{24\pi^2}\int_\Lambda d^5x\epsilon^{\mu\nu\rho\lambda\sigma}(A_\mu+\partial_\mu\theta)\partial_\nu(A_\rho+\partial_\rho\theta)\partial_\lambda(A_\sigma+\partial_\sigma\theta)\nonumber\\
&=&S_{CS}[A]+\frac{1}{8\pi^2}\int_\Lambda d^5x\epsilon^{\mu\nu\rho\lambda\sigma}\partial_\mu\theta\partial_\nu A_\rho \partial_\lambda A_\sigma\nonumber\\
&=&S_{CS}[A]+\frac{1}{32\pi^2}\int_\Lambda d(\theta F\wedge F)\nonumber\\
&=&S_{CS}[A]+\frac{1}{32\pi^2}\int_{\partial\Lambda} \theta F\wedge F,\hspace{5mm} A=(A_0,\ldots,A_4).
\label{3.41}
\end{eqnarray}

In the above calculation $\Lambda$ denotes the volume of the slab in (4+1) dimensions and $\partial\Lambda$ its boundary. We have used $dF=0$ in the third and Stokes' theorem in the last step.

One remarks the appearance of a surface term. In order to restore the gauge invariance of $S_{eff}$, it is therefore necessary to add a boundary action $\Gamma_{\partial\Lambda}(A|_{\partial\Lambda})$ which is not gauge invariant either but transforms in a way as to cancel the boundary term coming from the Chern-Simons action.

Luckily we have already examined some possible candidates earlier on. The functionals 

\begin{equation}
\Gamma_{r,l}[A]=-i\text{ln}\int \mathcal D\psi \mathcal D\bar\psi e^{i\int d^4x\bar\psi i\gamma^\mu(\partial_\mu-iA_\mu\frac{1}{2}(1\pm\gamma^5))\psi}\equiv-i\ln \text{det}(\D_A^{r,l}),
\label{3.42}
\end{equation}

which have been introduced in section 2.3.2 were shown in (\ref{2.47}) to transform under gauge transformations as

\begin{align}
\Gamma_{r,l}[A+d\theta]&=\Gamma_{r,l}[A]\mp\frac{1}{32\pi^2}\int \theta(F\wedge F),\hspace{5mm} A=(A_0,\ldots,A_3).
\label{3.43}
\end{align}

The integral in (\ref{3.43}) is over the boundary $\partial\Lambda$ of the (4+1)-dimensional slab, which consists of two copies of a (3+1)-dimensional space-time located at $x^4=L$ and $x^4=0$ respectively. Defining the boundary term as

\begin{align}
\Gamma_{\partial\Lambda}[A|_{\partial\Lambda}]&=\{\Gamma_l(A|_{x^4=L})+\Gamma_r(A|_{x^4=0})\}.
\label{3.44}
\end{align}

we find that the combination $-S_{CS}[A]+\Gamma_{\partial\Lambda}[A|_{\partial\Lambda}]$ is invariant under gauge transformations in (4+1) dimensions.

The effective action - which replaces (\ref{3.38}) - is then

\begin{equation}
\boxed{S_{eff}[A]=S_{EM}[A]-S_{CS}[A]+\Gamma_{\partial\Lambda}[A|_{\partial\Lambda}],}
\label{3.45}
\end{equation}

where the individual terms explicitly read

\begin{align}
S_{EM}[A]&=-\frac{1}{4L\alpha}\int_\Lambda d^5xF^{\mu\nu}F_{\mu\nu}\label{3.46}\\
S_{CS}[A]&=\frac{1}{96\pi^2}\int_\Lambda A\wedge F\wedge F\label{3.47}\\
\Gamma_{\partial\Lambda}[A|_{\partial\Lambda}]&
=-i\text{ln}\{\text{det}(\D_A^l|_{x^4=L})\text{det}(\D_A^r|_{x^4=0})\}\label{3.48}.
\end{align}

\subsection{Equations of motion}

The (4+1)-dimensional inhomogeneous Maxwell equations are obtained by substituting the action functional $S_{eff}$ defined in (\ref{3.45}) into (\ref{3.35}) and the homogeneous ones are $dF=0$. The full set of equations then becomes

\begin{eqnarray}
\delta F&=&\frac{L\alpha}{32\pi^2}\ast(F\wedge F)-L\alpha\frac{\delta\Gamma_{\partial\Lambda}}{\delta A}\label{3.49}\\
dF&=&0\label{3.50}.
\end{eqnarray}

Explicitly, equations (\ref{3.49}) and (\ref{3.50}) read:

\hspace{5mm} $\bullet$ Inhomogeneous Maxwell equations, $\delta F=\frac{L\alpha}{32\pi^2}\ast(F\wedge F)-L\alpha\frac{\delta\Gamma_{\partial\Lambda}}{\delta A}$:
\begin{align}
\vec{\nabla}\cdot\vec{E}+\partial_4E_4&=-\frac{L\alpha}{4\pi^2}\vec{V}\cdot\vec{B}-L\alpha\frac{\delta\Gamma_{\partial\Lambda}}{\delta A_0}\label{3.51}\\
\vec{\nabla}\times\vec{B}-\partial_0\vec{E}-\partial_4\vec{V}&=\frac{L\alpha}{4\pi^2}(E_4\vec{B}+\vec{V}\times\vec{E})-L\alpha\frac{\delta\Gamma_{\partial\Lambda}}{\delta \vec{A}}\label{3.52}\\
\partial_0E_4-\vec{\nabla}\cdot\vec{V}&=\frac{L\alpha}{4\pi^2}\vec{E}\cdot\vec{B} \label{3.53}\\
\intertext{\hspace{5mm}$\bullet$ Homogeneous Maxwell equations, $dF=0$:}
\vec{\nabla}\cdot\vec{B}&=0\label{3.54}\\
\vec{\nabla}\times\vec{E}+\partial_0\vec{B}&=0\label{3.55}\\
\vec{\nabla}E_4&=\partial_4\vec{E}+\partial_0\vec{V}\label{3.56}\\
\partial_4\vec{B}&=\vec{\nabla}\times\vec{V}\label{3.57}
\end{align}

Remarks:
\begin{enumerate}
\item In deriving the (4+1)-dimensional equations we assumed that $\vec{B}$ is $x^4$-independent (see (\ref{3.31})). From (\ref{3.57}) we now find that this condition is satisfied if $\vec{\nabla}\times\vec{V}=0$, which leads us to postulate that the vector field $V$ can be written as the gradient of some scalar field $\phi$:

\begin{equation}
\vec{V}=\vec{\nabla}\phi.
\label{3.58}
\end{equation} 

\item For an arbitrary $A(x)$, the boundary action $\Gamma_{\partial\Lambda}$ produces a current density on the (3+1) dimensional boundary of the slab. However, if $A(x)$ can be chosen such that

\begin{equation}
A(x)|_{\partial\Lambda}=0,
\label{3.59}
\end{equation}

then $\Gamma_{\partial\Lambda}$ does not contribute to the current density. Unfortunately, if $\vec{B}$ is $x^4$-independent (as it is the case for $\vec{V}=\vec{\nabla}\phi$), then any solution with non-vanishing $\vec{B}$-field is incompatible with the above boundary condition. So we either have to relax the condition on $\vec{V}$ or take care of the currents produced by $\Gamma_{\partial\Lambda}$.
\end{enumerate}

\subsection{Projection onto (3+1) dimensions}

In order to obtain the physical quantities observable in (3+1) dimensions as well as their equations of motion, one has to project the (4+1)-dimensional fields onto the the (3+1)-dimensional space-time identified with the planes $x^4=0$ and $x^4=L$. For arbitrary $x^4$-dependent fields it is not possible to express the equations of motion in terms of their averaged, $x^4$-independent counterparts. Some simplifying assumptions concerning the $x^4$-dependence of the (4+1)-dimensional fields are therefore inevitable.

We should also insist on the requirement that the left- and righthanded fermions propagating along the surfaces $x^4=L$ and $x^4=0$ respectively couple to the \textit{same} electromagnetic vector potential, that is 

\begin{equation}
A^{(4)}(x,L)=A^{(4)}(x,0),
\label{3.60}
\end{equation} 

where $A^{(4)}\equiv(A_0,\ldots,A_3)$ and $x\equiv(x^0,\ldots,x^3)$. This requirement is met if we assume that $A^{(5)}=(A^{(4)},A_4)$ is independent of $x^4$. In this particular case an averaging procedure for the fields $\vec{B}$, $\vec{E}$ and $\vec{V}$ is not necessary. The boundary action for $x^4$-independent fields is $\Gamma_{\partial\Lambda}^{(4)}[A^{(4)}]=-i\text{ln}\{\text{det}(\D_{A^{(4)}}^l)\text{det}(\D_{A^{(4)}}^r)\}$ and the boundary current now becomes $-\alpha\frac{\delta\Gamma_{\partial\Lambda}^{(4)}}{\delta A}$. Furthermore, all the partial derivatives $\partial_4$ appearing in (\ref{3.51})-(\ref{3.57}) can be replaced by zero. The equations of motion in (3+1) dimensions therefore become

\hspace{5mm} $\bullet$ ``Inhomogeneous Maxwell equations'':
\begin{align}
\vec{\nabla}\cdot\vec{E}&=-\frac{L\alpha}{4\pi^2}\vec{V}\cdot\vec{B}-\alpha\frac{\delta\Gamma_{\partial\Lambda}^{(4)}}{\delta A_0}\label{3.61}\\
\vec{\nabla}\times\vec{B}-\partial_0\vec{E}&=\frac{L\alpha}{4\pi^2}(E_4\vec{B}+\vec{V}\times\vec{E})-\alpha\frac{\delta\Gamma_{\partial\Lambda}^{(4)}}{\delta \vec{A}}\label{3.62}\\
\partial_0E_4-\vec{\nabla}\cdot\vec{V}&=\frac{L\alpha}{4\pi^2}\vec{E}\cdot\vec{B} \label{3.63}\\
\intertext{\hspace{5mm}$\bullet$ ``Homogeneous Maxwell equations'':}
\vec{\nabla}\cdot\vec{B}&=0\label{3.64}\\
\vec{\nabla}\times\vec{E}+\partial_0\vec{B}&=0\label{3.65}\\
\vec{\nabla}E_4&=\partial_0\vec{V}\label{3.66}\\
\vec{\nabla}\times\vec{V}&=0\label{3.67}.
\end{align}

Remarks:

\begin{enumerate}
\item The last two equations (\ref{3.66}) and (\ref{3.67}) can be satisfied by setting 

\begin{equation}
\vec{V}=\vec{\nabla}\phi,\hspace{5mm}E_4=\partial_0\phi,
\label{3.68}
\end{equation}

where $\phi$ is an arbitrary scalar field whose interpretation will be clarified in section 3.4. Substituting (\ref{3.68}) into (\ref{3.61})-(\ref{3.65}), we obtain the following system of equations

\begin{align}
\intertext{\hspace{5mm}$\bullet$ Inhomogeneous Maxwell equations and equation of motion for $\phi$:}
\vec{\nabla}\cdot\vec{E}&=-\frac{L\alpha}{4\pi^2}\vec{\nabla}\phi\cdot\vec{B}-\alpha\frac{\delta\Gamma_{\partial\Lambda}^{(4)}}{\delta A_0}\label{3.69}\\
\vec{\nabla}\times\vec{B}-\partial_0\vec{E}&=\frac{L\alpha}{4\pi^2}(\dot\phi\vec{B}+\vec{\nabla}\phi\times\vec{E})-\alpha\frac{\delta\Gamma_{\partial\Lambda}^{(4)}}{\delta \vec{A}}\label{3.70}\\
\Box\phi&=\frac{L\alpha}{4\pi^2}\vec{E}\cdot\vec{B}\label{3.71}\\
\intertext{\hspace{5mm}$\bullet$ Homogeneous Maxwell equations:}
\vec{\nabla}\cdot\vec{B}&=0\label{3.72}\\
\vec{\nabla}\times\vec{E}+\partial_0\vec{B}&=0.\label{3.73}
\end{align}

\item At the beginning of this chapter, we set out to derive an equation describing the time evolution of the difference in chemical potentials appearing in (\ref{3.11}). We now show that the system (\ref{3.61})-(\ref{3.67}) indeed provides such an equation and therefore generalizes the system (\ref{3.10})-(\ref{3.13}) obtained in section 3.1. Setting $V=0$, neglecting the term in $\Gamma_{\partial\Lambda}^{(4)}$ and introducing the notation

\begin{equation}
\mu_l-\mu_r=-LE_4
\label{3.74}
\end{equation}

consistent with the (4+1)-dimensional QH-analogy we find

\begin{itemize}
\item Inhomogeneous Maxwell equations and equation of motion for $\mu_l-\mu_r$:
\begin{eqnarray}
\vec{\nabla}\cdot\vec{E}&=&0\label{3.75}\\
\vec{\nabla}\wedge\vec{B}-\partial_0\vec{E}&=&-\frac{\alpha}{4\pi^2}(\mu_l-\mu_r)\vec{B}\label{3.76}\\
\partial_0(\mu_l-\mu_r)&=&-\frac{\alpha L^2}{4\pi^2}\vec{E}\cdot\vec{B}\label{3.77}
\end{eqnarray}
\item Homogeneous Maxwell equations and additional condition for $\mu_l-\mu_r$:
\begin{eqnarray}
\vec{\nabla}\cdot\vec{B}&=&0\label{3.78}\\
\vec{\nabla}\wedge\vec{E}+\partial_0\vec{B}&=&0\label{3.79}\\
\vec{\nabla}(\mu_l-\mu_r)&=&0.\label{3.80}
\end{eqnarray}
\end{itemize}   

Note that the condition of $x^4$-independence for $E$ and $B$ may be relaxed if the field $V$ is ignored. Indeed, a set of equations similar to (\ref{3.75})-(\ref{3.80}) is obtained by defining the electric field in (3+1) dimensions by $\vec{E}(x^0,\vec{x})=\frac{1}{L}\int_0^Ldx^4\vec{E}(x^0,\vec{x},x^4)$ and the difference in chemical potentials by $\mu_l-\mu_r\equiv-\int_0^Ldx^4E_4(x^0,\vec{x},x^4)$. However, we shall not pursue this possibility any further and content ourselves with the phenomena produced by $x^4$-independent fields. 
\end{enumerate}

\newpage

\section{Massive (4+1)-dimensional fermions confined to a slab}

After a first, more intuitive derivation of the (4+1)-dimensional theory based on an analogy with the quantum Hall effect, we will show in this section that the effective action (\ref{3.45}) may also be obtained in another way. While the introduction of a fifth dimension in the preceeding discussion could be regarded as a mathematical ``trick'', we now take this point of view more seriously and identify the (3+1)-dimensional world with a domain wall in a (4+1)-dimensional space-time. More precisely we admit the existence of a slab $\Lambda=(x|0\le x^4\le L)$ filled with (4+1)-dimensional fermions of mass $m$. Two domain walls at $x^4=0$ and $x^4=L$ are introduced by setting the fermion mass to $-M$ outside $\Lambda$. The calculation of the fermion determinant in the limit $M\rightarrow \infty$ produces the effective action (\ref{3.45}) up to terms of order $O(\frac{1}{m})$ and the massless chiral fermions in (3+1) dimensions appear as surface modes localized near the domain walls.     

\subsection{Calculation of the fermion determinant}

For the moment we ignore the existence of domain walls. Our aim is to calculate the fermionic effective action $S_{eff}[A,m]$, which is defined as

\begin{equation}
S_{eff}[A,m]=-i\text{ln}\int \mathcal D\psi \mathcal D\bar\psi e^{i\int d^5x\{\bar\psi(i\not D_A-m)\psi\}}=-i\ln \text{det}(\D_A+im)
\label{3.81}
\end{equation}

In the present context the symbol $i\D_A$ denotes the Dirac operator in (4+1) dimensions, $i\D_A=i\gamma^\mu(\partial_\mu-iA_\mu)$, $\mu=0,\ldots,4$, and $\gamma^4=i\gamma^5$, so that $\{\gamma^4,\gamma^4\}=-2=2g^{44}$. In order to evaluate the fermion determinant (\ref{3.81}) we define the operators $\A$ and $S$ through

\begin{align}
\A|x\rangle&=\A(x)|x\rangle=\gamma^\mu A_\mu(x)|x\rangle\label{3.82}\\
\langle x|S|y\rangle&=\int \frac{d^5k}{(2\pi)^5}\frac{\K+m}{k^2-m^2}e^{-ik(x-y)}.\label{3.83}
\end{align}

Since $\langle x|(i\D-m)S|y\rangle=(i\D^x-m)\langle x|S|y\rangle=\delta(x-y)=\langle x|y\rangle$, the following identities hold

\begin{align}
i(\D-m)S&=1\label{3.84}\\
(\D +im)(1+S\A)&=\D_A+im.\label{3.85}
\end{align}
  
Plugging (\ref{3.85}) into (\ref{3.81}) yields $S_{eff}[A,m]=-i\text{ln[det}(\D+im)\text{det}(1+S\A)]$. The first determinant is independent of $A$ and adds an irrelevant constant to the effective action, which we shall ignore. Using $\text{det}(1+S\A)=\text{expTrln}(1+S\A)$ and writing the logarithm as a power series we find

\begin{equation}
\boxed{S_{eff}[A,m]=i\sum_{n=1}^\infty\frac{(-1)^n}{n}\text{Tr}(S\A)^n.}
\label{3.86}
\end{equation}

The explicit expression for $\text{Tr}(S\A)^n$ reads

\begin{align}
\text{Tr}(S\A)^n&=\text{tr}\int dx\langle x|S\A\ldots S\A|x\rangle\nonumber\\
&=\text{tr}\int dx_1\ldots dx_nS(x_1-x_2)\A(x_2)\ldots S(x_n-x_1)\A(x_1)\nonumber\\
&=\int dx_1\ldots dx_nA_{\mu_1}(x_1)\ldots A_{\mu_n}(x_n)\int \frac{d^5k_1}{(2\pi)^5}\ldots \frac{d^5k_n}{(2\pi)^5}\nonumber\\
&\hspace{4mm}\times\text{tr}\left[\frac{\K_1+m}{k_1^2-m^2}\gamma^{\mu_1}\ldots\frac{\K_n+m}{k_n^2-m^2}\gamma^{\mu_n}\right]e^{-ik_1(x_1-x_2)}\ldots e^{-ik_n(x_n-x_1)},
\label{3.87}
\end{align}

where tr denotes the trace over the $\gamma$-matrices. An $i\epsilon$-prescription is implicitely contained in the definition of the fermion mass. We shall therefore write $-m^2$ rather than $-m^2+i\epsilon$ in expressions such as (\ref{3.87}). A notation consistent with the Feynman rules could be obtained by writing $S\A=S_F(-i\A)$. The right hand side of (\ref{3.87}) is then identified as the sum of the contributions of one-loop diagrams made of $n$ Feynman propagators $S_F=iS$ and $n$ vertices $-i\A$.

The contributions from the diagrams with $n\le 5$ are potentially divergent, while those from higher order diagrams vanish in the limit $m\rightarrow \infty$. We will now proceed to the calculation of the amplitudes for $n=1$, 2 and 3.

\subsection{Contribution in $A$}

\begin{figure}[htbp]
\centering
\includegraphics[width=0.25\textwidth]{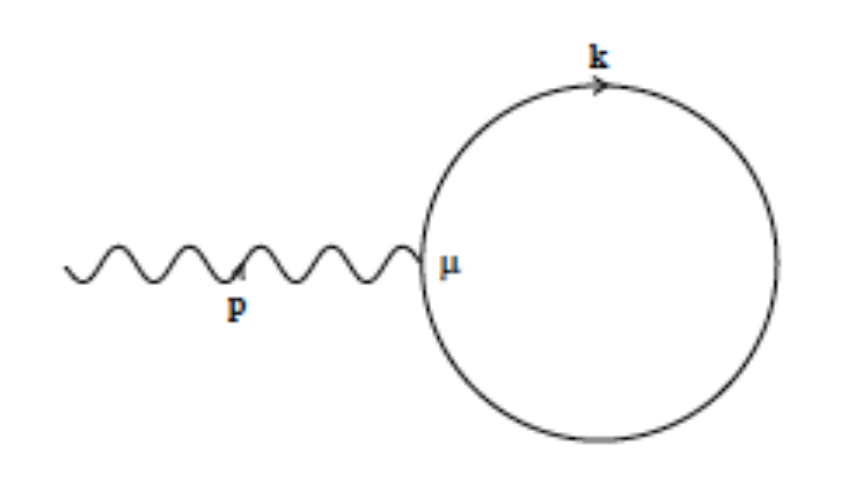}
\caption{Diagram corresponding to $\omega^\mu$.}
\label{n=1}
\end{figure}

$\text{Tr}(S\A)$ is easily seen to vanish. We will nonetheless treat this simple case systematically in order to warm up for the following calculations. From (\ref{3.87}) we find for $n$=1

\begin{equation}
\text{Tr}(S\A)=\int d^5x A_{\mu}(x) \int\frac{d^5k}{(2\pi)^5}\text{tr}\left[\frac{\K+m}{k^2-m^2}\gamma^{\mu}\right].
\label{3.88}
\end{equation}

The $k$-integral gives the amplitude corresponding to the diagram in figure \ref{n=1}:

\begin{equation}
\omega^\mu=\int\frac{d^5k}{(2\pi)^5}\text{tr}\left[\frac{\K+m}{k^2-m^2}\gamma^{\mu}\right].
\label{3.89}
\end{equation}

We now look at the different contributions found by evaluating the numerator in (\ref{3.89}). The term linear in $m$ vanishes since $\text{tr}\gamma^\mu=0$ and the remaining one is is odd in $k$, so $\int d^5k\text{tr}[...]$ yields zero. Hence $\text{Tr}(S\A)=0$, so that

\begin{equation}
S_{eff}^{(n=1)}=0.
\label{3.90}
\end{equation}

\subsection{Contribution in $A^2$}

\begin{figure}[htbp]
\centering
\includegraphics[width=0.4\textwidth]{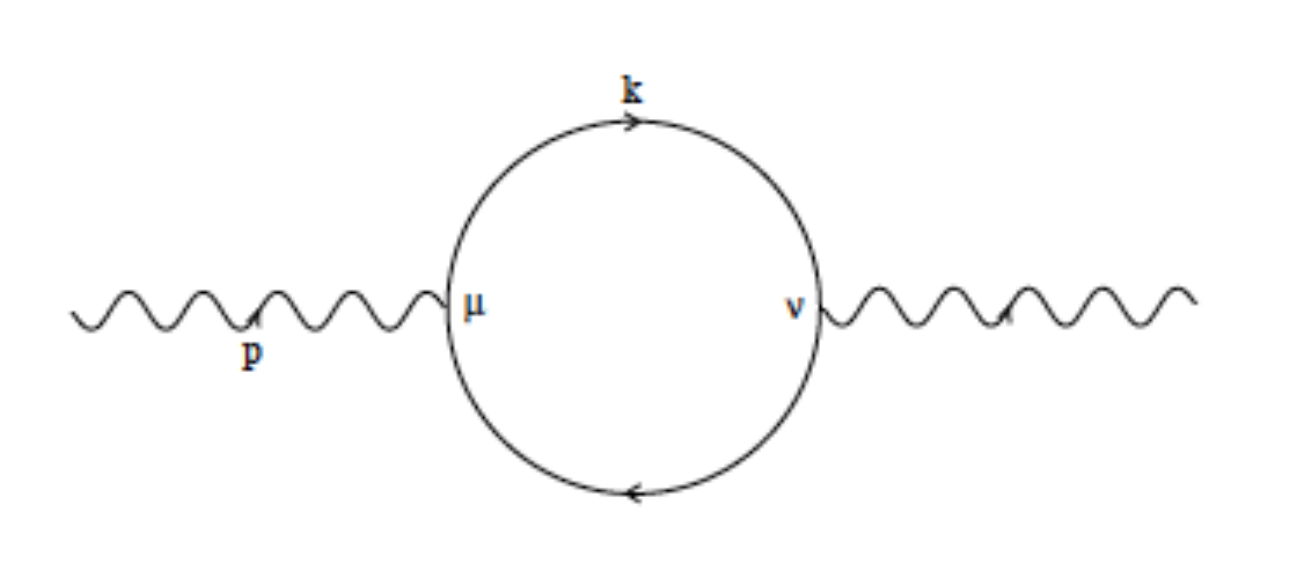}
\caption{Diagram corresponding to $\omega^{\mu\nu}$.}
\label{n=2}
\end{figure}

The calculation of the amplitude associated with the vacuum polarization diagram shown in figure \ref{n=2} is somewhat more involved. In the limit $m\rightarrow \infty$ it will lead to an expression  of the form $\int d^5x F^2$, with a divergent prefactor. This contribution may be combined with the Maxwell term to yield a finite result, if the bare charge of the fermions is appropriately redefined (charge renormalization).

For the actual calculation we proceed as before. Starting from equation (\ref{3.87}) for $n=2$ and changing variables according to $k_1\rightarrow k$, $k_2\rightarrow k-p$ one finds

\begin{equation}
\text{Tr}(S\A)^2=\int dx\int dyA_\mu(x)A_\nu(y)\int\frac{d^5p}{(2\pi)^5}\int\frac{d^5k}{(2\pi)^5}\text{tr}\left[\frac{\K+m}{k^2-m^2}\gamma^\mu\frac{\K-\p+m}{(k-p)^2-m^2}\gamma^\nu\right]e^{-ip(x-y)}.
\label{3.91}
\end{equation}

The next step is to define the function $\omega^{\mu\nu}$ as

\begin{equation}
\omega^{\mu\nu}(p)=\int\frac{d^5k}{(2\pi)^5}\text{tr}\left[\frac{\K+m}{k^2-m^2}\gamma^\mu\frac{\K-\p+m}{(k-p)^2-m^2}\gamma^\nu\right],
\label{3.92}
\end{equation}

which corresponds to the amplitude associated with the vacuum polarisation diagram shown in figure \ref{n=2}. The integral (\ref{3.92}) defining $\omega^{\mu\nu}$ - which looks quadratically divergent for large internal momentum $k$ - may be given a meaning using \textit{Pauli-Villars regularization}. This amounts to coupling the gauge field to additional spinor fields with a very large mass $\lambda_sm$ and (eventually) different statistics. Such a prescription implies the replacement

\begin{equation}
\omega^{\mu\nu}(p,m)\rightarrow\omega^{\mu\nu}(p,m)+\sum_{s=1}^SC_s\omega^{\mu\nu}(k,\lambda_sm),
\label{3.93}
\end{equation}

where the substitution is understood under the integral sign in (\ref{3.91}). The constants $C_s$ will be chosen in order to remove the divergence of the integral. The following calculation is based on the discussion of the analogous problem in (3+1) dimensions found in the book by Itzykson and Zuber \footnote{Chapter 7: Radiative corrections, p. 319-323} \cite{Zuber}.

Denoting the large masses $\lambda_sm$ collectively by $\Lambda$ we find

\begin{align}
\omega^{\mu\nu}(p,m,\Lambda)&=\int\frac{d^5k}{(2\pi)^5}\text{tr}\left\{\left[\frac{\K+m}{k^2-m^2}\gamma^\mu\frac{\K-\p+m}{(k-p)^2-m^2}\gamma^\nu\right]+\sum_{s=1}^SC_s(m\rightarrow\lambda_sm)\right\}\nonumber\\
&=4\int\frac{d^5k}{(2\pi)^5}\Bigg\{\frac{k^\mu(k-p)^\nu+k^\nu(k-p)^\mu-g^{\mu\nu}(k^2-kp-m^2)}{(k^2-m^2)((k-p)^2-m^2)}\nonumber\\
&\hspace{2.45cm}\left.+\sum_{s=1}^SC_s(m\rightarrow\lambda_sm)\right\}
\label{3.94}
\end{align}

By rewriting denominator of (\ref{3.94}) using the formula \cite{Zuber}

\begin{equation}
\frac{1}{k^2-m^2}=-i\int_0^\infty d\alpha e^{i\alpha(k^2-m^2)}
\label{3.95}
\end{equation}

and introducing auxiliary five-vectors $z_1$ and $z_2$, one can generate the integrand in (\ref{3.94}) by differentiation

\begin{align}
\omega_{\mu\nu}(p,m,\Lambda)&=
4\int_0^\infty d\alpha_1\int_0^\infty d\alpha_2\int\frac{d^5k}{(2\pi)^5}\Bigg\{\bigg[\frac{\partial}{\partial z_1^\mu}\frac{\partial}{\partial z_2^\nu}+\frac{\partial}{\partial z_1^\nu}\frac{\partial}{\partial z_2^\mu}-g_{\mu\nu}\Big(\frac{\partial}{\partial z_1}\frac{\partial}{\partial z_2}+m^2\Big)\bigg]\nonumber\\
&\hspace{.4cm}\times e^{i(\alpha_1(k^2-m^2)+\alpha_2((k-p)^2-m^2)+z_1k+z_2(k-p))}+\sum_{s=1}^SC_s(m\rightarrow\lambda_sm)\Bigg\}_{z_1=z_2=0}.
\label{3.96}
\end{align}

Integrating over $k$ (using Fresnel) and performing the required derivatives yields

\begin{align}
\omega_{\mu\nu}(p,m,\Lambda)&=\frac{e^{-\frac{3}{4}i\pi}}{8\pi^{\frac{5}{2}}}\int_0^\infty\int_0^\infty  \frac{d\alpha_1d\alpha_2}{(\alpha_1+\alpha_2)^{\frac{5}{2}}}\Bigg\{\bigg[\frac{2\alpha_1\alpha_2p_\mu p_\nu}{(\alpha_1+\alpha_2)^2}-g_{\mu\nu}\Big(\frac{\alpha_1\alpha_2p^2}{(\alpha_1+\alpha_2)^2}-\frac{i}{\alpha_1+\alpha_2}+m^2\Big)\bigg]\nonumber\\
&\hspace{.4cm}\times e^{i\{-m^2(\alpha_1+\alpha_2)+\frac{\alpha_1\alpha_2}{\alpha_1+\alpha_2}p^2\}}+\sum_{s=1}^SC_s(m\rightarrow\lambda_sm)\Bigg\}
\label{3.97}
\end{align}

The polynomial in $p$ appearing in the above integral can be rearranged to read

\begin{equation}
\bigg[\ldots\bigg]=2(p_\mu p_\nu-g_{\mu\nu}p^2)\frac{\alpha_1\alpha_2}{(\alpha_1+\alpha_2)^2}-g_{\mu\nu}\bigg[m^2-\frac{\alpha_1\alpha_2}{(\alpha_1+\alpha_2)^2}p^2-\frac{i}{\alpha_1+\alpha_2}\bigg].
\label{3.98}
\end{equation}

We will use the symbol $\Delta\omega$ to denote the contibution to $\omega_{\mu\nu}(p,m,\Lambda)$ coming from the second term on the right hand side of (\ref{3.98}) and the analogous terms involving $\lambda_sm$. Since Pauli-Villars regularization preserves gauge invariance and the latter term does not seem to exhibit this property, we shall admit that its contribution is zero. Under the assumption that $\Delta\omega$ vanishes, $\omega_{\mu\nu}(p,m,\Lambda)$ reduces to

\begin{align}
\omega_{\mu\nu}(p,m,\Lambda)&=\frac{e^{-\frac{3}{4}i\pi}}{4\pi^{\frac{5}{2}}}(p_\mu p_\nu-g_{\mu\nu}p^2)\omega(p^2,m,\Lambda),
\label{3.99}
\end{align}

where $\omega(p^2,m,\Lambda)$ is defined as

\begin{align}
\omega(p^2,m,\Lambda)&=\int_0^\infty\int_0^\infty d\alpha_1d\alpha_2 \frac{\alpha_1\alpha_2}{(\alpha_1+\alpha_2)^{\frac{9}{2}}}\sum_{s=0}^SC_s e^{i\left\{-m_s^2(\alpha_1+\alpha_2)+\frac{\alpha_1\alpha_2}{\alpha_1+\alpha_2}p^2\right\}}.
\label{3.100}
\end{align}

We used the notation $C_0=1$, $m_0^2=m^2$ and $m_s^2=\lambda_s^2m^2$. Introducing a factor $1=\int_0^\infty d\rho\delta(\rho-\alpha_1-\alpha_2)$ under the integral sign and changing the variables according to $\alpha_i=\rho\beta_i$ we find

\begin{equation}
\omega(p^2,m,\Lambda)=\int_0^1\int_0^1 d\beta_1d\beta_2\delta(1-\beta_1-\beta_2)\beta_1\beta_2\underbrace{\int_0^\infty\frac{d\rho}{\rho^{\frac{3}{2}}}\sum_{s=0}^SC_s e^{i\rho(-m_s^2+\beta_1\beta_2p^2)}}_{\equiv I_\rho}.
\label{3.101}
\end{equation}

\begin{figure}[htbp]
\centering
\includegraphics[width=0.4\textwidth]{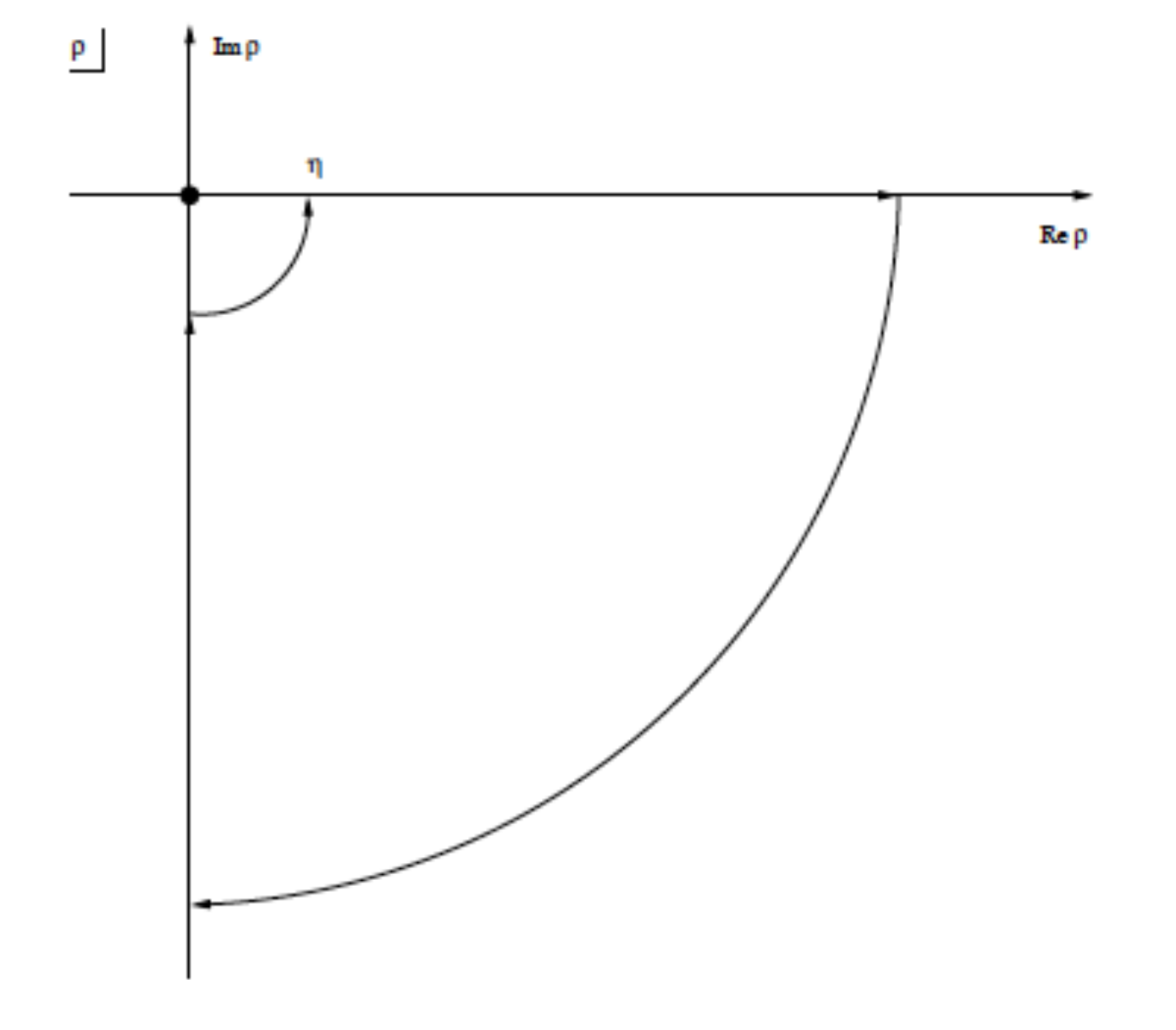}
\caption{Rotation of the integration contour in the complex $\rho$-plane.}
\label{rot_1}
\end{figure}

Since $0\le\beta_1,\beta_2\le 1$ and $\beta_1+\beta_2=1$ it follows that $\beta_1\beta_2\le\frac{1}{4}$. So if we pick $p^2\le4m^2$, then $m_s^2-\beta_1\beta_2p^2$ is positive and the integration contour in the complex $\rho$-plane can be rotated by $-\frac{\pi}{2}$ in such a way that the integral $I_\rho$ defined in (\ref{3.101}) reads

\begin{align}
I_\rho&=\lim_{\eta\rightarrow 0}\sum_{s=0}^SC_s\bigg[\frac{2i}{\eta^\frac{1}{2}}(e^{i\frac{\pi}{4}}-1)+\frac{1}{(-i)^\frac{1}{2}}\int_\eta^\infty\frac{d\rho}{\rho^{\frac{3}{2}}} e^{-\rho(m_s^2-\beta_1\beta_2p^2)}\bigg],
\label{3.102}
\end{align}

see figure \ref{rot_1}. The first term on the right hand side of \ref{3.102} is the contribution coming from the small quarter circle of radius $\eta$ (up to a correction of order O($\sqrt{\eta}$)) and the second term corresponds to the integration over the negative $\text{Im}\rho$-axis.

\vspace{5mm}
Changing the integration variable from $\rho$ to $\rho\sqrt{m_s^2-\beta_1\beta_2p^2}$ and integrating by parts, we obtain

\begin{align}
&I_\rho=\lim_{\eta\rightarrow 0}\sum_{s=0}^SC_s\bigg[\frac{2i}{\eta^\frac{1}{2}}(e^{i\frac{\pi}{4}}-1)+\frac{1}{(-i)^\frac{1}{2}}\sqrt{m_s^2-\beta_1\beta_2p^2}\int_{\eta(m_s^2-\beta_1\beta_2p^2)}^\infty\frac{d\rho}{\rho^{\frac{3}{2}}} e^{-\rho}\bigg]\nonumber\\
&=\lim_{\eta\rightarrow 0}\sum_{s=0}^SC_s\bigg[\frac{2i}{\eta^\frac{1}{2}}(e^{i\frac{\pi}{4}}-1)+\frac{1}{(-i)^\frac{1}{2}}\sqrt{m_s^2-\beta_1\beta_2p^2}\bigg(\frac{2}{\rho^\frac{1}{2}}e^{-\rho}\Big|_{\eta(m_s^2-\beta_1\beta_2p^2)}-2\int_0^\infty\frac{d\rho}{\rho^\frac{1}{2}}e^{-\rho}\bigg)\bigg]\nonumber\\
&=\lim_{\eta\rightarrow 0}\sum_{s=0}^SC_s\bigg[-\frac{2i}{\eta^\frac{1}{2}}-\frac{2}{(-i)^\frac{1}{2}}\sqrt{m_s^2-\beta_1\beta_2p^2}\sqrt{\pi}\bigg].
\label{3.103}
\end{align}

The constants $C_s$ have to be chosen such that the divergent term proportional to $\eta^{-\frac{1}{2}}$ disappears. This is achieved by setting 

\begin{equation}
\sum_{s=0}^SC_s\equiv 1+\sum_{s=1}^SC_s=0.
\label{3.104}
\end{equation}

Introducing the remaining term into (\ref{3.101}) then leads to 

\begin{equation}
\omega(p^2,m,\Lambda)=-\frac{2\sqrt{\pi}}{(-i)^{\frac{1}{2}}}\sum_{s=0}^SC_s\int_0^1 d\beta\beta(1-\beta)\sqrt{m_s^2-\beta(1-\beta)p^2}.
\label{3.105}
\end{equation}

Developing (\ref{3.105}) in powers of $p^2$ yields the result

\begin{equation}
\omega(p^2,m,\Lambda)=-\frac{\sqrt{\pi}}{3(-i)^{\frac{1}{2}}}|m|\sum_{s=0}^SC_s\lambda_s+O(\frac{p^2}{m}),
\label{3.106}
\end{equation}

where the $O(\frac{p^2}{m})$ also depends on the constants $C_s$ and $\lambda_s$, but unlike the first term - which corresponds to $\omega(0,m,\Lambda)$ - does not diverge if the regulator masses are taken to infinity. We now introduce the ultraviolet cutoff $\Lambda$ through the relation

\begin{equation}
\sum_{s=0}^SC_s\lambda_s\equiv 1+\sum_{s=1}^SC_s\lambda_s=\Lambda.
\label{3.107}
\end{equation}

Given the value of $\Lambda$, a possible choice compatible with (\ref{3.104}) and (\ref{3.107}) would be $S=2$, $C_1=-2$, $C_2=1$, $\lambda_1=\lambda$ and $\lambda_2=(\Lambda-1)+2\lambda$.

Upon substitution of (\ref{3.107}) and (\ref{3.106}) into (\ref{3.99}) we find 

\begin{align}
\omega_{\mu\nu}(p^2,m,\Lambda)&=\frac{i\Lambda|m|}{12\pi^{2}}(p_\mu p_\nu-g^{\mu\nu}p^2)+\ldots,
\label{3.108}
\end{align}

The additional contribution coming from the $O(\frac{p^2}{m})$ in (\ref{3.106}) has been indicated with triple dots since it vanishes in the limit $m\rightarrow\infty$. By introducing the leading term into (\ref{3.91}) one obtains

\begin{align}
\text{Tr}(S\A)^2&=-\frac{i\Lambda|m|}{24\pi^2}\int dx\int dyA_\mu(x)A_\nu(y)\int\frac{d^5p}{(2\pi)^5}\{2g^{\mu\nu}p^2-p^\mu p^\nu-p^\nu p^\mu\}e^{-ip(x-y)}\nonumber\\
&=-\frac{i\Lambda|m|}{24\pi^2}\int dx\int dyA_\mu(x)A_\nu(y)\left\{2g^{\mu\nu}\frac{\partial}{\partial x^\alpha}\frac{\partial}{\partial y_\alpha}-\frac{\partial}{\partial x_\mu}\frac{\partial}{\partial y_\nu}-\frac{\partial}{\partial x_\nu}\frac{\partial}{\partial y_\mu}\right\}\delta(x-y)\nonumber\\
&=-\frac{i\Lambda|m|}{24\pi^2}m\int d^5y \int d^5x (\partial^\mu A^{\nu}(x)-\partial^\nu A^\mu(x))(\partial_\mu A_{\nu}(y)-\partial_\nu A_\mu(y))\delta(x-y)\nonumber\\
&=-\frac{i\Lambda|m|}{24\pi^2}m\int d^5x F^{\mu\nu}(x)F_{\mu\nu}(x).
\label{3.109}
\end{align}

The contribution to the action $S_{eff}$ coming from $\text{Tr}(S\A)^2$ is found from (\ref{3.109}) and (\ref{3.86}). It reads

\begin{equation}
\boxed{S_{eff}^{(n=2)}=\frac{\Lambda|m|}{48\pi^2}\int d^5x F^{\mu\nu}F_{\mu\nu}+O(\frac{1}{m}).}
\label{3.110}
\end{equation} 

As we shall see in section 3.3.5, it is possible to redefine the bare charge $e^b$ appearing in the Lagrangian density in such a way, that the $S_{eff}^{(n=2)}$ and the Maxwell term with bare coupling constant $\alpha^b$ ($S_{EM}^b=-\frac{1}{4L\alpha^b}\int d^5xF^2$) add up to  $S_{EM}=-\frac{1}{4L\alpha}\int d^5xF^2$.

\subsection{Contribution in $A^3$}

\begin{figure}[htbp]
\centering
\includegraphics[width=0.4\textwidth]{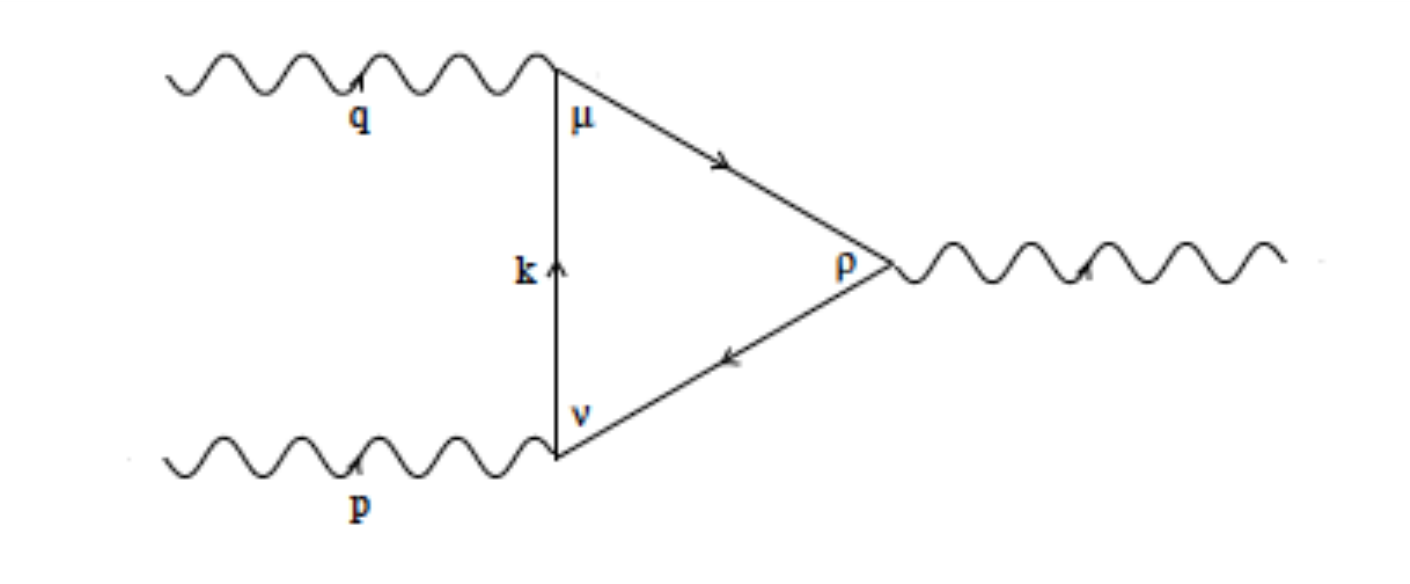}
\caption{Diagram corresponding to $\omega^{\mu\nu\rho}$.}
\label{n=3}
\end{figure}

$\text{Tr}(S\A)^3$ leads to the Chern-Simons 5-form $A\wedge F\wedge F$. Equation (\ref{3.87}) for $n=3$ and after a change of variables $k_2\rightarrow k$, $k_1\rightarrow k+q$, $k_3\rightarrow k-p$ reads

\begin{align}
\text{Tr}(S\A)^3&=\int d^5x \int d^5y \int d^5z A_{\mu}(x)A_\nu(y)A_\rho(z)\int\frac{d^5p}{(2\pi)^5}\int\frac{d^5q}{(2\pi)^5}\nonumber\\
&\times\int\frac{d^5k}{(2\pi)^5}\text{tr}\left[\frac{\K+\q+m}{(k+q)^2-m^2}\gamma^\mu\frac{\K+m}{k^2-m^2}\gamma^\nu\frac{\K-\p+m}{(k-p)^2-m^2}\gamma^\rho\right]e^{-iq(x-y)}e^{-ip(x-z)}.
\label{3.111}
\end{align}

The function $\omega^{\mu\nu\rho}(p,q)$ which (up to a factor of $-1$) corresponds to the amplitude associated with the triangle diagram shown in figure \ref{n=3} is

\begin{align}
\omega^{\mu\nu\rho}(p,q)&=\int\frac{d^5k}{(2\pi)^5}\text{tr}\left[\frac{\K+\q+m}{(k+q)^2-m^2}\gamma^\mu\frac{\K+m}{k^2-m^2}\gamma^\nu\frac{\K-\p+m}{(k-p)^2-m^2}\gamma^\rho\right].
\label{3.112}
\end{align}

Next, one tries to identify those terms in the nominator which do contribute to the final result. The term in $m^3$ vanishes since $\text{tr}(\gamma^\mu\gamma^\nu\gamma^\delta)=0$. Those in $m$ are necessarily of the form $\text{tr}(\not\!a\gamma^\mu\gamma^\nu\!\!\not\!b\gamma^\delta)=4\epsilon^{\alpha\mu\nu\beta\delta}a_\alpha b_\beta$ ($\gamma^4=i\gamma^5$) since the nonvanishing traces of this type must contain exactly one matrix $\gamma^4$. Furthermore the contributions linear in $k$ cancel and we cannot have $a=b=k$ because the $\epsilon$-tensor is completely antisymmetric. The only remaining possibility is $a=q$ and $b=-p$ that is 

\begin{equation}
\text{tr}(\q\gamma^\mu\gamma^\nu(-\p)\gamma^\delta)=-4\epsilon^{\alpha\mu\nu\beta\delta}q_\alpha p_\beta.
\label{3.113}
\end{equation}

The sum of the terms in $m^2$ and $m^0$ does not seem to meet the requirements of gauge invariance and Lorentz covariance, so we shall simply admit that this contribution vanishes. Under this assumption we find

\begin{align}
\omega^{\mu\nu\delta}(p,q)&=-4m\epsilon^{\alpha\mu\nu\beta\delta}q_\alpha p_\beta \omega(p,q,m),
\label{3.114}
\end{align}

where the function $\omega(p,q,m)$ is defined as

\begin{align}
\omega(p,q,m)&=\int\frac{d^5k}{(2\pi)^5}\frac{1}{((k+q)^2-m^2)(k^2-m^2)((k-p)^2-m^2)}\nonumber\\
&=2\int_0^1dx\int_0^{1-x}dy\int\frac{d^5k}{(2\pi)^5}\frac{1}{[k^2+2kP+M^2]^3},
\label{3.115}
\end{align}

The variables $P=qy-px$ and $M^2=q^2y+p^2x-m^2$ have been introduced for simplicity. The integral over $k$ is convergent and could be performed using one of the formulas listed in the appendix of \cite{Pokorski}. It might nonetheless be useful to do this calculation in more detail, were it only for earning a better understanding where such formulas come from.

Changing variables in (\ref{3.115}) according to $k\rightarrow k+P$ (which is legitimate for a convergent integral) and remembering the $i\epsilon$-prescription implicitely contained in the definition of $m$ we obtain

\begin{align}
\int\frac{d^5k}{(2\pi)^5}\frac{1}{[k^2+2kP+M^2]^3}&=\int\frac{d^5k}{(2\pi)^5}\frac{1}{[{k}^2-\{q^2(y^2-y)+p^2(x^2-x)-2qpxy+m^2-i\epsilon\}]^3}.
\label{3.116}
\end{align}

Since $k^2=k_0^2-\vec{k}^2$, the poles in the complex $k_0$-plane are located at

\begin{align}
k_0&=\pm[\vec{k}^2+\{q^2(y^2-y)+p^2(x^2-x)-2qpxy+m^2-i\epsilon\}]^{\frac{1}{2}}.
\label{3.117}
\end{align}

\begin{figure}[htbp]
\centering
\includegraphics[width=0.4\textwidth]{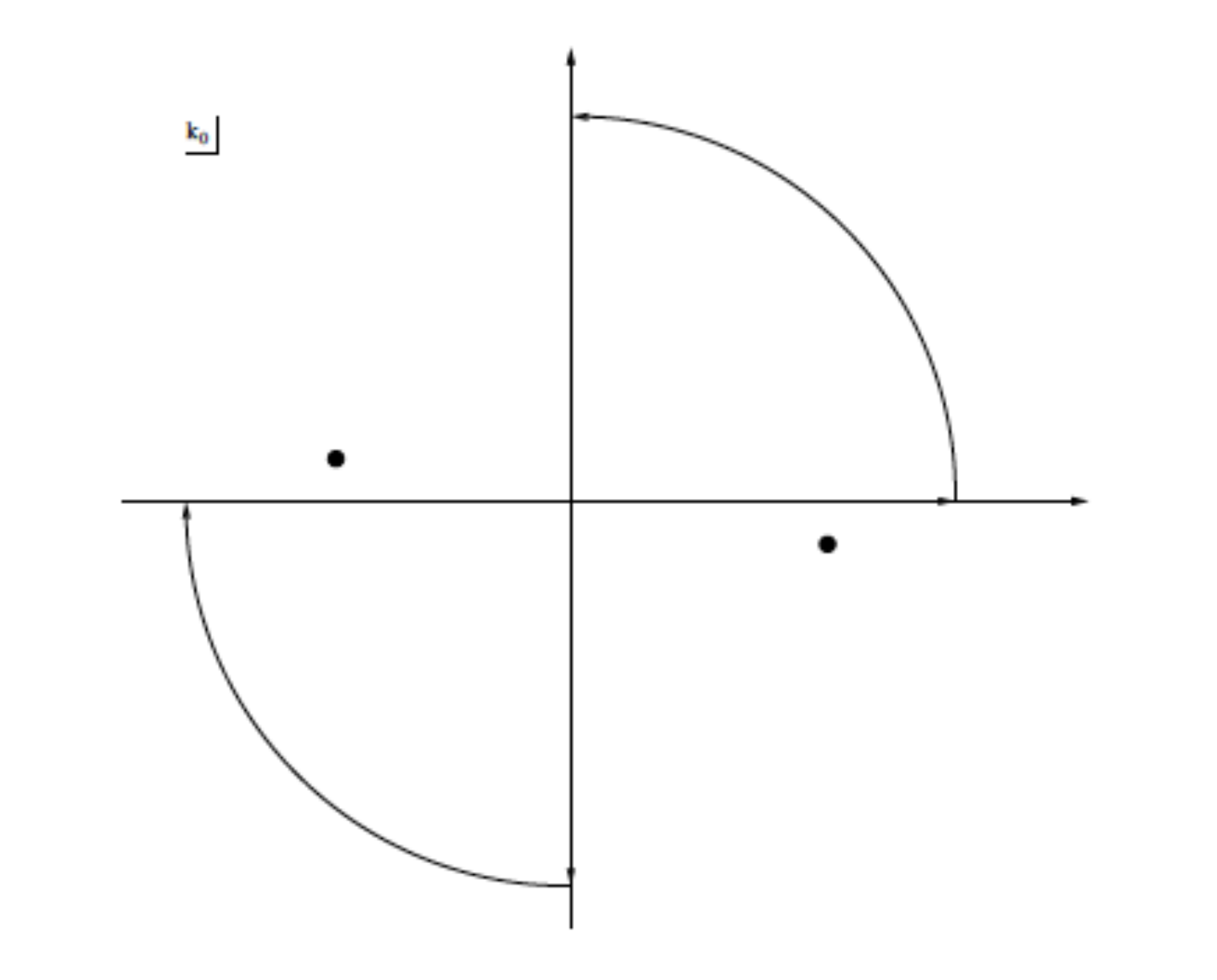}
\caption{Rotation of the integration contour in the complex $k_0$-plane.}
\label{rot_2}
\end{figure} 

They are illustrated in figure \ref{rot_2} by black dots. The integration contour can thus be rotated in the complex $k_0$-plane as indicated in the figure:

\begin{align}
\int\frac{d^5k}{(2\pi)^5}\frac{1}{[{k}^2-\{\ldots\}]^3}&=\int_{i\infty}^{-i\infty}\frac{dk_0}{2\pi}\int\frac{d^4\vec{k}}{(2\pi)^4}\frac{1}{[{k_0}^2-\vec{k}^2-\{\ldots\}]^3}\nonumber\\
&=-i\int_{-\infty}^{\infty}\frac{dk'_0}{2\pi}\int\frac{d^4\vec{k}'}{(2\pi)^4}\frac{1}{[-{k'}_0^2-\vec{k'}^2-\{\ldots\}]^3}\nonumber\\
&=i\int\frac{d^5k'}{(2\pi)^5}\frac{1}{[{k'}^2+\{\ldots\}]^3}
\label{3.118}
\end{align}

where in the second step we made the change of variables $k'_0=ik_0$, $\vec{k'}=\vec{k}$ (Wick rotation) and $\{\ldots\}$ stands for the expression in between the curly brackets appearing in (\ref{3.117}). This rotation being performed, we may again ignore the $i\epsilon$-term.
 
The last expression is an integral in 5-dimensional \textit{Euclidean space} which may be evaluated using the formula \cite{Pokorski}

\begin{equation}
\int\frac{d^nk}{(2\pi)^n}\frac{1}{(k^2+b^2)^\alpha}=\frac{1}{(4\pi)^\frac{n}{2}}\frac{b^{(\frac{n}{2}-\alpha)}\Gamma(\alpha-\frac{n}{2})}{\Gamma(\alpha)}.
\label{3.119}
\end{equation}

We therefore obtain a relatively simple expression for the integral over $k$

\begin{align}
\int\frac{d^5k}{(2\pi)^5}\frac{1}{[{k}^2-\{\ldots\}]^3}&=\frac{i}{64\pi^2}\frac{1}{\sqrt{q^2(y^2-y)+p^2(x^2-x)-2qpxy+m^2}},
\label{3.120}
\end{align}

and from (\ref{3.120}), (\ref{3.116}) and (\ref{3.115})

\begin{align}
\omega(p,q,m)&=\frac{i}{32\pi^2}\int_0^1dx\int_0^{1-x}dy\frac{1}{\sqrt{q^2(y^2-y)+p^2(x^2-x)-2qpxy+m^2}}\nonumber\\
&=\frac{i}{64\pi^2}\frac{1}{|m|}+O(\frac{1}{m^3}).
\label{3.121}
\end{align}

Substituting the above expression for $\omega$ into (\ref{3.114}) yields

\begin{equation}
\omega^{\mu\nu\delta}(p,q)=-\frac{i}{16\pi^2}\frac{m}{|m|}\epsilon^{\alpha\mu\nu\beta\delta}q_\alpha p_\beta+O(\frac{1}{m^2}),
\label{3.122}
\end{equation}

thus in the limit $m\rightarrow \infty$ we obtain

\begin{align}
\text{Tr}(S\A)^3&=\frac{-i}{16\pi^2}\text{sgn}(m)\int d^5x \int d^5y \int d^5z A_{\mu}(x)A_\nu(y)A_\delta(z)\nonumber\\
&\hspace{4mm}\times\int\frac{d^5p}{(2\pi)^5}\int\frac{d^5q}{(2\pi)^5}\epsilon^{\alpha\mu\nu\beta\delta}q_\alpha p_\beta e^{-iq(x-y)}e^{-ip(x-z)}\nonumber\\
&=\frac{i}{16\pi^2}\text{sgn}(m)\int d^5x \int d^5y \int d^5z \epsilon^{\alpha\mu\nu\beta\delta}A_{\mu}(x)A_\nu(y)A_\delta(z)\frac{\partial}{\partial y^\alpha}\frac{\partial}{\partial z^\beta}\delta(x-y)\delta(x-z)\nonumber\\
&=\frac{i}{16\pi^2}\text{sgn}(m)\int d^5x\epsilon^{\alpha\mu\nu\beta\delta}A_{\mu}\partial_\alpha A_\nu\partial_\beta A_\delta\nonumber\\
&=\frac{-i}{64\pi^2}\text{sgn}(m)\int A\wedge F\wedge F.
\label{3.123}
\end{align}

The contribution of $\text{Tr}(S\A)^3$ to the effective action $S_{eff}$ is found from (\ref{3.86}). It equals half the Chern-Simons action defined (for positive $m$) in (\ref{3.36})

\begin{equation}
\boxed{S_{eff}^{(n=3)}=-\frac{1}{2}\text{sgn}(m)S_{CS}=-\frac{1}{192\pi^2}\text{sgn}(m)\int A\wedge F\wedge F+O(\frac{1}{m^2}).}
\label{3.124}
\end{equation}

\subsection{Charge renormalization}

Infinities like the one encountered in the calculation of the vacuum polarization graph teach us that the parameters such as charge, mass, $\ldots$ appearing in the Lagrangian are not necessarily observable quantities. The Lagrangian is written in terms of the bare quantities $\psi^b$, $m^b$ and $e^b$. In our case

\begin{equation}
\mathcal{L}=\bar\psi^b\{i\gamma^\mu(\partial_\mu-iA_\mu)-m^b\}\psi^b-\frac{1}{4L\alpha^b}F_{\mu\nu}F^{\mu\nu},
\label{3.125}
\end{equation}

where $\alpha^b=(e^b)^2$ is the bare feinstructure constant. The reason why we do not introduce a bare gauge field $A^b$ is that the condition of gauge invariance requires the combination $\partial_\mu-iA_\mu$ to appear in the Lagrangian.

If the theory is renormalizable, then the bare quantities can be related to the physical ones by renormalization constants $\xi$, $\mu$ and $Z$ as

\begin{equation}
\psi^b=\xi^\frac{1}{2}(\Lambda)\psi,\hspace{5mm}m^b=\mu(\Lambda)m,\hspace{5mm}\alpha^b=\frac{1}{Z(\Lambda)}\alpha,
\label{3.126}
\end{equation} 

where the renormalization constants are functions of the ultraviolet cutoff $\Lambda$ (for a short treatment of this subject see for example the book by Collins-Martin-Squires \footnote{Section 2.6 on Renormalization}, \cite{Collins}). The Lagrangian (\ref{3.125}) may then be expressed in terms of the physical quantities as

\begin{equation}
\mathcal{L}=\xi(\Lambda)\bar\psi\{i\gamma^\mu(\partial_\mu-iA_\mu)-\mu(\Lambda)m\}\psi-Z(\Lambda)\frac{1}{4L\alpha}F_{\mu\nu}{F}^{\mu\nu}.
\label{3.127}
\end{equation}

Each of these renormalization constants can be written as a series in the effective coupling $\alpha=e^2\approx\frac{1}{137}$,

\begin{eqnarray}
\xi(\Lambda)&=&1+\xi_1(\Lambda)\alpha+\xi_2(\Lambda)\alpha^2\ldots\label{3.128}\\
\mu(\Lambda)&=&1+\mu_1(\Lambda)\alpha+\mu_2(\Lambda)\alpha^2\ldots\label{3.129}\\
Z(\Lambda)&=&1+Z_1(\Lambda)\alpha+Z_2(\Lambda)\alpha^2+\ldots\label{3.130}
\end{eqnarray}

The functions $\xi_n$, $\mu_n$ and $Z_n$ contain the divergences for $\Lambda\rightarrow \infty$, which are thus absorbed into the definition of the bare quantities in such a way that the physical quantities remain finite. The theory is called renormalizable if this can be done while keeping the form of the Lagrangian the same as the original.

As a consequence of these remarks, the fermion mass $m$ appearing in the calculation of the vacuum polarisation graph should be replaced by $\mu(\Lambda)m$ and the propagator $S$ by $\xi(\Lambda)S$. However, the most divergent term corresponds to order zero in the power series developments of $\mu$ and $\xi$, (that is $\mu(\Lambda)\approx 1$ and $\xi(\Lambda)\approx 1$) and therefore reads

\begin{equation}
S^{(n=2)}=\frac{\Lambda|m|}{48\pi^2}\int d^5x F_{\mu\nu}F^{\mu\nu}
\label{3.131}
\end{equation}

as in (\ref{3.110}). This term coming from the calculation of the vacuum polarization graph may be cancelled by an appropriate choice of the coefficient $Z_1(\Lambda)$, which determines the first order correction to the bare coupling constant $\alpha^b$. In fact, by adding the last term in (\ref{3.127}) to (\ref{3.131}) and developing $Z(\Lambda)$ to first order in $\alpha$, we obtain

\begin{equation}
S^{(n=2)}+S_{EM}^b=\frac{\Lambda|m|}{48\pi^2}\int d^5x F_{\mu\nu}F^{\mu\nu}-\frac{1}{4L\alpha}(1+Z_1(\Lambda)\alpha+\ldots)\int d^5x F_{\mu\nu}F^{\mu\nu}.
\label{3.132}
\end{equation}

Since the above sum should produce the result $S_{EM}=-\frac{1}{4L\alpha}\int d^5x F_{\mu\nu}F^{\mu\nu}$ this yields the identity

\begin{equation}
\left[\frac{\Lambda|m|}{48\pi^2}-\frac{1}{4L\alpha}(1+Z_1(\Lambda)\alpha+\ldots)\right]=-\frac{1}{4L},
\label{3.133}
\end{equation}
 
which determines the the coefficient $Z_1(\Lambda)$:

\begin{equation}
Z_1(\Lambda)=\frac{L\Lambda|m|}{12\pi^2}.
\label{3.134}
\end{equation}

The bare coupling constant $\alpha^b$ corrected to first order in the physical coupling $\alpha$ therefore reads

\begin{equation}
\boxed{\alpha^b=\alpha\Big(1+\frac{L\Lambda|m|}{12\pi^2}\alpha+\ldots\Big).}
\label{3.135}
\end{equation}

\subsection{Massless boundary modes}

If the massive (4+1)-dimensional fermions are confined to a slab of thickness $L$, then besides the contributions from the heavy bulk modes - which yield the Chern-Simons action $S_{CS}$ - there exist massless chiral boundary-modes localized near the edges at $x^4=L$ and $x^4=0$. These edge states are identified with the left- and righthanded fermions in (3+1)-dimensions. The effective action of these chiral modes corresponds to the term $\Gamma_{\partial\Lambda}$ appearing in (\ref{3.45}).

\begin{figure}[h!]
\noindent
\begin{minipage}[t]{0.45\linewidth}
\centering\epsfig{figure=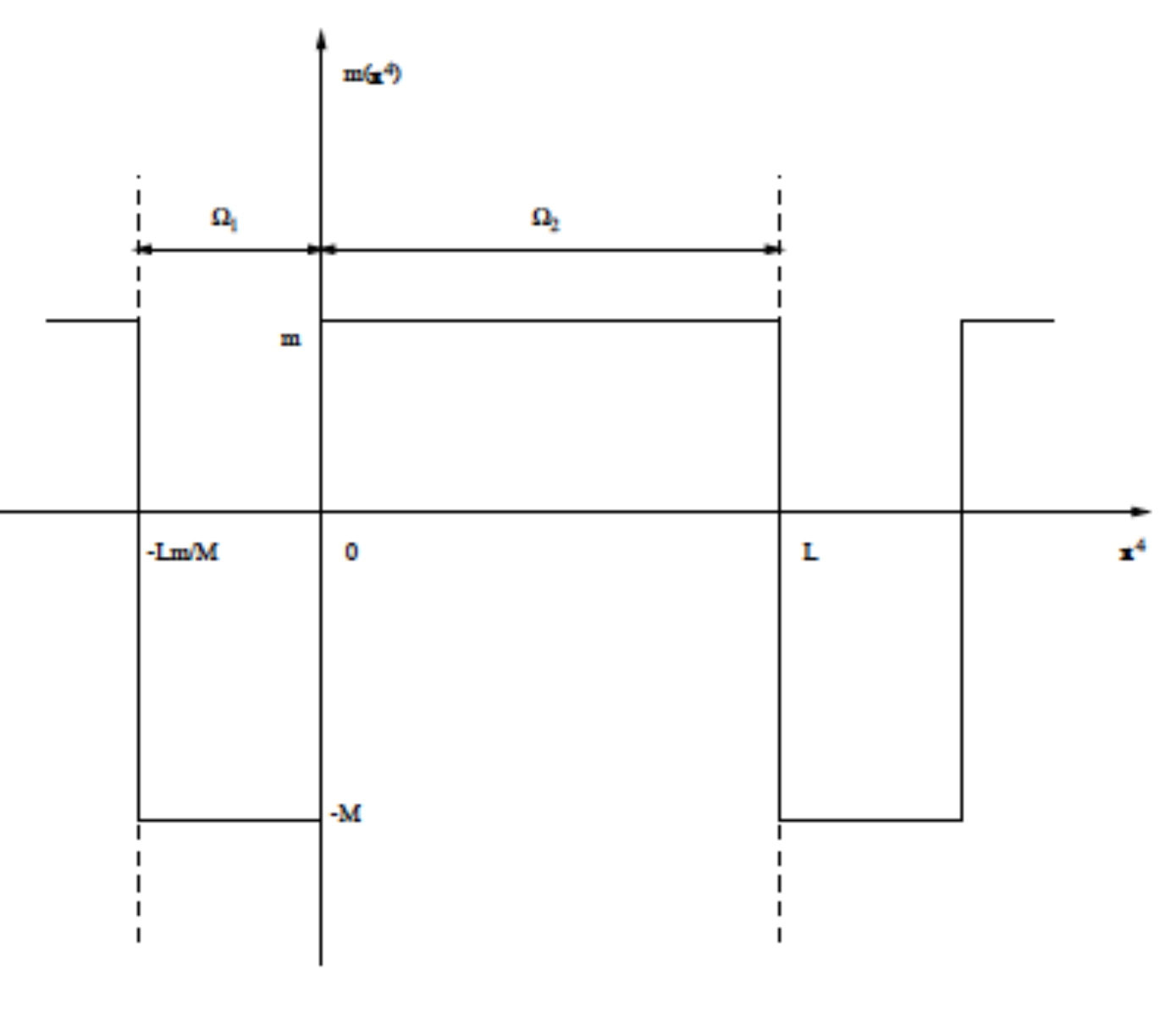,width=\linewidth,height=5.5cm}
\caption{Step function for the $x^4$-depen-dent mass and definition of the domains $\Omega_1$ and $\Omega_2$.}
\label{PBC}
\end{minipage} \hfill
\begin{minipage}[t]{0.45\linewidth}
\centering\epsfig{figure=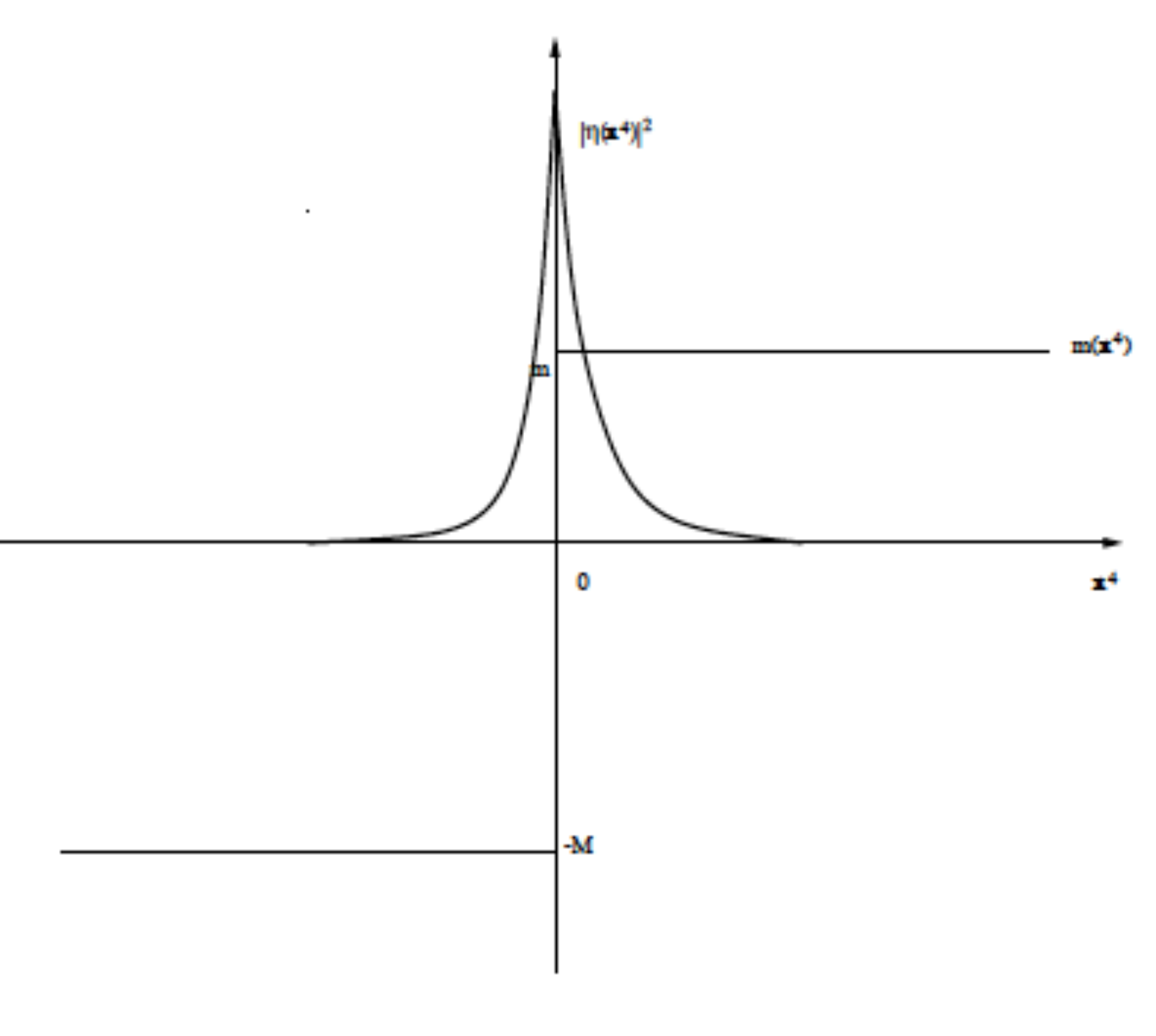,width=\linewidth,height=5.5cm}
\caption{Massless chiral zero mode loca-lized near the domain wall at $x^4=0$.}
\end{minipage}
\label{boundary}
\end{figure}

The confinement of the (4+1)-dimensional fermions to the slab could be modeled by choosing the fermion mass $m(x^4)$ as

\begin{equation}
m(x^4)=\left\{\begin{array}{rlr}m&,&0\le x^4 <L\\-M&,&-L\frac{m}{M}\le x^4<0\end{array} \right.,
\label{3.136}
\end{equation}  

and identifying the (3+1)-dimensional surface $x^4=-L\frac{m}{M}$ with $x^4=L$ (periodic boundary conditions). $M$ is eventually taken to infinity. The (4+1)-dimensional space-time is thus divided into two domains,

\begin{equation}
\Omega_1=(x,-L\frac{m}{M}\le x^4< 0),\hspace{5mm}\Omega_2=(x,0\le x^4\le L),
\label{3.137}
\end{equation}

see figure \ref{PBC}. Unfortunately, the calculation of the determinant $\text{det}(\D_{A^{(5)}}^{(5)}+im(x^4))$ is a difficult task. Papers related to this problem are \cite{Callan} and \cite{Kaplan}, but they mainly discuss the question of anomaly cancellation. An explicit calculation of the effective action in (2+1) dimensions for a single domain wall can be found in \cite{Chandrasekharan}. 

We shall content ourselves with the calculations of sections 3.2 - 3.4 for an $x^4$-independent mass, which are valid within the domains $\Omega_1$ and $\Omega_2$, as long as one does not approach the domain wall too closely. Making the simplifying assumption that formula (\ref{3.124}) remains valid up to the domain wall, we may again deduce the form of the boundary action $\Gamma_{\partial\Lambda}$ by looking at the gauge variation of the Chern-Simons term. Note that the sign of $S_{eff}^{(n=3)}$ depends on the sign of the fermion mass. For $m(x^4)$ defined in (\ref{3.136}) we therefore obtain

\begin{equation}
S_{eff}^{(n=3)}[A]=\frac{1}{2}S_{CS}^{(\partial\Omega_1)}[A]-\frac{1}{2}S_{CS}^{(\partial\Omega_2)}[A],
\label{3.138}
\end{equation}

where $S_{CS}^{(\Omega_i)}[A]=\frac{1}{96\pi^2}\int_{\Omega_i}A\wedge F\wedge F$, $i=1,2$. The variation of $S_{eff}^{(n=3)}$ under a gauge transformation $A\rightarrow A+d\theta$ is then found from (\ref{3.41}):

\begin{align}
S_{eff}^{(n=3)}[A+d\theta]&=S_{eff}^{(n=3)}[A]+\frac{1}{32\pi^2}\left[\frac{1}{2}\int_{\Omega_1}-\frac{1}{2}\int_{\Omega_2}\right]\theta(F\wedge F)\nonumber\\
&=S_{eff}^{(n=3)}[A]+\frac{1}{32\pi^2}\int_{x^4=0}\theta(F\wedge F)-\frac{1}{32\pi^2}\int_{x^4=L}\theta(F\wedge F).
\label{3.139}
\end{align}

Looking at (\ref{2.47}), we realise that gauge invariance may be restored by adding the boundary action

\begin{equation}
\Gamma_{\partial\Omega_2}[A|_{\partial\Omega_2}]=\Gamma_r[A|_{x^4=0}]+\Gamma_l[A|_{x^4=L}]=-i\text{lndet}\{(\D_A|_{x^4=L})(\D_A|_{x^4=0})\}.
\label{3.140}
\end{equation}

This is again the sum of the effective actions for lefthanded (3+1)-dimensional fermions localized at $x^4=L$ ($\Gamma_l[A|_{x^4=L}]$) and the effective action for righthanded fermions localized at $x^4=0$ ($\Gamma_r[A|_{x^4=0}]$). The action functional (\ref{3.45}) - multiplied by a factor of $\frac{1}{2}$ -  and the corresponding equations of motion (\ref{3.61})-(\ref{3.67}) may be obtained by an appropriate choice of the length parameter $l$ appearing in the Maxwell term $-\frac{1}{4l\alpha^b}\int F^2$. Indeed, let us renormalize the charge in such a way, that the Maxwell term and the contribution of the vacuum polarization diagram yields $-\frac{1}{4l\alpha}\int F^2$ inside $\Omega_2$. The coefficient multiplying the corresponding term in $\Omega_1$ will then be much larger, since $M\gg m$. This is equivalent to saying that the Chern-Simons and boundary currents associated with $\Omega_2$ (both independent of the fermion mass) will be strongly suppressed. Their contribution may thus be neglected when calculating the average over $x^4$. Since furthermore $\frac{m}{M}\ll L$, we obtain - after averaging - the equations of motion (\ref{3.61})-(\ref{3.67}) if $l=2L$.

The eigenfunctions of the 5-dimensional Dirac operator corresponding to the chiral boun-dary modes can be calculated (in the limit $m,M\rightarrow\infty$) if the 5-component vector potential $A^{(5)}$ is independent of the coordinate $x^4$. In this case the five dimensional Dirac operator can be written as

\begin{equation}
i\D_{A^{(5)}}^{(5)}-m(x^4)=i\D_A-\gamma^5\partial_4+i\gamma^5A_4-m(x^4),
\label{3.141}
\end{equation}

where $A=(A_0,\ldots,A_3)$ and $i\D_A$ denotes the four dimensional Dirac operator. We are looking for solutions $\Psi_r$ to the Dirac equation $(i\D_{A^{(5)}}^{(5)}-m(x^4))\Psi_r=0$, which are localized near the domain wall at $x^4=0$. To this end we write $\Psi_r=\psi_r(x)\eta(x,x^4)$, where $x=(x^0,\ldots,x^3)$ and $\psi_r$ describes a righthanded fermion in (3+1) dimensions, coupled to the exterior electromagnetic vector potential $A$:

\begin{equation}
i\D_A\psi_r=0, \hspace{5mm}\gamma^5\psi_r=\psi_r.
\label{3.142}
\end{equation}

The resulting equation for $\eta$ is not easy to solve (unless $A_4=0$) and we shall content ourselves with an approximate solution, which becomes exact in the limit $m,M\rightarrow\infty$. We propose the ansatz $\eta(x,x^4)=N(m,M)e^{-\int_0^{x^4} dym(y)-iA_4x^4}$ for the function $\eta$, so that

\begin{equation}
\Psi_r(x,x^4)=\psi_r(x)N(m,M)e^{-\int_0^{x^4} dym(y)-iA_4x^4}.
\label{3.143}
\end{equation}

$N(m,M)$ is a normalization constant. Applying the five dimensional Dirac operator to the function (\ref{3.143}) we find, using (\ref{3.142}) and (\ref{3.141})

\begin{equation}
[i\D_{A^{(5)}(x)}^{(5)}-m(x^4)]\Psi_r(x,x^4)=\gamma^\mu(\partial_\mu A_4(x))x^4\Psi_r(x,x^4).
\label{3.144}
\end{equation} 

The function $|\eta|^2=N(m,M)^2e^{-2\int_0^{x^4} dym(y)}$ is localized near the domain wall at $x^4=0$ and tends to the delta-function $\delta(x^4)$ in the limit $m,M\rightarrow\infty$. Hence $\Psi_r=\psi_r\eta$ is localized near $x^4=0$ and satisfies the Dirac equation in the latter limit, as $x^4\eta(x,x^4)\rightarrow 0$ (see figure \ref{boundary} for an illustration). 

The effective action for these righthanded boundary modes (in the limit $m,M\rightarrow \infty$) is formally given by

\begin{equation}
\int \mathcal D\psi_r\mathcal D\bar\psi_re^{i\int d^5x\bar\Psi_r(i\not D_{A^{(5)}}^{(5)}-m(x^4))\Psi_r}\longrightarrow\int \mathcal D\psi_r\mathcal D\bar\psi_re^{i\int d^4x\bar\psi_ri\not D_A\psi_r}=e^{i\Gamma_r[A]}.
\label{3.145}
\end{equation}

On the right hand side we recognise the effective action (\ref{2.40}) for righthanded (3+1)-dimensional fermions. 

\newpage

\section{Third attempt: Axion electrodynamics}

\subsection{Effective action}

A similar set of equations as the one derived from the (4+1)-dimensional theory, is obtained by coupling the (3+1)-dimensional fermions to an axion field\footnote{A few days before the completion of this report, we realised with some deception that the results derived in this section are not really new. More than a decade ago, M. S. Turner and L. M. Widrow have proposed that the axion field could provide a source term for large-scale magnetic fields, see \cite{Turner}.}. As it will turn out, the time derivative of the axion field corresponds to what we called $\mu_l-\mu_r$ in section 3.2.

For a short review on the subject of axions, see for example \cite{Kim}.
We will consider here the so called \textit{model independent axion}, first described by Witten \cite{Witten}. In string theory there appears an antisymmetric tensor field $B_{\mu\nu}$, which in (3+1) dimensions possesses one physical degree. The associated field strength is not just the curl of $B$, but is made gauge invariant by adding a Chern-Simons 3-form

\begin{equation}
H=dB-\frac{1}{32\pi^2}(A\wedge F),
\label{3.146}
\end{equation}

and therefore

\begin{equation}
dH=-\frac{1}{32\pi^2}(F\wedge F).
\label{3.147}
\end{equation}

Formula (\ref{3.146}) is valid in a flat space-time ($R=0$) and for a system in which the electromagnetic field is the only gauge field present. We shall accept these results without further justification. For more information see the book by Collins-Martin-Squires\footnote{Chapter 10: String Theories, in particular section 10.7 on Anomalies.}, \cite{Collins}. 

The equation of motion is 

\begin{equation}
\delta H=0,
\label{3.148}
\end{equation}

where $\delta$ is the co-differential. If we write $Y=\ast H$, this implies

\begin{equation}
\ast dY=\delta H=0,
\label{3.149}
\end{equation}

or simply $dY=0$. The latter equation is solved by setting $Y=da$. In (3+1) dimensions, $Y$ is a 1-form, so $a$ is a pseudo-scalar\footnote{From equation (\ref{3.152}) below it follows immediately that $\phi$ is a pseudo-scalar, since $\Box$ is a scalar and $\ast(F\wedge F)\sim \vec{E}\cdot\vec{B}$ a pseudo-scalar.}. We now define the axion field $\phi$ by

\begin{equation}
a=\frac{\phi}{l\alpha},
\label{3.150}
\end{equation}

where the parameter $l$ - with the dimension of length - has been introduced in order to obtain $[\phi]=\frac{1}{L}$. Hence the axion field $\phi$ is related to the dual of $H$ by 

\begin{equation}
\frac{1}{l\alpha}d\phi=\ast H.
\label{3.151}
\end{equation}

We shall now explain why $\phi$ is an axion field. Applying  the co-differential $\delta$ to equation (\ref{3.151}) we find $\delta d\phi=l\alpha\delta\ast H$, that is $\Box \phi=l\ast dH$. Using (\ref{3.147}) the latter equation yields

\begin{equation}
\boxed{\Box \phi=-\frac{l\alpha}{32\pi^2}\ast (F\wedge F),}
\label{3.152}
\end{equation} 

which is the Euler-Lagrange equation of motion corresponding to the action functional

\begin{equation}
S[A,\phi]=\int d^4x\frac{1}{2\alpha}(\partial_\mu\phi)(\partial^\mu\phi)-\frac{l}{32\pi^2}\int\phi(F\wedge F).
\label{3.153}
\end{equation}

In the second term we recognise the standard coupling of an axion to the gauge field A. This term may be understood as arising from coupling fermions to the axion as

\begin{equation}
H_{\mu\nu\rho}\bar\psi\gamma^\nu\gamma^\nu\gamma^\rho\psi\sim\frac{1}{l}\partial_\mu \phi\bar\psi\gamma^\mu\gamma^5\psi.
\label{3.154}
\end{equation}

A system of charged fermions coupled to an electromagnetic vector potential $A$ and to an axion field $\phi$ is therefore described by the action functional

\begin{equation}
S[A,\phi,\psi,\bar\psi]=\int d^4x\{\bar\psi i\D _A\psi+\frac{l}{2}\partial_\mu \phi\bar\psi\gamma^\mu\gamma^5\psi\}+\int d^4x\{-\frac{1}{4\alpha}F^{\mu\nu}F_{\mu\nu}+\frac{1}{2\alpha}(\partial^\mu \phi)(\partial_\mu \phi)\}.
\label{3.155}
\end{equation}

The $l$ in front of the $\partial_\mu \phi\bar\psi\gamma^\mu\gamma^5\psi$ term follows from dimensional consideration, whereas the factor of $\frac{1}{2}$ must be introduced in order to obtain  the equation of motion (\ref{3.152}). Carrying out the integral over the fermionic degrees of freedom we find the effective action

\begin{equation}
e^{iS_{eff}[A,\phi]}=\int \mathcal D\psi \mathcal D\bar\psi e^{iS[A,\phi,\psi,\bar\psi]}=\text{det}(\D _{A+\frac{l}{2}\partial_\mu \phi\gamma^5})e^{i\int d^4x\{-\frac{1}{4\alpha}F^{\mu\nu}F_{\mu\nu}+\frac{1}{2\alpha}(\partial^\mu \phi)(\partial_\mu \phi)\}}.
\label{3.156}
\end{equation}

In connection with the chiral anomaly we have seen in section 2.1 that

\begin{equation}
\text{det}(\D _{A+\frac{l}{2}\partial_\mu \phi\gamma^5})=e^{-\frac{il}{32\pi^2}\int \phi(F\wedge F)}\text{det}(\D_{A}).
\label{3.157}
\end{equation}

Hence the effective action may be written as

\begin{equation}
\boxed{S_{eff}[A,\phi]=W[A]-\frac{l}{32\pi^2}\int \phi(F\wedge F)+\int d^4x\{-\frac{1}{4\alpha}F^{\mu\nu}F_{\mu\nu}+\frac{1}{2\alpha}(\partial^\mu \phi)(\partial_\mu \phi)\},}
\label{3.158}
\end{equation}

which is expression (\ref{3.153}) up to the fermionic effective action $W[A]=-i\ln\text{det}(\D_A)$ and the Maxwell term. Note, that it is not necessary in this approach to assume that the fermions are massless. The same calculations go through in the case of fermions of mass $m$, except that the fermionic effective action then reads $W[A]=-i\ln\text{det}(\D_A+im)$. We shall henceforth consider this more realistic situation.

\subsection{Quantum fluctuations and axionic potential}

A transition amptitude from a configuration $(\phi_{in},A_{in})$ of the electromagnetic and the axion field at some very early time $t_1$ to a configuration $(\phi_{out},A_{out})$ at a much later time $t_2$ can be computed from the Feynman path integral

\begin{equation}
I=\int \mathcal D\phi\int \mathcal DA e^{iS_{eff}[A,\phi]},
\label{3.159}
\end{equation}

with boundary conditions $(\phi(t_1),A(t_1))=(\phi_{in},A_{in})$ and $(\phi(t_2),A(t_2))=(\phi_{out},A_{out})$ (the term $W[A]$ in $S_{eff}[A,\phi]$ also depends on the boundary conditions at times $t_1$, $t_2$ imposed on the fermion fields, which have been integrated out).

The integral over the gauge field configurations yields the effective action for the axion field, $S_{eff}[\phi]$, so that

\begin{align}
I&=\int \mathcal D \phi e^{iS_{eff}[\phi]},\label{3.160}\\
e^{iS_{eff}[\phi]}&=\int \mathcal D A e^{iS_{eff}[A,\phi]}.\label{3.161}
\end{align}

We evaluate the integral (\ref{3.160}) by using a semi-classical expansion based on the stationary phase method. The equation for the saddle point is

\begin{equation}
\frac{\delta S_{eff}[\phi]}{\delta \phi}=0,
\label{3.162}
\end{equation}

and we shall denote the solution of (\ref{3.162}) by $\phi_{cl}$. From (\ref{3.161}) one then finds the following expression for $S_{eff}[\phi_{cl}]$

\begin{equation}
S_{eff}[\phi_{cl}]=-i\ln \int \mathcal DAe^{iS_{eff}[A,\phi_{cl}]}.
\label{3.163}
\end{equation}

Again, we use the stationary phase method to evaluate the $A$-integral in (\ref{3.163}). Denoting the solution of the saddle point equation

\begin{equation}
\frac{\delta S_{eff}[A,\phi_{cl}]}{\delta A}=0
\label{3.164}
\end{equation}

by $A_{cl}^{(\phi_{cl})}$ and writing $A$ as the sum of $A_{cl}^{(\phi_{cl})}$ plus a fluctuation,

\begin{equation}
A=A_{cl}^{(\phi_{cl})}+\tilde A,
\label{3.165}
\end{equation}

one obtains

\begin{equation}
S_{eff}[\phi_{cl}]=S_{eff}[A_{cl}^{(\phi_{cl})},\phi_{cl}]-i\ln\int \mathcal D\tilde Ae^{\frac{i}{2}\langle\tilde A,\text{Hess}(A_{cl}^{(\phi_{cl})},\phi_{cl})\tilde A\rangle}+\ldots.
\label{3.166}
\end{equation}

We will now simplify the problem considerably by neglecting the terms of higher than \mbox{second} order in $A$ appearing in $W[A]=-i\ln\text{det}(\D_A+im)$ \footnote{Such difficulties may be circumvented by choosing a different approach, see \cite{J-F}. Since the axion field couples to all gauge fields $W$ through a term $\sim\int\phi(F_W\wedge F_W)$, one obtains an axionic potential (similar to the one which we shall derive) by integrating out additional gauge fields possibly present in the theoretical description.}. Under this condition $S_{eff}[A,\phi]$ only contains terms quadratic in $A$ and (\ref{3.166}) becomes

\begin{align}
S_{eff}[\phi_{cl}]&=S_{eff}[A_{cl}^{(\phi_{cl})},\phi_{cl}]-U[\phi_{cl}]\label{3.167}\\
-U[\phi_{cl}]&=-i\ln\int \mathcal D\tilde A\text{det}(\D _{\tilde A}+im)e^{-\frac{il}{32\pi^2}\int\phi(\tilde F\wedge \tilde F)}e^{i\int d^4x(-\frac{1}{4\alpha} \tilde F^{\mu\nu}\tilde F_{\mu\nu})}.\label{3.168}
\end{align} 

At this level of approximation we find from (\ref{3.162}), (\ref{3.164}) and (\ref{3.167}) that $\phi_{cl}$ and $A_{cl}^{(\phi_{cl})}$ are solutions of the equations of motion derived from the action functional 

\begin{align}
S'_{eff}[A,\phi]=S_{eff}[A,\phi]-U[\phi].
\label{3.169}
\end{align}

The quantum corrections resulting from the path integral over $A$ have produced the additional term $-U[\phi]$. Under suitable conditions, $U[\phi]$ plays the role of a potential and it is worthwile at this point to note some properties of the latter functional. To this end we consider an axion field $\phi=\theta$ independent of $x$ and denote the integral over the fluctuations by $e^{i\int d^4x\{-V[\theta]\}}$:

\begin{equation}
e^{i\int d^4x\{-V[\theta]\}}\equiv e^{i\{-U[\theta]\}}=\int \mathcal D A\text{det}(\D _{A}+im)e^{-\frac{il}{32\pi^2}\int \theta( F\wedge F)}e^{i\int d^4x(-\frac{1}{4\alpha} F^{\mu\nu} F_{\mu\nu})}.
\label{3.170}
\end{equation}

$V[\theta]$ will be called the \textit{axionic potential}. Changing to Euclidean space using (\ref{2.17}) and (\ref{2.18}) we find

\begin{equation}
e^{-\int d^4xV[\theta]}=\int \mathcal D A\text{det}(\D_{A}^E+m)e^{-\frac{il}{32\pi^2}\int \theta( F\wedge F)}e^{-\int d^4x(-\frac{1}{4\alpha}  F^{\mu\nu} F_{\mu\nu})}.
\label{3.171}
\end{equation}

Equation (\ref{3.171}) allows us to draw a certain number of conclusions concerning the form of the axionic potential $V[\theta]$:

\begin{enumerate}
\item First, we show that $e^{-\int d^4xV[\theta]}$ is real. This can be seen by changing $A_0(t,\vec{x})\rightarrow A'_0(t,\vec{x})=A_0(t,-\vec{x})$, $A_i(t,\vec{x})\rightarrow A'_i(t,\vec{x})=-A_i(t,-\vec{x})$, $i=1,2,3$ in (\ref{3.171}). Under such a transformation $\int\theta(F\wedge F)$ changes sign, whereas $\int F^2$ does not. The fermion determinant remains unaffected. If $\psi(t,\vec{x})$ is an eigenfunction of $\D_A^E$ for the eigenvalue $\lambda$, then $\psi'(t,\vec{x})=\gamma^0\psi(t,-\vec{x})$ is an eigenfunction of the $D_{A'}^E$ for the same eigenvalue. 

It is less evident to see whether or not $e^{-\int d^4xV[\theta]}$ is positive (and therefore $V[\theta]$ is real). Such is the case for $\theta=0$ and we shall admit that $V[\theta]$ remains positive at least for sufficiently small values of $\theta$.

\item The last exponential factor in the functional integral (\ref{3.171}) is obviously real and positive. Furthermore the determinant factor $\text{det}(\D_{A}^E+m)$ is also real and positive \cite{Vafa}.

It was mentioned in section 2.2 that $i\D_A^E$ is a hermitian operator, which therefore has real eigenvalues. Its non-zero eigenvalues are paired in a simple way. From $\{\D_A^E,\gamma^5\}=0$ if follows that if $i\D_{A}^E\psi=\lambda\psi$, then $i\D_{A}^E(\gamma^5\psi)=-\lambda(\gamma^5\psi)$. Hence if $\lambda$ is an eigenvalue, $-\lambda$ is also. The determinant $\text{det}(\D_A^E+m)$ is then real and positive:

\begin{equation}
\text{det}(\D _{A}^E+m)=\prod_\lambda(m-i\lambda)=\prod_{\lambda>0}(m^2+\lambda^2)>0.
\label{3.172}
\end{equation}

Using the fact that $e^{-\int d^4xV[\theta]}$ is real we find

\begin{align}
e^{-\int d^4xV[\theta]}&\equiv\left|\int D A\text{det}(\D _{A}^E+m)e^{-\frac{il}{32\pi^2}\int \theta( F\wedge F)}e^{-\int d^4x(-\frac{1}{4}  F^{\mu\nu} F_{\mu\nu})}\right|\nonumber\\
&\le\int D A\text{det}(\D _{A}^E+m)e^{-\int d^4x(-\frac{1}{4}  F^{\mu\nu} F_{\mu\nu})}=e^{-\int d^4xV[0]}.
\label{3.173}
\end{align}

Therefore $V[\theta]\ge V[0]$ and $\theta=0$ is the minimum of the axionic potential.

\item $V[\theta]$ is an even function of $\theta$. This follows immediately from  $V^*[\theta]=V[-\theta]$ (a direct consequence of (\ref{3.171}) - ``*'' denotes the complex conjugate) and the point 1 above.

\item The axionic potential $V[\theta]$ is a periodic function of $\theta$ and its qualitative shape is therefore as shown in figure \ref{axion}. 
\vspace{10mm}
\begin{figure}[htbp]
\centering
\includegraphics[width=0.7\textwidth]{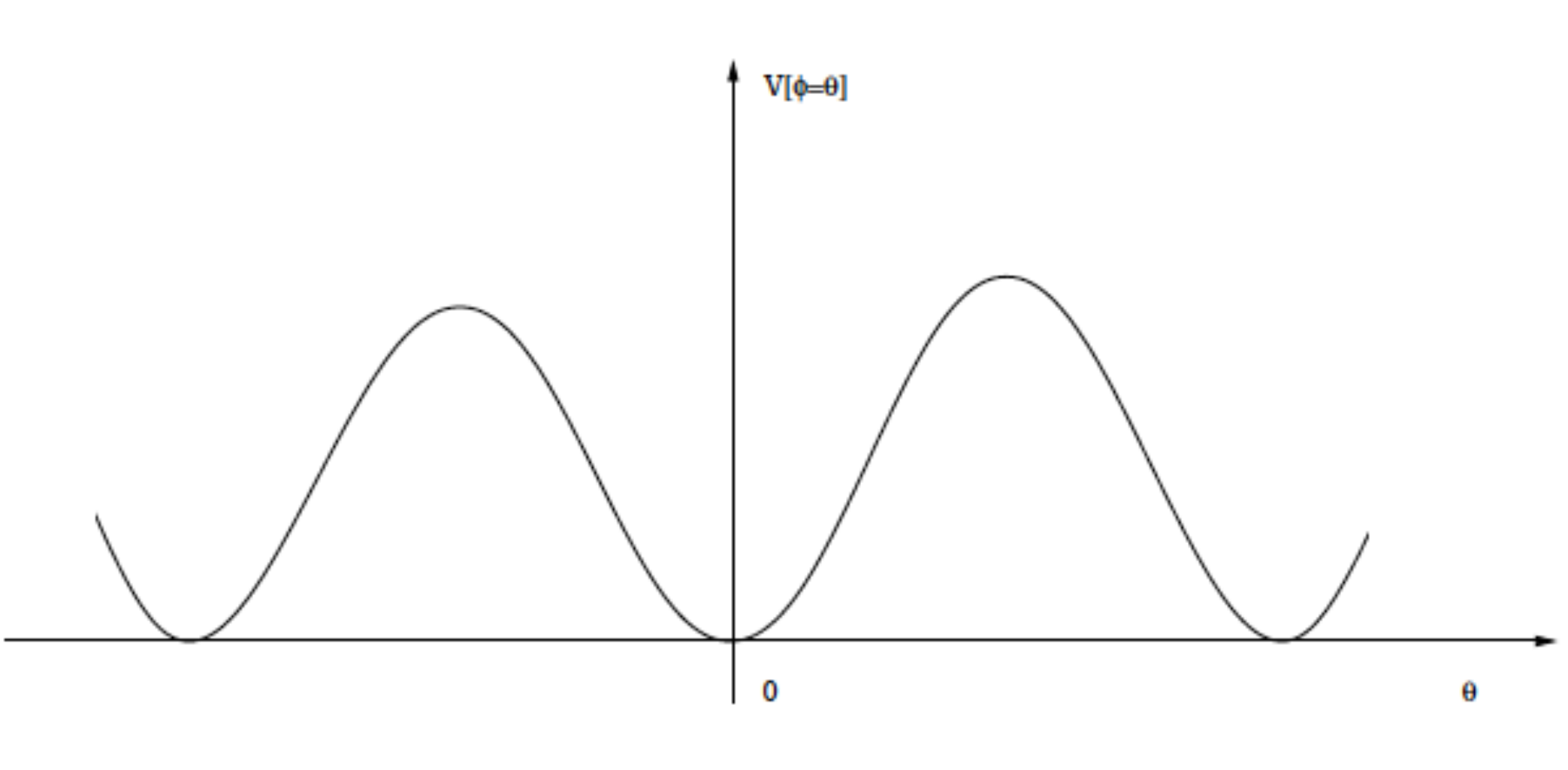}
\caption{Schematic view of the axionic potential $V[\phi=\theta]$ with a minimum at $\theta=0$.}
\label{axion}
\end{figure}

The period of the axionic potential can be obtained from the \textit{index theorem} \cite{A-G}. Using (\ref{2.28}) and (\ref{2.32}) we find

\begin{equation}
-\frac{1}{32\pi^2}\int F\wedge F=\int d^4x\mathcal{A}(x)=\int d^4x\sum_n\psi_n^\dag(x)\gamma^5\psi_n(x).
\label{3.174}
\end{equation}

The $\psi_n$ are eigenfunctions of the Euclidean Dirac operator corresponding to the eigenvalue $\lambda_n$. Since $\gamma^5\psi_n$ is orthogonal to $\psi_n$ if $\lambda_n\ne 0$, only the eigenfunctions corresponding to the eigenvalue $\lambda=0$ contribute to the integral in (\ref{3.174}). This leaves

\begin{equation}
-\frac{1}{32\pi^2}\int F\wedge F=\int d^4x\left(\sum_{i=1}^{n_+}v_i^\dag(x)v_i(x)-\sum_{i=1}^{n_-}u_i^\dag(x)u_i(x)\right)=n_+-n_-,
\label{3.175}
\end{equation}

which is the number of positive minus the number of negative chirality zero modes. The $u_i's$ satisfy $\D_A^Eu_i=0$, $\gamma_lu_i=u_i$ while $\D_A^Ev_i=0$ and $\gamma_rv_i=v_i$. This notation is consistent with the one introduced in chapter 2. The value of $n_+-n_-$ depends on the vector potential $A$, but since it always equals some integer we conclude that the period of the axionic potential $V[\theta]$ is

\begin{equation}
\Delta \theta=\frac{2\pi}{l}.
\end{equation}
\label{3.176}
\end{enumerate}

\subsection{Equations of motion}

The remarks in section 3.4.2 were aimed at explaining the qualitative properties of the potential $V[\theta]$. Some knowledge of its shape is necessary for finding sensible approximate solutions to the equations of motion, which we shall now derive. Recall that by taking into account the quantum fluctuations in the $A$-integral, we have found the following expression for the effective action in Minkowski space (see (\ref{3.169}), (\ref{3.168}) and (\ref{3.158})):

\begin{equation}
\boxed{S'_{eff}[A,\phi]=W[A]-\frac{l}{32\pi^2}\int\phi(F\wedge F)+\int d^4x\left\{-\frac{1}{4\alpha}F^{\mu\nu}F_{\mu\nu}+\frac{1}{2\alpha}(\partial^\mu \phi)(\partial_\mu \phi)\right\}-U[\phi],}
\label{3.177}
\end{equation}

with 

\begin{align}
W[A]&=-i\text{ln}\text{det}(\D_A+im)\label{3.178}\\
-U[\phi]&=-i\text{ln}\int \mathcal D A\text{det}(\D _{A}+im)e^{-\frac{il}{32\pi^2}\int \phi( F\wedge F)}e^{i\int d^4x(-\frac{1}{4\alpha} F^{\mu\nu} F_{\mu\nu})}.
\label{3.179}
\end{align}

The equations of motion which govern the evolution of the electromagnetic- and the axion field are obtained from the effective action (\ref{3.177}) by calculating 

\begin{equation}
\frac{\delta S'_{eff}}{\delta A_{\mu}}=0,\hspace{3mm}\mu=0\ldots 3,\hspace{3mm}\text{and}\hspace{3mm}\frac{\delta S'_{eff}}{\delta \phi}=0. 
\label{3.180}
\end{equation}

Adding to these the homogeneous Maxwell equations $dF=0$, we find the following system ($\tilde F^{\mu\nu}=\frac{1}{2}\epsilon^{\mu\nu\rho\lambda}F_{\rho\lambda}$ denotes the dual of the field strenght tensor)

\begin{align}
\partial_\mu F^{\mu\nu}&=
-\frac{l\alpha}{4\pi^2}\partial_\mu \phi\tilde F^{\mu\nu}-\alpha\frac{\delta W}{\delta A}\label{3.181}\\
\square \phi&=-\frac{l\alpha}{32\pi^2}\ast(F\wedge F)-\alpha U'(\phi)\label{3.182}\\
dF&=0.\label{3.183}
\end{align}

Explicitly, these equations read:

\begin{align}
\vec{\nabla}\cdot\vec{E}&=-\frac{l\alpha}{4\pi^2}\vec{\nabla}\phi\cdot\vec{B}-\alpha\frac{\delta W}{\delta A_0}\label{3.184}\\
\vec{\nabla}\times\vec{B}-\partial_0\vec{E}&=\frac{l\alpha}{4\pi^2}(\dot \phi\vec{B}+\vec{\nabla}\phi\times\vec{E})-\alpha\frac{\delta W}{\delta \vec{A}}\label{3.185}\\
\square \phi&=\frac{l\alpha}{4\pi^2}\vec{E}\cdot\vec{B}-\alpha U'(\phi)\label{3.186}\\
\vec{\nabla}\cdot\vec{B}&=0\label{3.187}\\
\vec{\nabla}\times\vec{E}+\partial_0\vec{B}&=0.\label{3.188}
\end{align}

\newpage

\section{Relation between the (4+1)-dimensional and the axion field approach}

By comparing the equations of motion written in (\ref{3.69})-(\ref{3.73}) with those derived in the last section, one remarks certain similarities. Thus, before proceeding to the study of these systems in the following chapter, we wish to explore the relation between the (4+1)-dimensional theory developed in section 3.2 and the system obtained by coupling (3+1)-dimensional fermions to an axion field. In particular we will show that the two formulations are equivalent if one considers  $x^4$-independent fields (more precisely $x^4$-independent vector potential $A$) in the former theory and massless fermions in the latter. In this case the component $A_4$ of the (4+1)-dimensional vector potential plays the role of the axion field $\phi$ and the thickness of the slab is related to the lenght parameter introduced in (\ref{3.150}).

In fact, let us start with the effective action (\ref{3.45}) derived in (4+1) dimensions

\begin{equation}
S_{eff}^{(5)}[A]=\int d^5x\left\{-\frac{1}{4L\alpha}F^{\mu\nu}F_{\mu\nu}-\frac{1}{24\pi^2}\epsilon^{\mu\nu\rho\lambda\sigma}A_\mu\partial_\nu A_\rho\partial_\lambda A_\sigma\right\}+\Gamma_{\partial\Lambda}^{(5)}[A|_{\partial\Lambda}],
\label{3.189}
\end{equation}

where

\begin{equation}
\Gamma_{\partial\Lambda}^{(5)}[A|_{\partial\Lambda}]=-i\text{ln}\{\text{det}(\D_A^l|_{x^4=L})\text{det}(\D_A^r|_{x^4=0})\}
\label{3.190}
\end{equation}

and replace $A_4$ by $\phi$ and $\partial_4$ by $0$. Integrating over $x^5$ simply produces a factor of $L$ multiplying the first term on the right hand side of (\ref{3.190}). The Maxwell term $-\frac{1}{4\alpha}F^{\mu\nu}F_{\mu\nu}$ splits up into a contribution $-\frac{1}{4\alpha}F^{\alpha\beta}F_{\alpha\beta}$ coming from the first four components of the vector potential and a kinetic energy term $\frac{1}{2\alpha}(\partial^\mu\phi)(\partial_\mu\phi)$ for the field $\phi\equiv A_4$, whereas the Chern-Simons 5-form produces a familiar looking coupling between $\phi$ and the four component gauge field $A=(A_0,\ldots,A_3)$:

\begin{equation}
S_{eff}^{(4)}[A,\phi]=\int d^4x\left\{-\frac{1}{4\alpha}F^{\alpha\beta}F_{\alpha\beta}+\frac{1}{2\alpha}(\partial^\mu\phi)(\partial_\mu\phi)\right\}-\frac{L}{32\pi^2}\int \phi(F\wedge F)+\Gamma_{\partial\Lambda}^{(4)}[A].
\label{3.191}
\end{equation} 

$S_{eff}^{(4)}[A,\phi]$ denotes the effective action for the fields $A$ and $\phi\equiv A_4$ in (3+1) dimensions. Since $A$ is $x^4$-independent, the expression for the boundary action $\Gamma_{\partial\Lambda}^{(4)}[(A_0,\ldots,A_3)]\equiv\Gamma_{\partial\Lambda}^{(5)}[(A_0,\ldots,A_4)|_{\partial\Lambda}]$ simplifies to

\begin{equation}
\Gamma_{\partial\Lambda}^{(4)}[A]=-i\text{ln}\{\text{det}(\D_A^l)\text{det}(\D_A^r)\},
\label{3.192}
\end{equation}

an expression, which may be transformed even further. To this end we recall the explicit form of the operators $\D_A^l$ and $\D_A^r$ in the chiral representation (see (\ref{2.42})-(\ref{2.43})):

\begin{equation}
\D_A^l=\left(\begin{array}{cc}0&\pa\\\hat{\pa}_A&0\end{array}\right),\hspace{.5cm}\D_A^r=\left(\begin{array}{cc}0&\pa_A\\\hat{\pa}&0\end{array}\right),
\label{3.193}
\end{equation}

where $\hat {\pa}=\hat \sigma\cdot\partial=-\sigma^0\partial_0-\vec\sigma\cdot\vec\nabla$, $\pa= \sigma\cdot\partial=\sigma^0\partial_0-\vec\sigma\cdot\vec\nabla$ and a subscript $A$ denotes the covariant derivative. Inspired by a similar argument in \cite{A-G} we formally rewrite the product of the two determinants on the right hand side of (\ref{3.192}) as

\begin{equation}
\text{det}(\D_A^l)\text{det}(\D_A^r)=\text{det}(-\Box)\text{det}(\hat {\pa}_A \pa_A)=-\text{det}(-\Box)\text{det}(\D_A).
\label{3.194}
\end{equation}

The prefactor $-\text{det}(-\Box)$ can be ignored, since it merely adds a (diverging) constant to the effective action. In this sense we identify 

\begin{equation}
\Gamma_{\partial\Lambda}^{(4)}[A]=-i\text{ln}\{\text{det}(\D_A^l)\text{det}(\D_A^r)\}\simeq-i\text{lndet}(\D _A)=W[A]
\label{3.195}
\end{equation}

The effective action obtained from the (4+1)-dimensional theory with $x^4$-independent fields therefore reads

\begin{equation}
S_{eff}^{(4)}[A,\phi]=\int d^4x\left\{-\frac{1}{4\alpha}F^{\alpha\beta}F_{\alpha\beta}+\frac{1}{2\alpha}(\partial^\mu\phi)(\partial_\mu\phi)\right\}-\frac{L}{32\pi^2}\int \phi(F\wedge F)+W[A].
\label{3.196}
\end{equation} 

Upon comparision with (\ref{3.158}) we conclude that this functional is identical to $S_{eff}[A,\phi]$ obtained by coupling (3+1)-dimensional massless fermions to an axion field. The (4+1)-dimensional theory with $x^4$-independent vector potential and the axion field formulation coincide if we identify the lenght parameter $l$ introduced in (\ref{3.150}) with the thickness $L$ of the slab. 

We end this chapter with a remark concerning the physical interpretation of the axion field. Our first attempt to devise a (4+1)-dimensional theory was based on an analogy with the quantum Hall effect. In that context we defined the quantity $(\mu_l-\mu_r)(x)$ as being the potential difference generated by the 4-component of an electric field at the space-time point $x$. The interpretation of $\mu_{l,r}$ as chemical potentials of the left- and righthanded fermions was suggested by the QH-analogy. In the case of an $x^4$-independent vector potential, 

\begin{equation}
\mu_l-\mu_r=-LE_4=-L\partial_0A_4,
\label{3.197}
\end{equation}

see (\ref{3.74}). Identifying $A_4$ with the axion field $\phi$, we find from (\ref{3.197}) that \textit{the time derivative of the axion field plays the role of a space-time dependent ``difference in chemical potentials'' between the left- and righthanded fermions}.

\chapter{Cosmic evolution}

The purpose of this report was to present a mechanism which could explain the generation of large cosmic magnetic fields in the early universe. In the last chapter we have discussed several models from which equations of motion were derived. The present chapter is now devoted to the study of these systems of equations. We shall be looking for special solutions and try to solve the equations obtained by linearising the system around these special solutions. Our hope is of course to find unstable states in the sense that these linearised equations predict a growing (electro-)magnetic field.

We have shown that the axion field theory is equivalent to the (4+1)-dimensional formulation in the case of an $x^4$-independent vector potential. The equations of motion derived from the (4+1)-dimensional theory in turn generalise the system of equations obtained in section 3.1. We shall therefore concentrate on the set of equations (\ref{3.184}) through (\ref{3.188}). Even though we are merely looking for special solutions, a certain number of simplifying hypotheses are inevitable. Especially the contribution $\frac{\delta W}{\delta A}$ will be neglected throughout this chapter. The argument in favour of this simplification goes as follows: Once the divergent contribution corresponding to the vacuum polarization graph has been absorbed into the Maxwell term through charge renormalization, the remaining contributions are of higher than second order in the electromagnetic vector potential $A$. Thus, if we restrict ourselves to fairly \textit{small} electromagnetic fields, we may neglect $W[A]$. Furthermore, if we only consider an axion field that varies \textit{slowly} in space-time, then we may omit all contributions to $U[\phi]$ involving derivatives, $\partial_\mu\phi$, of the axion field. 

The system of equations which we shall consider therefore reads

\begin{align}
\vec{\nabla}\cdot\vec{E}&=-\frac{l\alpha}{4\pi^2}\vec{\nabla}\phi\cdot\vec{B}\label{4.1}\\
\vec{\nabla}\times\vec{B}-\partial_0\vec{E}&=\frac{l\alpha}{4\pi^2}(\dot \phi\vec{B}+\vec{\nabla}\phi\times\vec{E})\label{4.2}\\
\square \phi&=\frac{l\alpha}{4\pi^2}\vec{E}\cdot\vec{B}-\alpha U'[\phi]\label{4.3}\\
\vec{\nabla}\cdot\vec{B}&=0\label{4.4}\\
\vec{\nabla}\times\vec{E}+\partial_0\vec{B}&=0.\label{4.5}
\end{align}

\section{Space independent solutions}

\subsection{Solution without axionic potential}

If we neglect the axionic potential $U$ and furthermore assume that $\phi$ is space-independent, then equations (\ref{4.1})-(\ref{4.5}) simplify to

\begin{align}
\vec{\nabla}\cdot\vec{E}&=0\label{4.6}\\
\vec{\nabla}\times\vec{B}-\partial_0\vec{E}&=-\frac{\alpha}{4\pi^2}\mu\vec{B}\label{4.7}\\
\partial_0\mu&=-\frac{l^2\alpha}{4\pi^2}\vec{E}\cdot\vec{B}\label{4.8}\\
\vec{\nabla}\cdot\vec{B}&=0\label{4.9}\\
\vec{\nabla}\times\vec{E}+\partial_0\vec{B}&=0,\label{4.10}
\end{align} 

where we have set $\dot\phi=-\frac{1}{l}\mu$ in order to make it apparent how the system (\ref{3.75})-(\ref{3.79}) obtained in section 3.2 appears as a particular case of the axion field theory. 

A possible special solution to equations (\ref{4.6})-(\ref{4.10}) is $F=F^{(0)}=0$, $\mu=\mu^{(0)}=const$. We may linearise the system of equations around this solution by writing 

\begin{eqnarray}
B&=&B^{(0)}+\epsilon B^{(1)}=\epsilon B^{(1)}\label{4.11}\\
E&=&E^{(0)}+\epsilon E^{(1)}=\epsilon E^{(1)}\label{4.12}\\
\mu&=&\mu^{(0)}+\epsilon\mu^{(1)}\label{4.13}
\end{eqnarray}

and ignoring terms quadratic in $\epsilon$:

\begin{align}
\vec{\nabla}\cdot\vec{E}^{(1)}&=0\label{4.14}\\
\vec{\nabla}\times\vec{B}^{(1)}-\partial_0\vec{E}^{(1)}&=-\frac{\alpha}{4\pi^2}\mu^{(0)}\vec{B}^{(1)}\label{4.15}\\
\vec{\nabla}\cdot\vec{B}^{(1)}&=0\label{4.16}\\
\vec{\nabla}\times\vec{E}^{(1)}+\partial_0\vec{B}^{(1)}&=0\label{4.17}\\
(\partial_0\mu^{(1)}&=0).\label{4.18}
\end{align}

No particular assumption is made concerning the form of $B^{(1)}$, $E^{(1)}$ and $\mu^{(1)}$. Dropping the superscript ``$(1)$'' which indicates that we are dealing with small perturbations and introducing the notation $\mu^{(0)}\equiv\mu_l-\mu_r$ we rediscover in (\ref{4.14})-(\ref{4.17}) the system of equations derived in section 3.1 for constant chemical potentials $\mu_l$, $\mu_r$ and charge density $\langle j^0\rangle_{\beta,\vec{\mu}}=0$. These equations are linear in the fields $\vec{E}$ and $\vec{B}$ with constant coefficients. Such a system can be solved by means of Fourier transformation. Using lowercase $\vec{e}$ and $\vec{b}$ for the (spatially) Fourier transformed fields

\begin{xalignat}{2}
\vec{e}(t,\vec{k})&=\int d^3x \vec{E}(t,\vec{x})e^{-i\vec{k}\cdot\vec{x}}&\qquad \vec{b}(t,\vec{k})&=\int d^3x \vec{B}(t,\vec{x})e^{-i\vec{k}\cdot\vec{x}},
\label{4.19}
\end{xalignat}

the system to be solved becomes

\begin{xalignat}{2}
\vec{k}\cdot\vec{e}&=0&\qquad i\vec{k}\times\vec{e}+\partial_0\vec{b}&=0\label{4.20}\\
\vec{k}\cdot\vec{b}&=0&\qquad i\vec{k}\times\vec{b}-\partial_0\vec{e}&=-\frac{\alpha}{4\pi^2}(\mu_l-\mu_r)\vec{b}.\label{4.21}
\end{xalignat}

If we choose $\vec{k}=(0,0,k)$ in the 3-direction, then it follows from the above equations that only the 2,3-components of $\vec{e}$ and $\vec{b}$ can be non-zero. The time evolution of these remaining four components is determined by the differential equation

\begin{equation}
\partial_0\left(\begin{array}{c}e_1\\e_2\\b_1\\b_2\end{array}\right)=
\left(\begin{array}{cccc}0&0&\frac{\alpha}{4\pi^2}(\mu_l-\mu_r)&-ik\\0&0&ik&\frac{\alpha}{4\pi^2}(\mu_l-\mu_r)\\
                         0&ik&0&0\\-ik&0&0&0\end{array}\right)
\left(\begin{array}{c}e_1\\e_2\\b_1\\b_2\end{array}\right).
\label{4.22}
\end{equation}

The eigenvalues of the above matrix are

\begin{equation}
(\text{Eigenvalues})^2=\left\{-k^2\pm k\frac{\alpha}{4\pi^2}(\mu_l-\mu_r)\right\},
\label{4.23}
\end{equation}

from which it follows that a real, positive eigenvalue - and hence an exponentially growing solution of equation (\ref{4.22}) - exists for 

\begin{equation}
\boxed{|\vec{k}|\le\frac{\alpha}{4\pi^2}|\mu_l-\mu_r|.}
\label{4.24}
\end{equation}

In anticipation of this result we mentioned on several occasion that if in the early universe there existed a slight asymmetry in the chemical potentials of left- and right-handed fermions (or equivalently a space-independent axion field which was growing at a constant rate), this might have lead to the generation of large, cosmic magnetic fields.








\subsection{Oscillating axion field and parametric resonance}

Since $U[\phi=\theta]$ for $\theta=const$ is a periodic function of $\theta$ (see section 3.4.2), the space-independent solution of the equation $\Box\phi=-\alpha U'[\phi]$ is either oscillating around $\phi=0$ or linearly increasing/decreasing with periodic modulations superimposed. In both cases the time derivative $\dot \phi$ is a periodic function of time, with vanishing mean value for the oscillating $\phi$ and strictly positive/negative for the increasing/decreasing fields.
\vspace{10mm}

We therefore consider the equations of motion

\begin{align}
\vec{\nabla}\cdot\vec{E}&=0\label{4.25}\\
\vec{\nabla}\times\vec{B}-\partial_0\vec{E}&=\frac{l\alpha}{4\pi^2}\dot \phi\vec{B}\label{4.26}\\
\vec{\nabla}\cdot\vec{B}&=0\label{4.27}\\
\vec{\nabla}\times\vec{E}+\partial_0\vec{B}&=0,\label{4.28}
\end{align} 

where $\dot\phi$ is space-independent and periodic in time. Again, we will work with the Fourier transformed versions of the above equations and choose $k=(0,0,k)$. From equations (\ref{4.27}) and (\ref{4.28}) it then follows that the 3-components of the transformed fields $\vec b(x^0,\vec{k})$ and $\vec e(x^0,\vec{k})$ vanish. The system of equations for the remaining four components is

\begin{equation}
\partial_0\left(\begin{array}{c}e_1\\e_2\\b_1\\b_2\end{array}\right)=
\left(\begin{array}{cccc}0&0&q&-ik\\0&0&ik&q\\
                         0&ik&0&0\\-ik&0&0&0\end{array}\right)
\left(\begin{array}{c}e_1\\e_2\\b_1\\b_2\end{array}\right),
\label{4.29}
\end{equation}

where we have put $q=-\frac{l\alpha}{8\pi^2}\dot \phi$ for simplicity. The $4\times 4$ matrix in (\ref{4.29}) - let us call it $A$ - can be brought to the following block-diagonal form by an appropriate change of basis: 

\begin{equation}
P^{-1}AP=
\left(\begin{array}{cccc}0&1&0&0\\-k^2-2kq&0&0&0\\
                         0&0&0&1\\0&0&-k^2+2kq&0\end{array}\right),\hspace{0mm}
P=
\frac{1}{2}\left(\begin{array}{cccc}i(1-k)&2i&i(1-k)&2i\\-1+k&-2&1-k&2\\
                         -i(1+k)&0&i(1+k)&0\\1+k&0&1+k&0\end{array}\right).
\label{4.30}
\end{equation}

We shall henceforth denote the $2\times 2$ block in the upper left corner of the matrix $P^{-1}AP$ by $M$ and the other one by $N$. 

At this point we remember that $q$ is a periodic function of $x^0=t$. We will now consider a particular example and choose $q=\cos(t)$, which could be regarded as the solution obtained from a parabolic potential approximating $V(\phi)$ near $\phi=0$. This approximation is valid for small oscillations. 

The problem which we are studying is then equivalent to the \textit{Mathieu equation}

\begin{equation}
\frac{d^2y}{dt^2}+(a+2b \cos(t))y=0,
\label{4.31}
\end{equation}

which written as a system of first order differential equations reads

\begin{equation}
\partial_0\left(\begin{array}{c}y\\z\end{array}\right)=
\left(\begin{array}{cc}0&1\\-a-2b\cos(t)&0\end{array}\right)
\left(\begin{array}{c}y\\z\end{array}\right).
\label{4.32}
\end{equation}

\begin{figure}[p]
\noindent
\begin{minipage}[b]{0.5\linewidth}
\centering\epsfig{figure=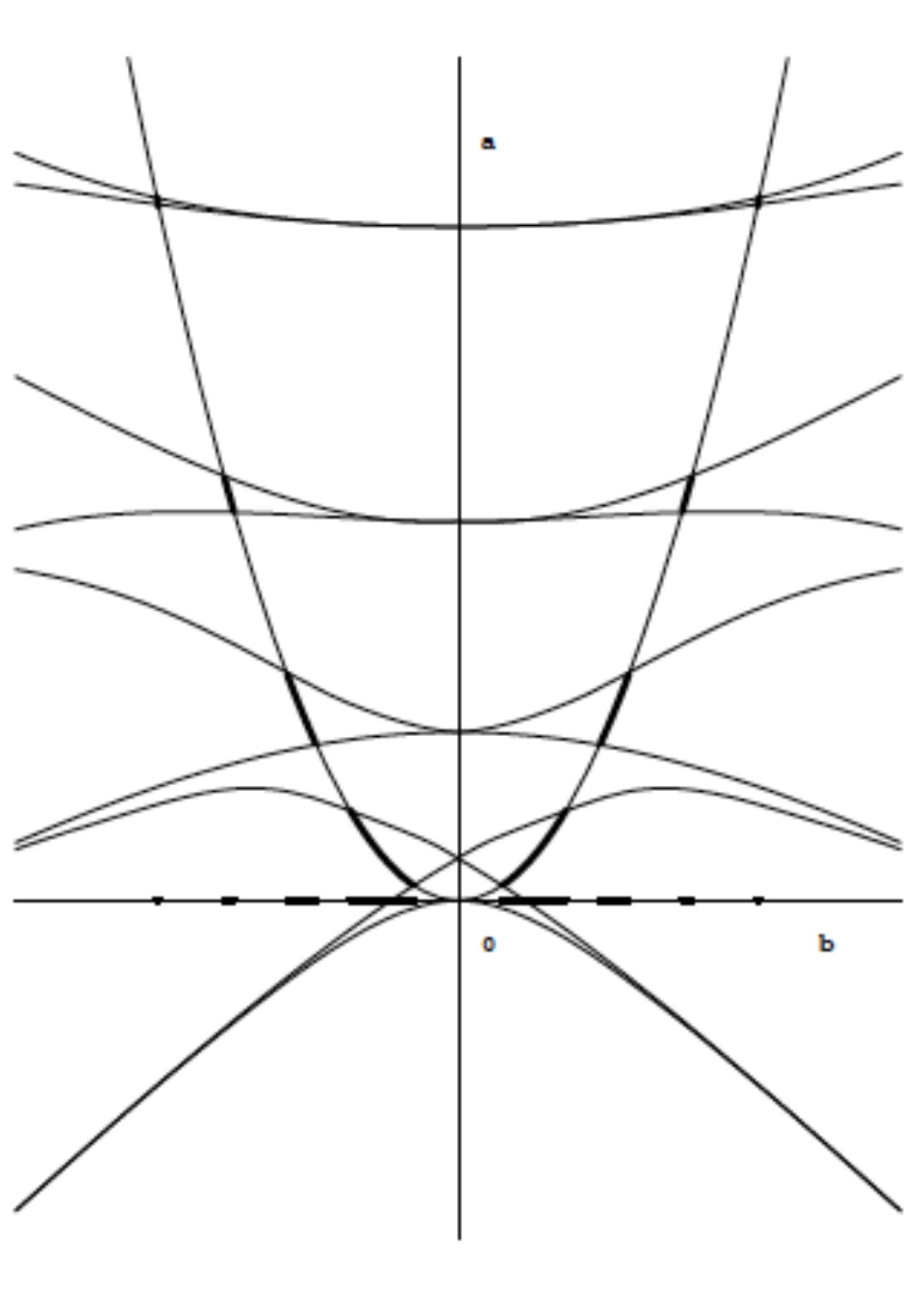,width=0.8\linewidth}\\
\small{$a=b^2$}
\end{minipage} \hfill
\begin{minipage}[b]{0.5\linewidth}
\centering\epsfig{figure=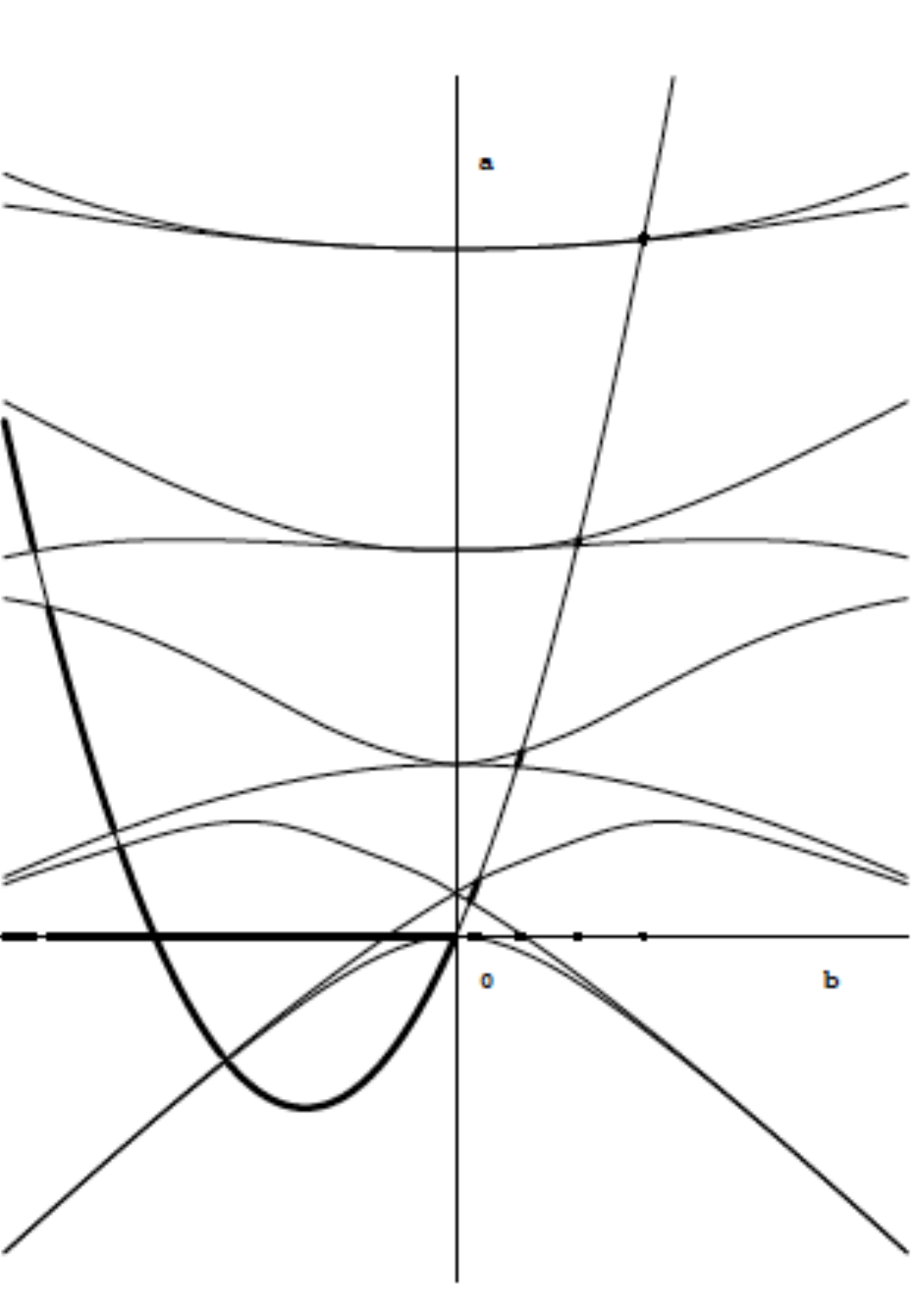,width=0.8\linewidth}\\
\small{$a=b^2+2b$}
\end{minipage}
\caption{Graphical method for the determination of the instable $k$-intervals. The thin lines show the stability boundaries of the Mathieu equation (\ref{4.31}) as a function of the parameters $a$ and $b$. On these boundaries, the solution is periodic. The heavy lines show the intersection of the parabola $a=b^2$ ($a=b^2+2b$) with the instable regions. The corresponding instable $k$-intervals are found after projection onto the $b$-axis, which is identified with $k$ for $b>0$ and $-k$ for $b<0$.}
\label{mathieu_1}
\end{figure}

\begin{figure}[p]
\noindent
\begin{minipage}[b]{0.5\linewidth}
\centering\epsfig{figure=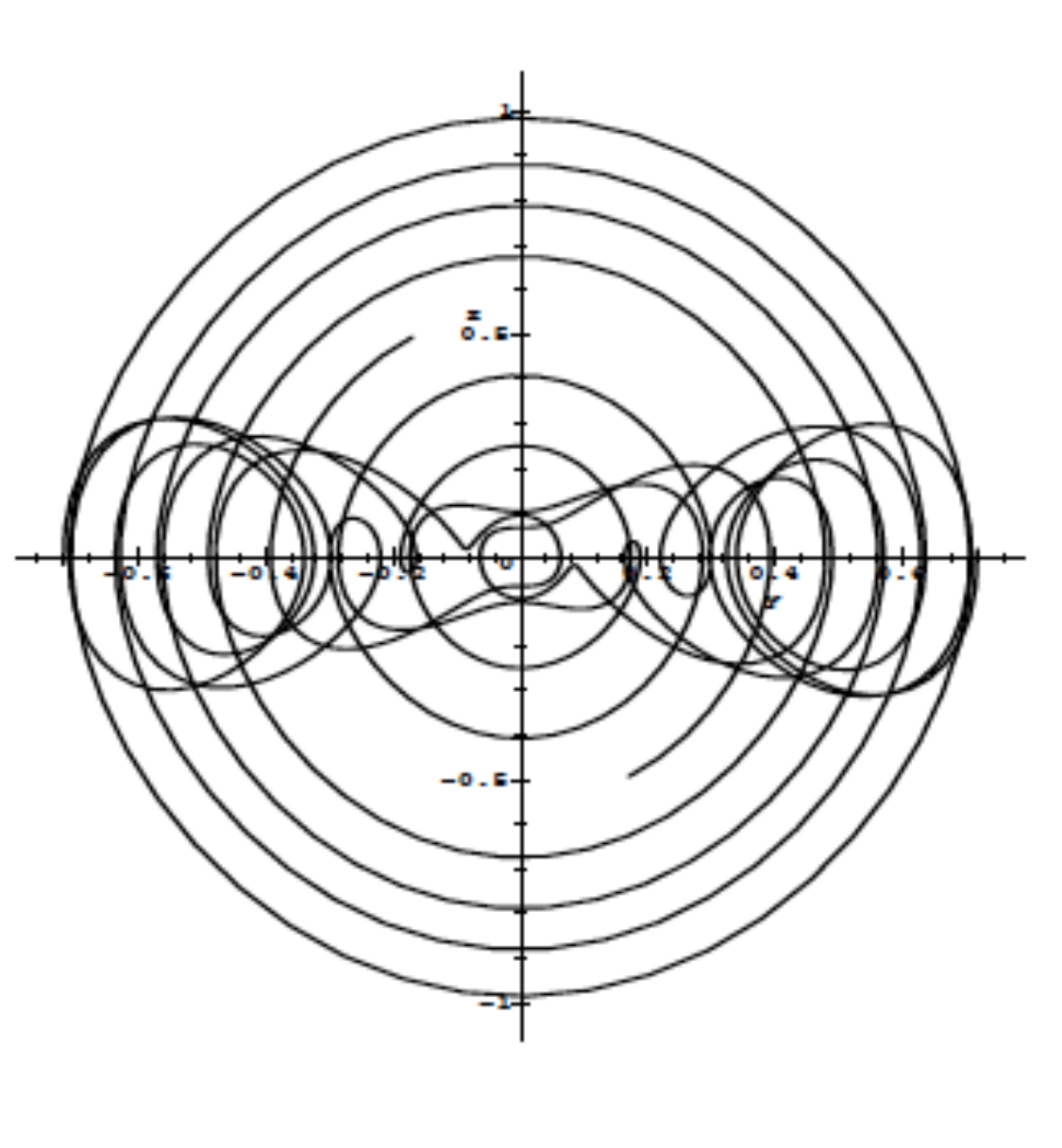,width=0.9\linewidth,height=7cm}
\small{$k=0.8$}
\end{minipage} \hfill
\begin{minipage}[b]{0.5\linewidth}
\centering\epsfig{figure=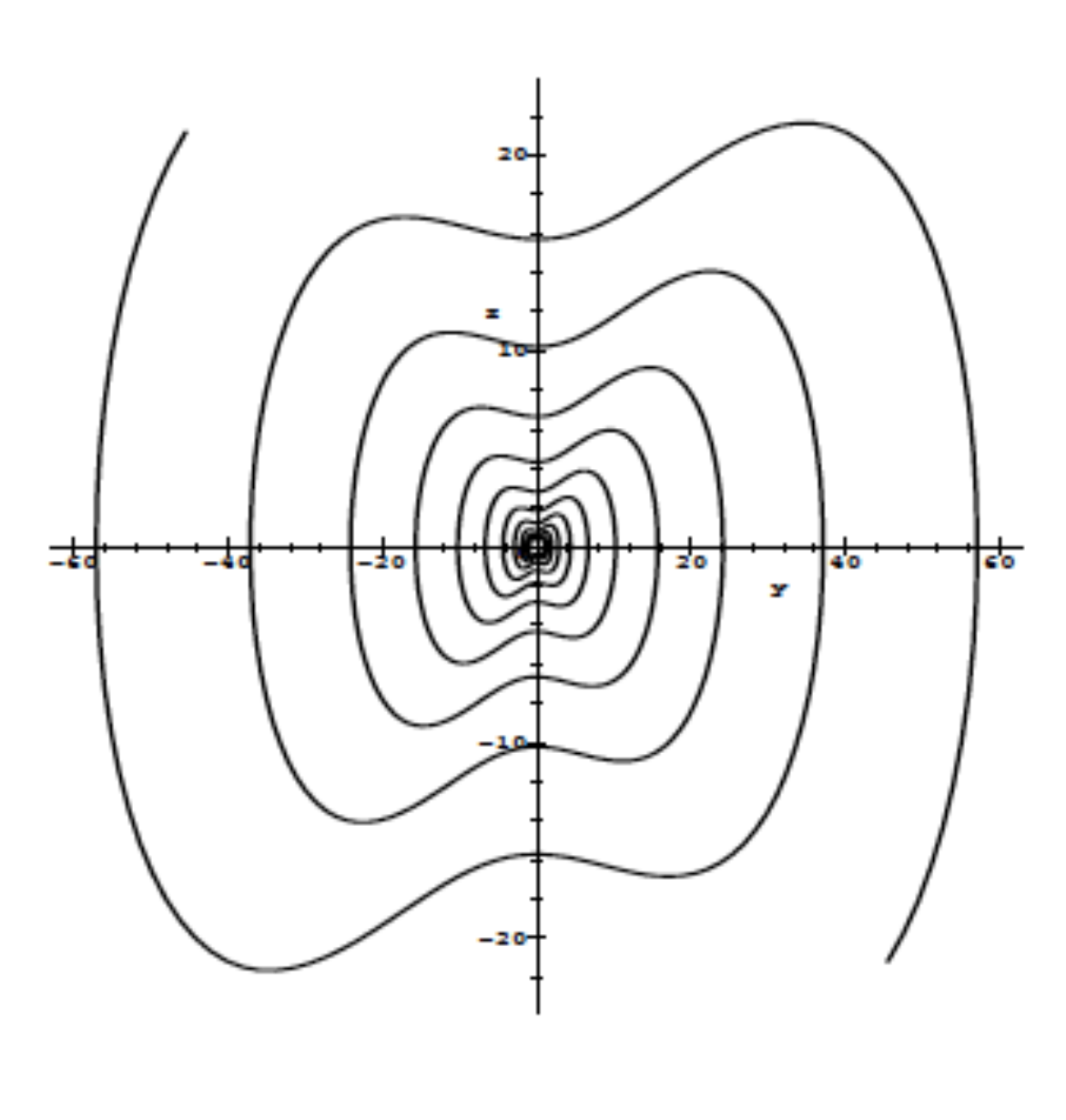,width=0.9\linewidth,height=7cm}
\small{$k=0.2$}
\end{minipage}
\caption{The figure on the left shows a stable, but non-periodic solution of the differential equation $\partial_0(y,z)=M(y,z)$, where $M$ is the $2\times 2$ matrix defined in the text following equation (\ref{4.30}). The initial positions at $t=0$ have been chosen $(y=0.7,z=0)$ and $(y=-0.7,z=0)$.
On the right hand side we plot a solution for a different value of the parameter $k=|\vec{k}|$, which illustrates how the corresponding Fourier mode grows by parametric resonance.}
\label{mathieu_2}
\end{figure}

Depending on the parameter values, the solution to the Mathieu equation (\ref{4.31}) is stable or unstable. A plot of the stability boundaries as a function of the parameters $a$ and $b$ can be found for example in \cite{Mechel} (see also figure \ref{mathieu_1}). 

We now compare the $2\times 2$-matrix in (\ref{4.32}) with the $2\times 2$ blocks $M$ and $N$ appearing in the matrix (\ref{4.30}), which determines the time evolution of the electromagnetic field. The matrix $M$ with $q=\cos(t)$ yields the Mathieu equation for the parameters $a=k^2$ and $b=k$, while the values corresponding to $N$ are $a=k^2$ and $b=-k$. The values of $k=|\vec{k}|$ for which the Fourier coefficients of the electromagnetic field grow by parametric resonance can thus be determined graphically from the plot in figure \ref{mathieu_1}. The instable $k$-intervals are obtained by intersecting the parabola $a=b^2$ with the unstable regions and subsequent projection onto the $b$-axis.

Similarly, for the periodic function $q=\cos(t)+d$ (d a constant), which for $|d|\ge 1$ could be regarded as the solution corresponding to a monotonically increasing/decreasing axion field $\phi(t)$, one obtains the intervals of instability by intersecting the instable regions in figure \ref{mathieu_1} with the parabola $a=b^2+2db$. As before, we have $b=k$ for positive $b$ and $-b=k$ for negative $b$, so the relevant intervals are found by reflecting the $(-b)$-axis at the origin and taking the union of the contributions from both semi-axes. 

We conclude that in either case the solutions will be unstable for certain intervals of $k$ and therefore the magnetic field is growing. The mechanism however is a new one. In the example discussed in section 4.1.1, the growth resulted from exponentially increasing Fourier coefficients (for small enough values of $k$), while this time we found an infinite number of intervals, for which the Fourier coefficients grow by parametric resonance.

It seems reasonable that this general picture remains valid if we consider more complicated periodic functions $q$. In particular, since the instable regions for $q=\cos(t)$ were found to be intervals and not isolated points, they are stable to small perturbations of the periodic function.

\section{Space dependent special solutions of finite energy}

Equations (\ref{4.1})-(\ref{4.5}) are Lagrangian equations of motion. They were derived from the action functional (\ref{3.177}) by setting $W=0$. The Lagrangian density does not depend on time explicitely. Therefore, there exists a conserved energy functional $\mathcal{E}[A,\phi]$. The solutions discussed in the preceeding section were unrealistic in the sense that they corresponded to an axion field (and growing electromagnetic fields) of infinite energy. The instabilities in the time evolution of the electromagnetic field are due to a reshuffling of energy from axionic to electromagnetic degrees of freedom \cite{J-F}.

In this last section we shall argue that the mechanism for the generation of seed magnetic fields based on growth by parametric resonance also works in systems of finite energy. 
The energy density derived from the action functional (\ref{3.177}) contains terms in $\dot\phi^2$, $(\vec{\nabla}\phi)^2$, $\vec{E}^2$ and $\vec{B}^2$.
Hence $\vec{E}$, $\vec{B}$ and $\partial_\mu\phi$ have to fall off at infinity if the total energy stored in the fields is to be finite. This explains the necessity to consider space-dependent solutions.

\subsection{Sperically symmetric solutions with magnetic monopoles}

A wealth of special solutions could be derived by allowing for magnetic monopoles in the early universe. Writing $\partial_\mu \tilde F^{\mu\nu}=\tilde J_m^\nu$, where $\tilde J_m^\mu=(\rho_m,\vec{J}_m)$ is the magnetic current density, the system of equations to be solved becomes

\begin{align}
\partial_\mu F^{\mu\nu}&=-\frac{l\alpha}{4\pi^2}[\phi\partial_\mu F^{\mu\nu}+\partial_\mu\phi\tilde F^{\mu\nu}]\label{4.33}\\
\Box\phi&=-\frac{l\alpha}{32\pi^2}\ast(F\wedge F)-\alpha U'[\phi]\label{4.34}\\
\partial_\mu \tilde F^{\mu\nu}&=\tilde J_m^\nu.\label{4.35}
\end{align}

In vector notation, (\ref{4.33}) reads

\begin{align}
\vec{\nabla}\cdot\vec{E}&=-\frac{l\alpha}{4\pi^2}\{\phi\vec{\nabla}\cdot\vec{B}+\vec{\nabla}\phi\cdot\vec{B}\}\label{4.36}\\
\vec{\nabla}\times\vec{B}-\partial_0\vec{E}&=\frac{l\alpha}{4\pi^2}\{(\vec{\nabla}\times\vec{E}+\partial_0\vec{B})\phi+\partial_0\phi\vec{B}+\vec{\nabla}\phi\times\vec{E}\}.\label{4.37}
\end{align}

Choosing a sherically symmetric axion field $\phi(r)$, the electric field $\vec{E}$ parallel to the magnetic field $\vec{B}$ and pointing in a radial direction, we can solve equations (\ref{4.36}) and (\ref{4.37}) by the ansatz

\begin{equation}
\vec{E}=-\frac{l\alpha}{4\pi^2}\phi\vec{B}.
\label{4.38}
\end{equation}

Substituting (\ref{4.38}) into (\ref{4.34}) yields

\begin{equation}
\Box{\phi}=-\frac{l^2\alpha^2}{16\pi^4}B^2\phi-\alpha U'[\phi].
\label{4.39}
\end{equation}

Given the form of the axionic potential $U$, an acceptable special solution for $\phi$ must be constructed such that

\begin{enumerate}
\item $B^2$ defined through (\ref{4.39}) is positive
\item \{$\phi,\vec{B}=B\vec e_r,\vec{E}=E\vec e_r$\} corresponds to a special solution of finite energy.
\end{enumerate}
 
However, all of these radial solutions require the presence of more or less strange looking distributions of magnetic (and electric) charge, which is the reason why we do not want to pursue this idea any further here.

\subsection{The sine-Gordon equation and an approximate solution of finite energy}

After these purely mathematical considerations, let us recall what we actually intended to explain: the growth of seed magnetic fields in a universe with no electromagnetic fields initially present. Therefore all the energy is initially stored in the axion field. These physical considerations lead us to look for particular solutions corresponding to small (better: vanishing) electromagnetic fields and to an axion field which is localized in space. 

We shall try to find spherically symmetric solutions for $\phi$, so that equation (\ref{4.3}) after linearisation around $\vec{E}=\vec{B}=0$ becomes

\begin{equation}
\left(\partial_0^2-\frac{2}{r}\frac{\partial}{\partial r}-\frac{\partial^2}{\partial r^2}\right)\phi=-\alpha U'[\phi].
\label{4.40}
\end{equation}

In section 3.4.2 we have shown that $U[\phi]$ is periodic in $\phi$ (for $\phi=\theta$ independent of $x$) and has a minimum at $\phi=0$. In order to work with a concrete example, we choose

\begin{equation}
\alpha U'[\phi]=\frac{a}{b}\sin(b\phi),
\label{4.41}
\end{equation}

where $a,b>0$. One idea would be to restrict ourselves to oscillations of small amplitude and to approximate $\frac{a}{b}\sin(b\phi)$ by $a\phi$, thereby obtaining the equation of motion

\begin{equation}
\left(\partial_0^2-\frac{1}{r}\frac{\partial^2}{\partial r^2}r\right)\phi=-a\phi.
\label{4.42}
\end{equation}

Assuming a time dependence of the form $e^{i\omega t}$ and setting $\phi=\frac{\varphi}{r}$ we find

\begin{equation}
\frac{\partial^2}{\partial r^2}\varphi=(a-\omega^2)\varphi.
\label{4.43}
\end{equation}

For $\omega^2>a$ there exists a solution which is finite at the origin and falls off like $\frac{1}{r}$ for \mbox{large $r$:}

\begin{equation}
\phi(r)=\frac{A}{r}\cos(\omega t)\sin(\sqrt{\omega^2-a}r).
\label{4.44}
\end{equation}

The constant $A$ must be chosen such that $A\sqrt{\omega^2-a}\ll 1$ since otherwise the approximation $\sin(\phi)\approx\phi$ is not valid. Unfortunately, the total energy associtated with the particular solution (\ref{4.44}) is infinite and we have to cut off the function at a certain value of $r$. The latter procedure comes down to introducing a thin double shell of positive/negative magnetic (and electric) charge inside of which the axion field is oscillating.
\begin{figure}[h]
\noindent
\begin{minipage}[b]{0.5\linewidth}
\centering\epsfig{figure=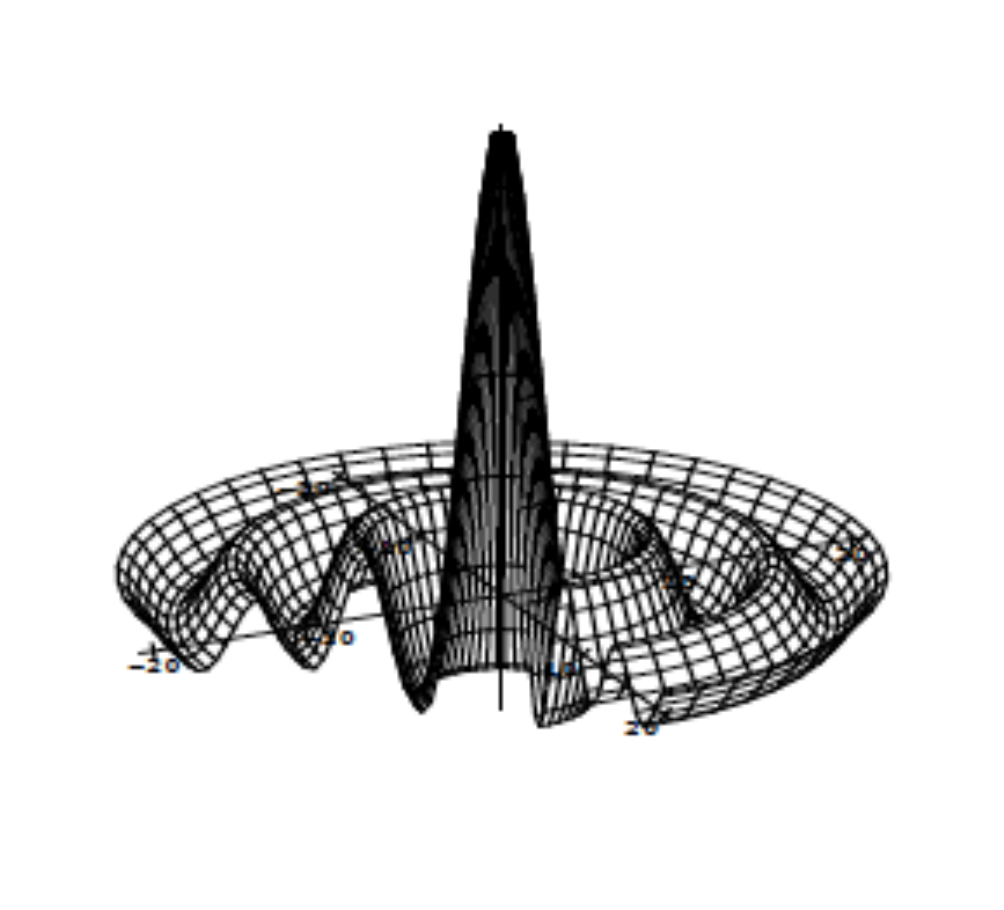,width=\linewidth,height=6.5cm}
\end{minipage} \hfill
\begin{minipage}[b]{0.5\linewidth}
\centering\epsfig{figure=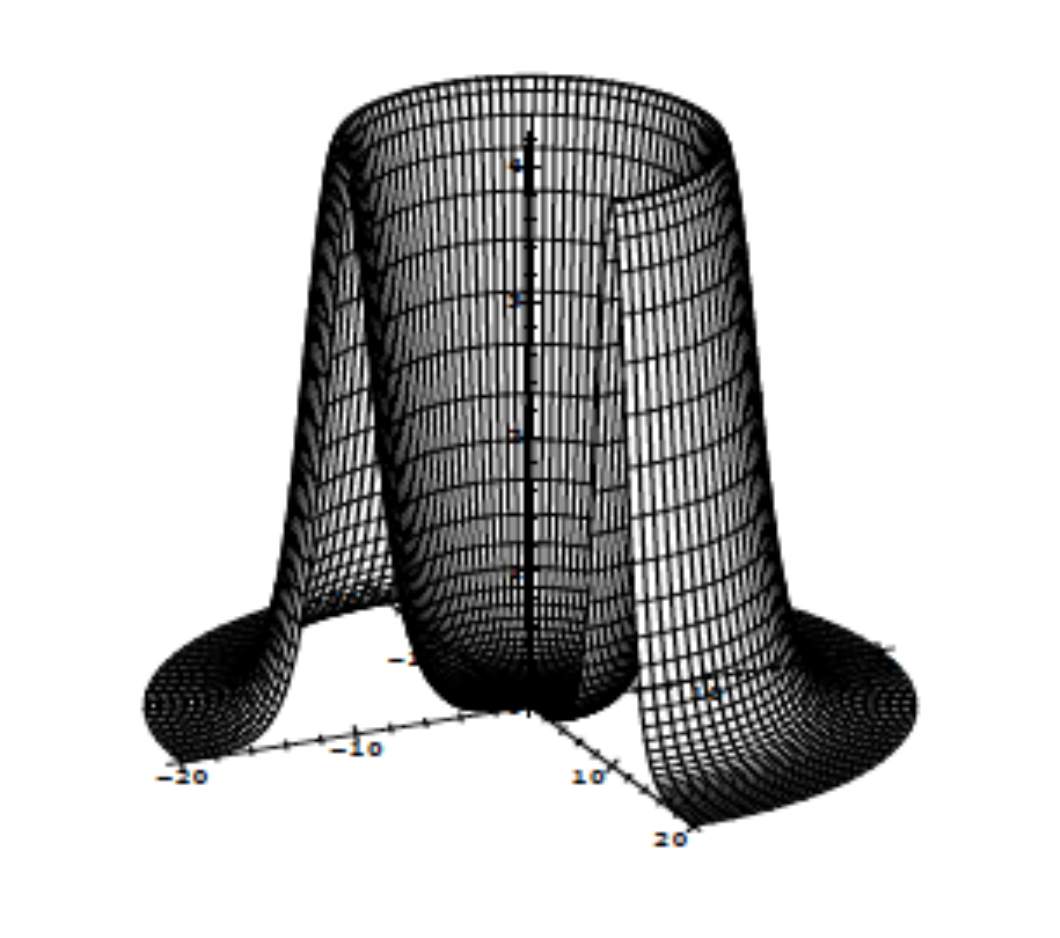,width=\linewidth,height=6.5cm}
\end{minipage}
\caption{The figure on the left shows the particular solution (\ref{4.44}) for $\sqrt{\omega^2-a}=1$ and $\omega t=\pi$. On the right is a plot of the approximate solution $\phi_{DHN}(r-10)$ for the parameters $a=b=1$, $\omega=\frac{1}{2}$ and $\omega t=\pi$.}
\label{DHN}
\end{figure} 

Another possibility is to keep the potential of the form (\ref{4.41}), but to neglect the term $-\frac{2}{r}\frac{\partial}{\partial r}$ in the Laplacian. This simplification is justified if $\phi$ is appreciable only far from the origin. The equation for $\phi$ then becomes the so-called \textit{sine-Gordon equation} (for a spherically symmetric function):\vspace{5mm}
\begin{equation}
(\partial_0^2-\partial_r^2)\phi=-\frac{a}{b}\sin(b\phi),
\label{4.45}
\end{equation}

whose analytic solution can be found in \cite{DHN} (see also \cite{Collins}). It reads

\begin{equation}
\phi_{DHN}(r,t)=\frac{4}{b}\text{arctan}\left(\frac{\eta\sin(\omega t)}{\cosh(\eta\omega r)}\right),\hspace{5mm}\eta=\frac{\sqrt{a-\omega^2}}{\omega},\hspace{5mm}0\le\omega\le\sqrt{a},
\label{4.46}
\end{equation}

and is translationally invariant (the subscript ``DHN'' stands for Dashen - Hasslacher - Neveu). An (approximate) special solution for $\phi$ which clearly satisfies the requirement of finite total energy is therefore

\begin{equation}
\phi(r,t)=\phi_{DHN}(r-r_0,t)
\label{4.47}
\end{equation}

for $r_0$ large enough. Of course, (\ref{4.47}) does not exactly satisfy equation (\ref{4.40}) for the potential (\ref{4.41}) and needs to be corrected in the vincinity of $r=0$ (where the neglected $\frac{2}{r}\frac{\partial}{\partial r}$-term is large) and $r=r_0$ (where $\phi_{DHN}(r-r_0)$ is large). However, it is probably not worth putting much effort into fixing this little deficiency, since our choice (\ref{4.41}) for the functional derivative of the axionic potential was purely arbitrary.

The purpose of this last section was merely to argue that solutions of finite energy may also give rise to growing magnetic fields. We believe that this can be seen from the above solution - even though time has not premitted us to attack the difficult problem of sol-ving the (linearised) system of partial differential equations with (now) space-dependent coefficients. In the neighbourhood of $r=r_0$, where the space-dependence of $\phi$ can be neglected, we find ourselves in the situation of section 4.1.2, which suggests that in the region of space, where the oscillating axion field is appreciable, the (electro-)magnetic field will again grow by parametric resonance.

\appendix

\chapter{General solution for the continuity equation in (3+1) dimensions}

In section 3.1.1 we have mentioned that the the general solution for the continuity equation $\partial_\mu J^\mu=0$ in (3+1)-dimensional Minkowski space-time reads

\begin{equation}
J^0=\vec{\nabla}\cdot\vec{A},\hspace{5mm}\vec{J}=-\partial_t\vec{A},
\label{A.1}
\end{equation}

where $\vec{A}$ denotes some vector field and $x^0\equiv t$. This not so obvious statement will be justified below. Assuming that $J^0$ falls off sufficiently rapidly at infinity we can write 

\begin{equation}
\vec{\nabla}\cdot\vec{A}=J^0.
\label{A.2}
\end{equation}

Equation (\ref{A.2}) determines the vector field $\vec{A}$, whose explicit expression is

\begin{equation}
\vec{A}(x)=-\vec{\nabla}_{\vec{x}}\int d^3y\frac{1}{4\pi|\vec{x}-\vec{y}|}J^0(t,\vec{y})+\vec{B}(x),\hspace{5mm}\vec{\nabla}\cdot\vec{B}(x)=0.
\label{A.3}
\end{equation}

The first term on the right hand side of (\ref{A.3}) is a particular solution of the inhomogeneous equation (\ref{A.2}), whereas the divergenceless field $\vec{B}(x)$ is the general solution of the homogeneous equation $\vec{\nabla}\cdot\vec{A}=0$. Imposing the condition $\partial_t\vec{A}=-\vec{J}$ yields an equation for $\vec{B}$:

\begin{equation}
\vec{J}(x)=-\partial_t\vec{A}(x)=-\vec{\nabla}_{\vec{x}}\int d^3y\frac{1}{4\pi|\vec{x}-\vec{y}|}\vec{\nabla}\cdot\vec{J}(t,\vec{y})-\partial_t\vec{B}(x),
\label{A.4}
\end{equation}

where we have made use of the continuity equation in order to replace $\partial_0J^0$ by $-\vec{\nabla}\cdot\vec{J}$. Equation (\ref{A.4}) can be solved by means of Fourier transformation. Denoting the spacially Fourier transformed fields by lower case $j$ and $b$, we obtain $\vec{j}(t,\vec{k})=\frac{\vec{k}}{|\vec{k}|^2}(\vec{k}\cdot\vec{j}(t,\vec{k}))-\partial_t\vec{b}(t,\vec{k})$, that is

\begin{equation}
-\partial_t\vec{b}(t,\vec{k})=\vec{j}(t,\vec{k})-\frac{\vec{k}}{|\vec{k}|^2}(\vec{k}\cdot\vec{j}(t,\vec{k}))=\vec{j}_{\text{transversal}}(t,\vec{k}).
\label{A.5}
\end{equation}

Applying the inverse Fourier transformation to equation (\ref{A.5}) yields $\partial_t\vec{B}(t,\vec{x})=-\vec{J}_{\text{transversal}}(t,\vec{x})$, so the vector field $\vec{B}(x)$ can be defined as

\begin{align}
\vec{B}(t,\vec{x})&=-\int_0^td\tau\vec{J}_{\text{transversal}}(\tau,\vec{x})\label{A.6}\\
\vec{J}_{\text{transversal}}(t,\vec{x})&=\vec{J}(t,\vec{x})-\partial_t\vec{\nabla}_{\vec{x}}\int d^3y\frac{1}{4\pi|\vec{x}-\vec{y}|}J^0(t,\vec{y}).\label{A.7}
\end{align}

Since $\vec{J}_{\text{transversal}}(t,\vec{x})$ is divergenceless we have $\vec{\nabla}\cdot\vec{B}(x)=0$, as it should. The substitution of (\ref{A.7}) and (\ref{A.6}) into (\ref{A.3}) yields the almost trivial result

\begin{equation}
\boxed{\vec{A}(t,\vec{x})=-\int_0^td\tau\vec{J}(\tau,\vec{x})-\vec{\nabla}_{\vec{x}}\int d^3y\frac{1}{4\pi|\vec{x}-\vec{y}|}J^0(0,\vec{y}),}
\label{A.8}
\end{equation} 

which proves that any current satisfying the continuity equation can be expressed in the form (\ref{A.1}).

\chapter{Gauge invariance and boundary currents in (2+1) dimensions}

\section{Gauge invariance}

In (2+1) dimensions, it is possible to write down the explicit expression for the effective action. The determinant associated with chiral fermions in (1+1) dimensions has been calculated in \cite{Leutwyler} (for the more general case of a curved space). In this appendix, we shall use the corresponding formula for $\Gamma_{\partial\Lambda}$ (see (\ref{B.3}) below) and show that the action functional obtained thereby is gauge invariant.

The effective action for the quantum Hall sample discussed in section 3.2.1 is given by 

\begin{equation}
S_{eff}[A]=-S_{CS}[A]+\Gamma_{\partial\Lambda}[A|_{\partial_\Lambda}],
\label{B.1}
\end{equation} 

where 

\begin{align}
S_{CS}[A]&=-\frac{\sigma_H}{4}\int_\Lambda A\wedge F\label{B.2}\\
\Gamma_{\partial\Lambda}[a=A|_{\partial\Lambda}]&=-\frac{\sigma_H}{4}\int_{\partial\Lambda}d^2x\left\{a_+(x)a_-(x)-a_-(x)\int_{\partial\Lambda}d^2y\frac{1}{\Box}(x-y)\partial_+^2a_-(y)\right\}.\label{B.3}
\end{align}

The following notations have been used in the formula for $\Gamma_{\partial\Lambda}$ (recall that the boundary of the Hall sample is given by $\partial\Lambda=\{(x|x^2=0)\cup(x|x^2=L)\}$):

\begin{xalignat}{3}
\partial_+&=\partial_0+\partial_1&\qquad \partial_-&=\partial_0-\partial_1&\qquad \Box&=\partial_+\partial_-\label{B.4}\\
a_+&=A_0+A_1&\qquad a_-&=A_0-A_1&\qquad&\label{B.5}
\end{xalignat}

and $\frac{1}{\Box}(x-y)$ is the operator defined through the relation

\begin{equation}
\Box_x\frac{1}{\Box}(x-y)=\delta(x-y).
\label{B.6}
\end{equation}

Under a gauge transformation $A\rightarrow A+d\theta$ the contribution $S_{CS}$ transforms as

\begin{equation}
S_{CS}[A+d\theta]=S_{CS}[A]+\frac{\sigma_H}{4}\int_{\partial\Lambda} \theta F.
\label{B.7}
\end{equation}

The latter result follows from a calculation similar to the one performed in section 3.2.3. For simplicity one may suppose that the gauge field $\theta(x)$ vanishes on the lower boundary of the Hall sample, that is 

\begin{equation}
\theta(x)|_{x^2=0}=0.
\label{B.8}
\end{equation}

In this case the only contribution to the integral over $\partial\Lambda$ comes from the upper boundary $x^2=L$ and the second term on the right hand side of (\ref{B.7}) may be written explicitly as

\begin{equation}
\frac{\sigma_H}{4}\int_{\partial\Lambda} \theta F=\frac{\sigma_H}{2}\int_{x^2=L}dx^0dx^1 \theta(\partial_0A_1-\partial_1A_0). 
\label{B.9}
\end{equation} 

The boundary action $\Gamma_{\partial\Lambda}$ transforms as follows:

\begin{align}
\Gamma_{\partial\Lambda}[(A+d\theta)|_{\partial\Lambda}]&=\Gamma_{\partial\Lambda}[A|_{\partial\Lambda}]-\frac{\sigma_H}{4}\int_{\partial\Lambda}d^2x\Big\{a_+(\partial_-\theta)+(\partial_+\theta) a_++(\partial_+\theta)(\partial_-\theta)\nonumber\\
&\hspace{4mm}-a_-\frac{1}{\Box}\ast(\partial_+^2\partial_-\theta)-(\partial_-\theta)\frac{1}{\Box}\ast\partial_+^2a_--(\partial_-\theta)\frac{1}{\Box}\ast(\partial_+^2\partial_-\theta)\Big\}.
\label{B.10}
\end{align}

The symbol ``$\ast$'' denotes a convolution product. For $\theta(x)$ specified in (\ref{B.8}) this becomes

\begin{align}
\Gamma_{\partial\Lambda}[(A+d\theta)|_{\partial\Lambda}]&=\Gamma_{\partial\Lambda}[A|_{\partial\Lambda}]-\frac{\sigma_H}{4}\int_{\partial\Lambda}d^2x\Big\{a_+(\partial_-\theta)+(\partial_+\theta) a_++(\partial_+\theta)(\partial_-\theta)\nonumber\\
&\hspace{4mm}-a_-(\partial_+\theta)-(\partial_+\theta)a_--(\partial_-\theta)(\partial_+\theta)\Big\}.
\label{B.11}
\end{align}

where we have used $\partial_+\partial_-=\Box$, formula (\ref{B.6}) and $\partial_x\frac{1}{\Box}(x-y)=-\partial_y\frac{1}{\Box}(x-y)$. The expression in between the curly brackets in (\ref{B.11}) reduces to 

\begin{equation}
\{...\}=-2(A_0\partial_1\theta-A_1\partial_0\theta).
\label{B.12}
\end{equation}

After partial integration one therefore obtains

\begin{equation}
\frac{\sigma_H}{4}\int_{\partial\Lambda}d^2\xi\{...\}=-\frac{\sigma_H}{2}\int_{x^2=L}dx^0dx^1 \theta(\partial_0A_1-\partial_1A_0).
\label{B.13}
\end{equation}

From (\ref{B.7}), (\ref{B.9}), (\ref{B.11}) and (\ref{B.13}) we find

\begin{align}
S_{CS}[A+d\theta]&=S_{CS}[A]+\frac{\sigma_H}{2}\int_{x^2=L}dx^0dx^1 \theta(\partial_0A_1-\partial_1A_0)\label{B.14}\\
\Gamma_{\partial\Lambda}[(A+d\theta)|_{\partial\Lambda}]&=\Gamma_{\partial\Lambda}[A|_{\partial\Lambda}]+\frac{\sigma_H}{2}\int_{x^2=L}dx^0dx^1 \theta(\partial_0A_1-\partial_1A_0),
\label{B.15}
\end{align}

which proves that the effective action defined in (\ref{B.1}) is gauge invariant.

\section{Boundary currents}

For an arbitrary $A(x)$, the boundary action $\Gamma_{\partial\Lambda}$ yields a boundary current $j^\mu=\frac{\delta\Gamma_{\partial\Lambda}}{\delta A_\mu}$. Using the explicit formula (\ref{B.3}) for the boundary action of a (2+1)-dimensional quantum Hall sample one finds

\begin{eqnarray}
j_+&\equiv& j^0+j^1=\frac{\delta \Gamma_{\partial\Lambda}}{\delta a_+}=a_-\label{B.16}\\
j_-&\equiv& j^0-j^1=\frac{\delta \Gamma_{\partial\Lambda}}{\delta a_-}=a_+-\frac{2}{\Box}\ast(\partial_+^2a_-)\label{B.17}
\end{eqnarray}

and the boundary current density is therefore given by

\begin{eqnarray}
j^0&=&a_0-\frac{1}{\Box}\ast(\partial_+^2a_-)\label{B.18}\\
j^1&=&-a_1+\frac{1}{\Box}\ast(\partial_+^2a_-).\label{B.19}
\end{eqnarray}

Similarly in (4+1) dimensions $\Gamma_{\partial\Lambda}$ produces a current density on the (3+1) dimensional boundary of the slab.

\end{document}